\documentclass[a4paper,superscriptaddress,11pt]{article}
\usepackage{authblk}

\title{The Hamiltonian constraint in Polymer Parametrized field theory}
\author[a,b,c,]{Alok Laddha}
\author[c]{Madhavan Varadarajan}

\affil[a]{Institute for Gravitation and the Cosmos\\
 Pennsylvania State University,  University Park, PA 16802-6300, U.S.A} 
\affil[b]{Chennai Mathematical Institute,\\ SIPCOT IT Park, Padur PO, Siruseri 603103, India}
\affil[c]{Raman Research Institute,\\ Bangalore-560 080, India}

\begin{document}

\maketitle

\thispagestyle{empty}

\let\oldthefootnote\thefootnote
\renewcommand{\thefootnote}{\fnsymbol{footnote}}
\footnotetext{Email:alok@gravity.psu.edu,madhavan@rri.res.in}
\let\thefootnote\oldthefootnote

\begin{abstract}
Recently, a generally covariant reformulation of 2 dimensional flat spacetime
free scalar field theory known as Parameterised Field Theory
was quantized using Loop Quantum Gravity (LQG) type `polymer' representations. Physical states were constructed, 
without intermediate regularization structures, by averaging over the 
group of gauge transformations generated by the constraints, the constraint algebra being a Lie algebra.
We consider classically equivalent combinations of these constraints corresponding to a diffeomorphism and
a Hamiltonian constraint, which, as in gravity,  define a Dirac algebra. Our treatment of the quantum 
constraints parallels that 
of LQG and obtains the following results, expected to be of use in the construction of the quantum dynamics of LQG:
(i) the (triangulated) Hamiltonian constraint 
acts only on vertices, its construction involves some of the same
ambiguities as in LQG and its  action on diffeomorphism 
invariant states admits a continuum limt 
(ii)if the regulating holonomies are in  representations tailored to the
edge labels of the state, all previously obtained physical states
 lie in the kernel of the Hamiltonian constraint, 
(iii) the commutator of two (density weight 1) Hamiltonian
constraints as well as the operator correspondent of their classical Poisson bracket
 converge to zero in the continuum limit 
defined by diffeomorphism invariant states, and
vanish  on the Lewandowski- Marolf (LM) habitat 
(iv) the rescaled density 2 Hamiltonian constraints and their
commutator are ill defined on the LM habitat despite the well 
defined- ness of the operator correspondent of their classical Poisson bracket there 
(v) there is a new habitat which supports a non-trivial representation of the Poisson- Lie algebra of 
density 2 constraints

%

\end{abstract}

\section{Introduction}
One of the key open problems in canonical LQG is a satisfactory treatement of the Hamiltonian constraint operator.
Problems stem from the 
tension between the local nature of the Hamiltonian constraint and the non- local nature of some of the 
basic operators used in its construction. As a result, intermediate regularization structures have to be introduced 
and the final operator definition depends on the (infinitely manifold) choice of the regulating structures.
In our opinion not much progress has been made on this issue in the canonical theory.
Confronted with such a situation, we believe that the availablity of a good toy model would go a long way in 
testing various proposals for the Hamiltonian constraint and suggest avenues to restrict the choices in its construction.

In recent work \cite{alokme1,alokme2} we have developed just such a toy model which we call Polymer Parameterised
Field Theory. 
In \cite{alokme2} we used polymer representations and Group Averaging techniques to construct the physical Hilbert space
of the model and demonstrated that the quantization did encode the right classical limit. 
Here we use the arena provided
by Reference \cite{alokme2} to explore issues related to the definition of the Hamiltonian constraint operator.

Polymer Parameterised Field Theory is  an LQG- type ``polymer'' quantization of classical Parameterised Field Theory
(PFT) on the Minkowskian cylinder. PFT was introduced by Dirac \cite{dirac} and its use as a
toy model for quantum gravity was pioneered by Kucha{\v r}.
PFT is just free field theory on flat spacetime, cast in a diffeomorphism invariant disguise. It offers an elegant 
description of free scalar field evolution on {\em arbitrary} (and in general curved) foliations of the 
background spacetime by treating the `embedding variables' which describe the foliation as dynamical variables
to be varied in the action in addition to the scalar field. Specifically, let $X^A= (T,X)$ denote inertial coordinates
on 2 dimensional flat spacetime. In PFT, $X^A$ are parametrized by a new set of arbitrary coordinates
 $x^{\alpha}=(t,x)$
such that for fixed $t$, the embedding variables $X^A(t,x)$ define a spacelike Cauchy slice of flat spacetime.
General covariance of PFT ensues from the arbitrary choice of $x^{\alpha}$ and implies that in its canonical 
description, evolution from one slice of an arbitrary foliation to another is generated by a pair of constraints.
Its field theoretic nature, general covariance and the fact that the dynamics of the true degrees of freedom is just that 
of a free scalar field make PFT a good toy model for gravity.

In \cite{alokme2} we took advantage of
the simplicity of PFT when expressed in terms of light cone variables $X^{\pm}(x)= T(x)\pm X(x)$ and right and left moving
matter variables to solve the quantum theory. Specifically, we used density weight 2 constraints, $H_+$ and $H_-$.
$H_+$ generates
the dynamics of left moving matter fields and advances the foliation along `$+$' null direction and $H_-$ generates the 
dynamics  for right movers and advances the foliation  along the `$-$' direction.
However, one can equally well use as constraints, the generators of motions along the
Cauchy slice and normal to the Cauchy slice (the flat spacetime metric defines this normal).
We shall refer to these, for obvious reasons, as the diffeomorphism and Hamiltonian constraints of the model.
These constraints are appropriate combinations of the density weight 2 ones  and 
their Poisson brackets yield a Dirac algebra
exactly as is the case for the spatial diffeomorphism and Hamiltonian constraints of 4-d gravity. Specifically, the 
Possion bracket between 2 Hamiltionian constraints yields the diffeomorphism constraint smeared by a vector field
which involves the induced spatial metric on the Cauchy slice. The spatial metric is constructed
 from the canonical embedding data and hence the constraint algebra is a Dirac algebra.
In contrast the density 2 constraints form a Lie algebra and this fact was used in Reference \cite{alokme2} 
to construct physical states by Group Averaging \cite{grpavg}. The Group Averaging technique only uses the structure of
the group of 
(unitary representations of)  finite gauge transformation and, hence,  does not require
any auxilliary regularization structures thus yielding an unambiguous construction of the physical Hilbert space of 
the theory.

Here we follow the strategy used in LQG to construct the quantum dynamics of the model 
and isolate the space of physical states.
In doing so we find
a remarkably close  structural similarity to corresponding constructions in LQG. 
We solve the diffeomorphism constraint by group averaging.
Then we construct  the Hamiltonian constraint operator at finite triangulation, 
show that it has a finite action on states in the kinematic Hilbert space and that its dual action
admits a well defined continuum limit on diffeomorphism invariant states.
Holonomies  play a crucial role in the construction and there is an ambiguity in the choice of their  representation 
just as in LQG \cite{perez}. The inverse square root of the determinant of the spatial metric also plays an essential
role in the finiteness of the action of the Hamiltonian constraint and we show that, similar to LQG, it acts only at
vertices.

Next we enquire if the physical states constructed in \cite{alokme2} also solve the diffeomorphism and Hamiltonian 
constraints as defined above. It happens to be  straightforward to see that these states solve the diffeormorphism 
constraint. However, contrary to the expectations in LQG,  they are {\em not} normalizable in the Hilbert space
inner product obtained through the Group Averaging procedure applied to the diffeomorphism constraint.
While we do not claim the existence of a proof, it does seem unlikely from the structure of the Hamiltonian constraint
and that of the physical states of Reference \cite{alokme2} that an LQG type choice of 
regulating holonomies in a fixed weight representation would result in a Hamiltonian constraint operator which 
annihilates all the physical states of Reference \cite{alokme2}. What we do show is that 
there does exist a regularization choice in which  
the holonomy labels are chosen to {\em depend on the edge labels of the state} in  just the right way so that 
all physical states of \cite{alokme2} are annhilated by the Hamiltonian constraint. 

Next we turn our attention to the quantum constraint algebra. 
We evaluate the commutator between the smeared density weight one Hamiltonian constraints using the topology provided by 
the arena of
diffeomorphism invariant states along the lines of Thiemann's seminal work \cite{tthh}. We find that this commutator
vanishes as in LQG. We then 
introduce a habitat similar to that of Lewandowski and Marolf \cite{donjurek,lmhabitat2} and 
construct the smeared density weight one Hamiltonian constraint as an operator on this `LM' habitat.
We show that, as in LQG,
the commutator between the 
density weight one Hamiltonian constraints as well as the operator corresponding to
the classical right hand side of the corresponding Poisson bracket, all vanish on this `LM' habitat.
Our computations clearly indicate that the constraint algebra trivialises in this manner due to the 
density weight one character of the Hamiltonian constraint. Hence, we turn to an analysis of slightly
``more singular'' operators obtained by rescaling the density weight 1 operator by the 
determinant of the spatial metric to obtain a density weight 2 Hamiltonian constraint.
We show that
neither this density 
weight 2 Hamiltonian constraint nor the commutator of two such constraints is   well defined on the  LM habitat. 
However 
the operator corresponding to the Poisson bracket of the corresponding classical quantities,
{\em is} well defined on the LM habitat and, in this sense, 
the algebra of these constraints is anomalous on the LM habitat.
\footnote{We also show that there is a subspace of the LM habitat on which the commutator  trivialises
but the operator correspondent of its Poisson bracket is non- trivial, thus reinforcing our view
that the algebra is anomalous on the LM habitat.}

Finally,  we introduce a new habitat geared
to the physical state space constructed in \cite{alokme2}; these are ``vertex smooth'' generalizations of the 
physical states in \cite{alokme2}. We show that the smeared density weight two Hamiltonian (and 
spatial diffeomorphism) constraints are well defined on this
new habitat and satisfy the correct constraint algebra. 

All these results have important 
repercussions for LQG. Chief among them are (i) one should consider the possibility of allowing the representations of 
regulating holonomies to be state dependent
(ii) the lack of weak continuity of operators on the kinematic
Hilbert space is not necessarily a hindrance to defining their generators on an appropriate space of 
distributions through the mechanisms of triangulation and continuum limit of dual actions
(iii) in order to analyse the quantum constraint algebra, 
it may be profitable to look beyond smeared density weight one operators towards ones which are ``more singular''
i.e. of higher density weight. We shall discuss these points as well as other possible lessons for LQG.
in the concluding section of this paper.
There is still much to be learnt about the structure of the constraint algebras in this model. The
close structural similarity with LQG ensures  
that the lessons learnt will provide strategies to probe the constraint algebra in LQG at a deeper level than the seminal 
works of \cite{tthh,donjurek}.
\footnote{Indeed, this work has already motivated a definition of the operator corresponding to 
(the finite triangulation approximant of ) the curvature of 
the Ashtekar- Barbero connection in such a way as to render a satisfactory definition of the diffeomorphism constraint
in LQG \cite{alokmelqg}.}

The layout of the paper is as follows. Since we have already reviewed the necessary material in \cite{alokme2},
in the interests of brevity, we shall not do so again. Instead, in section 2 we  provide a quick and not necessarily
complete list of essential definitions so that the reader can follow the broad thrust of this paper 
For a 
detailed understanding, familiarity with \cite{alokme2} is necessary and will be assumed.
In section 3 we restrict attention to a certain physically relevant 
 superselected subspace of states in the kinematic Hilbert space and 
construct solutions to the diffeomorphism constraint by Group Averaging. The group averaging
procedure automatically defines an inner product on these solutions and the Cauchy completion
of their finite span yields the diffeomorphism invariant Hilbert space ${\cal H}_{diff}$.
We show that none of the solutions 
of \cite{alokme2} lie in ${\cal H}_{diff}$.
In section 4 we define the operators
 corresponding to the spatial  volume and the inverse square root of the determinant of the 
metric. In section 5 we construct the action of the Hamiltonian constraint on diffeomorphism invariant states.
As in LQG, a regulated operator on the kinematic Hilbert space 
involving a choice of `small edge' holonomies is constructed and the  continuum
limit of its action is obtained on the space of diffeomorphism invariant states.
Section 6 is devoted to the algebra of quantum constraints using the arena of diffeomorphism invariant states
along the line of Thiemann's seminal work \cite{tthh}.
We probe the constraint algebra on the LM habitat in section 7 and on the new habitat in 
section 8. Section 9  is devoted to a discussion of our results with a view to LQG.

\noindent{\bf Note on notation}:A minor change with respect to \cite{alokme2} is that, here, we only use
objects which respect the zero mode constraint. Accordingly, 
our notation is the same as that of \cite{alokme2} except that (i) we continue to 
use $l_e$ to denote matter charge labels after the imposition of the zero mode constraint; in contrast, in 
\cite{alokme2}, $l_e$ was used for matter charges before imposition of the constraint and $\Delta l_e$ was
used after solving the zero mode constraint, and 
(ii)we will omit the subscript $\lambda^{\pm}$ of \cite{alokme2} which is relevant only 
for objects which do not respect the zero mode constraint.

Finally, we would like to bring to the attention of the reader a recent paper by Thiemann \cite{ttpft} which touches
on issues similar to those discussed in this paper.

\section{Brief Review of Polymer PFT}

\subsection{Classical Theory}
Cauchy slices are oriented circles  coordinatized by the angular coordinate $x\in[0,2\pi]$, 
with the direction of angular increase agreeing with the orientation
of the circle.
\footnote{In \cite{alokme1,alokme2} we fixed an angular coordinate system once and for all. Here we allow any
positively oriented angular coordinate system ranging between $0$ and $2\pi$ such that the coordinate values 
$0$ and $2\pi$ label the same point on the Cauchy slice i.e. the Cauchy slice has a preferred point.\label{footnote1}}
 
Inertial time and space coordinates on the flat spacetime are $T,X$. Null coordinates are $X^{\pm}=T\pm X$.
The length of the $T=$ constant
circles in the flat spacetime is $L$. The scalar field is $f$. \\

\noindent
{ Canonically conjugate embedding variables} :$(X^{+}(x), \Pi_{+}(x)),(X^{-}(x), \Pi_{-}(x))$, 
$X^{\pm}(2\pi)= X^{\pm}(0) \pm 2\pi$\\

\noindent
{ Matter variables} : $Y^{\pm}(x):= \pi_f \pm f^{\prime}$, $\;\{f(x), \pi_f (y)\}= \delta (x,y)$\\ $\{Y^+,Y^-\}=0,
\{Y^{\pm}(x),\ Y^{\pm}(y)\}\ =\ \pm\ (\partial_{x}\delta(x,y)\ -\ \partial_{y}\delta(y,x)\ )$\\

\noindent
{ Density weight 2 constraints}:
${{H_{\pm}}}(x)\ =\ [\ \Pi_{\pm}(x)X^{\pm'}(x)\ \pm\
\frac{1}{4}Y^{\pm}(x)^2\ ].$ \\
The constraint algebra is isomorphic to the Lie Algebra of vector fields on the 
circle. \\

\noindent
{ Diffeormorphism constraint} : $C_{diff}$ generates spatial diffeomorphisms.
\begin{equation}
C_{diff}(x)\ =\ H_+ + H_- =\left[\Pi_{+}(x)X^{+'}(x)\ +\ \Pi_{-}(x)X^{-'}(x)\ +\ \pi_{f}(x)f^{'}(x)\right].
\label{diffconstraint}
\end{equation}
\\

\noindent
{ Hamiltonian constraint} : $C_{ham}$ generates evolution normal to the Cauchy slice,
\begin{eqnarray}
C_{ham}(x)&=& \frac{1}{\sqrt{X^{+'}(x)X^{-'}(x)}}(H_+ -H_-)\nonumber \\
&=&\frac{1}{\sqrt{X^{+'}(x)X^{-'}(x)}}\left[\Pi_{+}(x)X^{+'}(x)\ -\ \Pi_{-}(x)X^{-'}(x)\ +\ \frac{1}{4}(\pi_{f}^{2}+f^{' 2})\right]\nonumber\\
\label{hamconstraint}
\end{eqnarray}
\\

\noindent 
{ Constraint algebra}:
The Poisson algebra generated by 
 $C_{diff}$ and $C_{ham}$ is the  Dirac algebra:
\begin{equation}\label{eq:structure}
\begin{array}{lll}
\{C_{diff}[\vec{N}],\ C_{diff}[\vec{M}]\}\ =\ C_{diff}[\vec{N},\vec{M}]\\
\vspace*{0.1in}
\{C_{diff}[\vec{N}],\ C_{ham}[M]\}\ =\ C_{ham}[{\it L}_{\vec{N}}M]\\
\vspace*{0.1in}
\{C_{ham}[N],\ C_{ham}[M]\}\ =\ C_{diff}[\vec{\beta}(N,M)]
\end{array}
\end{equation}
wherein $\vec{N},\vec{M}$ are shift vectors, $N,M$ are lapse functions and  
the structure function $\beta^{a}(N,M):=\ q^{ab}(N\nabla_{b}M\ -\ M\nabla_{b}N)$ in (\ref{eq:structure}) is defined by the induced spatial metric $q_{ab}$, 
\begin{equation}
q_{ab}dx^a dx^b = -X^{+\prime}X^{-\prime}(dx)^2
\label{qab}
\end{equation}

\subsection{Quantum Theory}

A {\em charge network} $s$ is a finite collection, $\gamma (s)$, of coloured, non-overlapping (except at vertices) edges,
$e$, 
which
span the range of the angular coordinate $x$, (i.e. $[0,2\pi]$), the colours being referred to as charges, and the 
collection of edges being referred to as a graph. Charge network labels depend only on equivalence classes of graphs, similar to
the situation for spin networks in LQG \cite{alm^2t}. $s=s_1+s_2$ is the charge network label
associated to a fine enough graph underlying  both $s_1$ and $s_2$. An edge $e$ of this graph
 is coloured by the sum of the 
charges of $s_1$  and $s_2$ which colour $e$. Charge network states are in correspondence with charge networks and 
constitute an orthonormal basis similar to spin network states in LQG.

\subsubsection{Embedding Sector}

\noindent 
Charge network : $s^{\pm}\ =\ \{\gamma(s^{\pm}), (k_{e_{1}^{\pm}}^{\pm},...,k_{e_{n}^{\pm}}^{\pm})\}$
where $k_{e_{I}^{\pm}}^{\pm}$ are embedding charges whose range is specified by 
 $k_{e_{I}^{\pm}}^{\pm}\in \frac{2\pi L}{\hbar A}{\bf Z}\ \forall I$. Here
$A$ is a fixed,  positive, integer- valued Barbero-Immirizi like parameter.
It is useful to define the `minimum length increment', $a$, as $a:=\frac{2\pi L}{ A}$.
\\

\noindent Elementary variables : $X^{\pm}(x)$,
$T_{s^{\pm}}[\Pi_{\pm}]:= \exp[-i\sum_{e\in \gamma(s^{+})}k_{e^{\pm}}^{\pm}\int_{e^{\pm}}\Pi_{\pm}]$.
\\

\noindent Representation : $T_{s_{1}^{\pm}}$ denotes an embedding charge network state. $\hat{X}^{\pm}(x),\hat{T}_{s^{\pm}}$
denote the operators corresponding to the classical quantities 
$X^{\pm}(x),T_{s^{\pm}}[\Pi_{\pm}]$. Their action is given by
\begin{equation}\label{eq:33}
\hat{T}_{s^{\pm}}T_{s_{1}^{\pm}}\ =\ T_{s^{\pm}+s_{1}^{\pm}}\;\;\;\;\hat{X}^{\pm}(x)T_{s^{\pm}}\ :=\ \lambda_{x,s^{\pm}}T_{s^{\pm}}, 
\end{equation}
where, for $\gamma (s^{\pm})$ with $n^{\pm}$ edges,
\begin{equation}\label{eq:33a}
\begin{array}{lll}
\lambda_{x,s^{\pm}}:=\ \hbar k_{e_{I^{\pm}}^{\pm}}^{\pm}\ if\ x\in \textrm{Interior}(e_{I^{\pm}}^{\pm})\ 1\leq I^{\pm}\leq n^{\pm}\\
\vspace*{0.1in}
\hspace*{0.3in}:=\ \frac{\hbar}{2} (k_{e_{I^{\pm}}^{\pm}}^{\pm}\
+\ k_{e_{(I+1)^{\pm}}^{\pm}}^{\pm})\ \textrm{if}\ x\in e_{I^{\pm}}^{\pm}\cap e_{(I^{\pm}+1)}^{\pm}\ 1\leq I^{\pm}\leq (n^{\pm}-1)\\
\end{array}
\end{equation}
\begin{equation}\label{eq:33b}
\begin{array}{lll}
\hspace*{0.3in}:=\ \frac{\hbar}{2} (k_{e_{n^{\pm}}^{\pm}}^{\pm}\ \mp\ \frac{1}{\hbar}2\pi L\ +\ k_{e_{1}^{\pm}}^{\pm})\ \textrm{if}\ x=0\\
\vspace*{0.2in}
\hspace*{0.3in}:=\ \frac{\hbar}{2} (k_{e_{1}^{\pm}}^{\pm}\ \pm\ \frac{1}{\hbar}2\pi L\ +\ k_{e_{n^{\pm}}^{\pm}}^{\pm})\ \textrm{if}\ x=2\pi
\end{array}
\end{equation}

\subsubsection{Matter Sector}
\noindent Charge-network :
$s^{\pm}\ =\  \{\gamma(s^{\pm}), 
( l_{e_{1}^{\pm}}^{\pm},...,l_{e_{n}^{\pm}}^{\pm})\}\ ,\sum_{I=1}^{n^{\pm}} l_{e_{I}^{\pm}}^{\pm}=0, 
 l_{e_{I}^{\pm}}^{\pm}\in \epsilon{\bf Z}\ 
\forall\ I$. Here $\epsilon$ is a fixed (real, positive) parameter
 with dimensions $(ML)^{-\frac{1}{2}}$. $\epsilon$ is also a Barbero- Immirizi like parameter.
The zero sum condition on the matter charges stems from technicalities related to the scalar field zero mode
\cite{alokme2}, an understanding of which, is 
not essential for the discussion here.
\\

\noindent Elementary variables : $W_{s^{\pm}_{\lambda^{\pm}}}[Y^{\pm}]\ =
\exp[i\sum_{e^{\pm}\in \gamma(s^{\pm})} l_{e^{\pm}}^{\pm}
\int_{e^{\pm}}Y^{\pm}]$\\
\\


\noindent Weyl algebra
\footnote{The definition of the Weyl algebra follows in the standard way 
from the Poisson brackets between $Y^{\pm}(x), Y^{\pm}(y)$ 
and an application of the Baker- Campbell- Hausdorff Lemma \cite{bakercampbellhausdorff}}
 of operators:\\
$\hat{W}(s^{\pm})\hat{W}(s'^{\pm})\ =\exp[-i\frac{\hbar}{2}\alpha(s^{\pm}, s'^{\pm})]
\hat{W}(s^{\pm} + s'^{\pm})$.\\
Here the exponent in the phase-factor $\alpha(s^{\pm},s'^{\pm})$ is given by,
\begin{equation}
\alpha(s^{\pm},s'^{\pm})\ :=\ \sum_{e^{\pm}\in \gamma(s^{\pm})}
\sum_{e^{\prime \pm}\in \gamma(s^{\prime \pm})}( l_{e^{\pm}}^{\pm})(l_{e^{\prime \pm}}^{\pm})
\alpha(e^{\pm},e^{\prime \pm}),
\label{defalphass'}
\end{equation}
Here $\alpha(e^{\pm}, e^{\prime \pm})\ =\ (\kappa_{e'^{\pm}}(f(e^{\pm}))-\kappa_{e'^{\pm}}(b(e^{\pm})))-(\kappa_{e^{\pm}}(f(e^{\prime \pm}))-\kappa_{e^{\pm}}(b(e^{\prime \pm})))$.\\ 
Here $f(e)$, $b(e)$ are the final and initial points of the edge e respectively. $\kappa_{e}$ is defined as,
\begin{equation}
\begin{array}{lll}
\kappa_{e}(x)\ =\ 1\ \textrm{if}\ x\ \textrm{is in the interior of e}\\
\hspace*{1.0in} =\ \frac{1}{2}\ \textrm{if}\ x\ \textrm{is a boundary point of e}
\end{array}
\end{equation}
\\

\noindent Representation : $\hat{W}(s^{\pm})W(s'^{\pm})\ =\ \exp(\frac{-i\hbar}{2}\alpha(s^{\pm}, s'^{\pm}))W(s^{\pm}+
s'^{\pm})$.\\

\subsubsection{Kinematic Hilbert Space}
The kinematic Hilbert space ${\cal H}_{kin}$ is the product of the plus and minus sectors,${\cal H}^{\pm}_{kin}$,  
each of which is 
a product of the appropriate embedding and matter sectors.
${\cal H}^{\pm}_{kin}$ is spanned by an orthornormal basis of   charge network states. 

A charge network state in 
${\cal H}^{\pm}_{kin}$ is denoted by  
$|{\bf s}^{\pm}\rangle :=T_{s^{\pm}}\otimes W(s^{\prime\pm})$.

The label ${\bf s}^{\pm}$ is specified by
${\bf s}^{\pm}_{\lambda^{\pm}}:= 
\{\gamma {(\bf s}^{\pm}), (k^{\pm}_{e_1^{\pm}},
 l^{\pm}_{e_1^{\pm}}),...,(k^{\pm}_{e_{n^{\pm}}^{\pm}},l^{\pm}_{e_{n^{\pm}}^{\pm}})\}$.
Here we have used the equivalence of charge networks to set 
$\gamma {(\bf s}^{\pm}):=\gamma(s^{\pm})=\gamma(s^{\prime\pm})$ so that each edge of the charge network is labelled by 
an embedding  charge and a matter charge.

\subsubsection{Unitary Representation of gauge transformations}
Finite gauge transformations generated by the density 2 constraints act, essentially, as  2 {\em independent}
 diffeomorphisms
of the spatial manifold, one which acts only on the `+' fields and one which acts only on the `-' fields. Consequently,
in analogy to spatial diffeomorphisms in LQG,  
their action on charge networks is to appropriately `drag' them around the circle. However, due to the 
quasi periodic nature of $X^{\pm}$ it is more appropriate to think of these diffeomorphisms as being periodic
diffeormorphisms of the real line. Consequently  the action of these gauge transformations in quantum theory 
also keeps track of `factors of $2\pi$' when embedding charge edges `go past $x=2\pi$'.

More precisely, 
the action of finite gauge transformations is specified by introducing the notion of
an extension of a charge network $s$ to the real line. Such an extension is labelled by the graph
$\gamma (s)_{ext}$ which covers the real line and by charge labels on each edge of $\gamma (s)_{ext}$.
Let $T_N(x)\in R$ denote a rigid translation of the point $x\in[0,2\pi]$ by $2N\pi$ so that 
$T_N(\gamma (s) )$ spans $[2N\pi, 2(N+1)\pi]$. Then $\gamma (s)_{ext} = \cup_{N\in\mathbf{Z}}\ T_{N}(\gamma (s))$.
For the  embedding charge network $s^{\pm}$ we define the {\em  quasiperiodic extension} ${\bar s}^{\pm}_{ext}$
by specifying the embedding charges on $T_N(\gamma (s) )$ by $k^{\pm}_{T_N(e)}:= k_e^{\pm}\pm 2N\pi \frac{L}{\hbar} $
for every edge $e\in \gamma (s)$. Similarly, 
for the  matter charge network $s^{\pm}$ we define the {\em periodic extension} $s^{\pm}_{ext}$ by 
setting $l^{\pm}_{T_N(e)}:= l_e^{\pm}$. 

The action of periodic diffeomorphisms, $\phi$, of the real line on ${\bar s}^{\pm}_{ext}$, $s^{\pm}_{ext}$
is defined by mapping $\gamma (s)_{ext}$ to $\phi (\gamma (s)_{ext})$ and setting 
$k^{\pm}_{\phi (e)}= k^{\pm}_e$,$l^{\pm}_{\phi (e)}= l^{\pm}_e$ for every edge $e\in \gamma (s)_{ext}$.

Then  unitary representation of the gauge group is given by,
\begin{equation}\label{eq:56}
\begin{array}{lll}
\hat{U}^{\pm}(\phi^{\pm})T_{s^{\pm}}\ :=\ T_{\phi(\overline{s}^{\pm}_{ext})\vert_{[0,2\pi]}}\\
\vspace*{0.1in}
\hat{U}^{\mp}(\phi^{\mp})T_{s^{\pm}}\ :=\ T_{s^{\pm}}\\
\vspace*{0.1in}
\hat{U}^{\pm}(\phi^{\pm})W(s^{\prime\pm})\ :=\ W((\phi^{\pm})(s^{\prime\pm}_{ ext})\vert_{[0,2\pi]}).\\
\vspace*{0.1in}
\hat{U}^{\mp}(\phi^{\mp})W(s^{\prime\pm})\ :=\ W(s^{\prime\pm})
\end{array}
\end{equation}
Denoting, $T_{s^{\pm}}\otimes W(s^{\prime\pm})$ by $|{\bf s}^{\pm}\rangle$ and 
$T_{\phi(\overline{s}^{\pm}_{ext})\vert_{[0,2\pi]}}\otimes W((\phi^{\pm})(s^{\prime\pm}_{ ext})\vert_{[0,2\pi]})$ by
$|{\bf s}^{\pm}_{ \phi^{\pm}}\rangle$ , the above equations can be written in a compact form as,
\begin{equation}
|{\bf s}^{\pm}_{\phi^{\pm}}\rangle := {\hat U}^{\pm}(\phi^{\pm})|{\bf s}^{\pm}\rangle .
\label{bfsphi}
\end{equation}

\subsubsection{Physical Hilbert Space}
Physical states are obtained by group averaging the action of the finite gauge transformations discussed in the 
previous section. Henceforth we restrict attention to a physically relevant superselected sector of the physical 
Hilbert space. This sector is obtained by group averaging a superselected subspace, ${\cal D}_{ss}$ of 
${\cal H}_{kin}$, ${\cal D}_{ss}={\cal D}^+_{ss} \otimes {\cal D}^-_{ss}$.

${\cal D}^{\pm}_{ss}$ is defined as follows.
Fix a pair of graphs $\gamma^{\pm}$ with A edges. 
Place the embedding charges 
$\vec{k}^{\pm}$ such that $k_{e_{I^{\pm}}^{\pm}}^{\pm}-k_{e_{I^{\pm}-1}^{\pm}}^{\pm}=\frac{2\pi}{A\hbar}\ \forall\ I^{\pm}$.
Consider the set of all charge-network states\\ 
$\{|{\bf s}^{\pm}\rangle\ =\ 
|\gamma^{\pm},\ \vec{k}^{\pm}, (l_{e_{1}^{\pm}}^{\pm},...,l_{e_{A}^{\pm}}^{\pm})\rangle\}$, where
$l_{e_{I}^{\pm}}^{\pm}\in {\bf Z}\epsilon$ are allowed to take all possible values subject to the zero sum condition
$\sum_{I}l_{e_I^{\pm}}^{\pm}=0$.
Let ${\cal D}_{ss}^{\pm}$ be finite span of charge network states of the type 
$\{|{\bf s}^{\pm}_{\lambda^{\pm}\ \phi^{\pm}}\rangle\ \forall\ \phi^{\pm}\}$.

The action of the Group Averaging map $\eta^{\pm}$ on a charge network state in ${\cal D}^{\pm}_{ss}$ yields the 
distribution, 
\begin{equation}\label{eq:120}
\begin{array}{lll}
\eta^{\pm} (|{\bf s}^{\pm}\rangle) & = & \sum_{{\bf s}^{\pm}\in [{\bf s}^{\pm}]}
    <\ {\bf s}^{\prime\pm}|\\
\vspace*{0.1in}
& = & \sum_{\phi^{\pm}\in Diff_{[{\bf s}^{\pm}]}^{P}\mathbf{R}}<{\bf s}_{\phi^{\pm}}^{\pm}| .
\end{array}
\end{equation}
Here $[{\bf s}^{\pm}]$ is the equivalence class defined by
$[{\bf s}^{\pm}]\: =\ \{ {\bf s}^{\prime\pm}\vert {\bf s}^{ \prime\pm}\ 
=\  {\bf s}_{\phi^{\pm}}^{\pm} \; {\rm for\ some\ }\phi^{\pm}\}$, 
and 
$Diff_{[\bf{s}^{\pm}]}^{P}\mathbf{R}$ is a set of gauge transformations such that
for each  ${\bf s}^{\prime\pm}\in\ [{\bf s}^{\pm}]$ there is precisely one gauge transformation  in the set which maps 
${\bf s}^{\pm}$ to ${\bf s}^{\prime\pm}$.
The space of such  gauge invariant distributions comes equipped with 
the inner product
\begin{equation}
<\eta^{\pm}(|{\bf s}_{1}^{\pm}\rangle), \eta^{\pm}(|{\bf s}_{ 2}^{\pm}\rangle)>_{phys}=
\eta^{\pm}(|{\bf s}_{1}^{\pm}\rangle) [|{\bf s}_{2}^{\pm}\rangle ]. 
\label{physip}
\end{equation}
which can be used to complete $\eta^{\pm}({\cal D}_{ss}^{\pm})$ to the Hilbert space ${\cal H}_{phy}^{ss\pm}$.
We shall restrict attention to ${\cal H}_{phy}^{ss}:={\cal H}_{phy}^{ss+}\otimes{\cal H}_{phy}^{ss-}$.

\section{The Hilbert space of diffeomorphism invariant distributions}
In section 3.1 we show that the solutions of \cite{alokme2} are invariant under the unitary 
action of spatial diffeomorphisms. In section 3.2 we 
 construct the Hilbert space of diffeomorphism invariant distributions, ${\cal H}_{diff}$, by group averaging. 
In section 3.3 we restrict attention 
to a physically relevant superselected subspace of ${\cal H}_{diff}$ and  show that 
(a basis of) this subspace is in correspondence with  quantum matter states on discrete Cauchy slices of the 
flat spacetime. We then use this correspondence to
show that no solution of \cite{alokme2} is normalizable in ${\cal H}_{diff}$.

\subsection{Diffeomorphism invariance of the solutions of \cite{alokme2}}
The diffeomorphism constraint (see equation (\ref{diffconstraint}))  generates spatial diffeomorphisms of the 
circular Cauchy slice. As indicated in section 2.2.4, due to the quasiperiodicity of $X^{\pm}$ it is useful to think 
of diffeomorphisms of the circle in terms of  periodic diffeomorphisms of the real line. 

Let $\phi$ be a periodic diffeomorphism of the real line so that $\phi (x+2\pi m)= \phi (x) +2\pi m, \;m\in {\bf Z}$.
From equation (\ref{diffconstraint}), recall that $C_{diff}= H_+ +H_-$. Also recall that $\{H_+, H_-\}=0$. 
It follows from section 2.2.4 that the unitary action of the finite spatial diffeomorphism labelled by $\phi$ on 
any charge network state $\vert {\bf s}^+, {\bf s}^-\rangle := \vert {\bf s}^+\rangle\otimes\vert {\bf s}^-\rangle$
is given by 
\begin{equation}
\vert {\bf s}_{\phi}^+, {\bf s}_{\phi}^-\rangle := \vert {\bf s}_{\phi}^+\rangle\otimes\vert {\bf s}_{\phi}^-\rangle
= {\hat U}^+(\phi)\vert {\bf s}^+\rangle\otimes  {\hat U}^-(\phi)\vert {\bf s}^-\rangle .
\label{udiff}
\end{equation}
Thus, spatial diffeomorphisms correspond to those gauge transformations for which $\phi^+ = \phi^- =\phi$.
Since the solutions of \cite{alokme2} are invariant under the action of {\em all} gauge transformations, they are,
in particular, invariant under the action of spatial diffeomorphisms.

\subsection{The construction of ${\cal H}_{diff}$.}

Spatial diffeormorphism invariant distributions are constructed by the action of the group averaging map,
$\eta_{diff}$ on the dense space of finite linear combinations of charge network states as follows.
Let $[{\bf s}^+, {\bf s}^-]$ be the orbit of ${\bf s}^+, {\bf s}^-$ under all spatial diffeomorphisms
so that $[{\bf s}^+, {\bf s}^-]$ is the set of all distinct charge network labels obtained by the action of
spatial diffeomorphisms on ${\bf s}^+, {\bf s}^-$. Then
\begin{equation}
\eta_{diff} (\vert {\bf s}^+\rangle\otimes\vert {\bf s}^-\rangle )=
\eta_{[{\bf s}^+, {\bf s}^-]}\sum_{{\bf s}^{\prime +}, {\bf s}^{\prime -}\in [{\bf s}^+, {\bf s}^-]}
                              \langle {\bf s}^{\prime +}\vert\otimes\langle {\bf s}^{\prime -}\vert .
\label{diffavg}
\end{equation}.
Here $\eta_{[{\bf s}^+, {\bf s}^-]}$ is a constant which depends only on the orbit of 
${\bf s}^+, {\bf s}^-$. The arbitrariness in the choice of this constant can be reduced by requiring  that 
$\eta_{diff}$ commute with all diffeomorphism invariant observables. Specifically,
consider the discrete time translation operator  ${\hat T}_n, n\in {\bf Z},$ which translates $t_e$ by an amount 
$na$ and leaves $x_e,l^{\pm}_e$ untouched i.e. 
\begin{eqnarray}
{\hat T}_n \vert {\bf s}^+, {\bf s}^-\rangle &=:&
\vert {\bf s}_n^+, {\bf s}_n^-\rangle , \nonumber \\
{\bf s}_n^{\pm}&=& (\gamma ({\bf s}^{\pm}), 
(k^{\pm}_{e^{\pm}_{I^{\pm}}}+ \frac{2\pi nL}{\hbar A}, l^{\pm}_{e^{\pm}_{I^{\pm}}},I^{\pm}=1,..,N^{\pm})) .
\end{eqnarray}
It is straightforward to check that ${\hat T}_n$ commutes with all finite gauge transformations and, hence, 
with diffeomorphisms. Requiring its commutativity with $\eta_{diff}$ implies that 
\begin{equation}
\eta_{[{\bf s}_n^+, {\bf s}_n^-]}= \eta_{[{\bf s}^+, {\bf s}^-]}.
\label{difft}
\end{equation}
While one could attempt to further restrict the choice of  $\eta_{[{\bf s}^+, {\bf s}^-]}$, 
we shall not do so here in view of the fact that our subsequent considerations are independent of
any such further restrictions.

\subsection{Quantum  Cauchy data from states in ${\cal H}_{diff}$}

We restrict attention to states in the superselected sector ${\cal D}_{ss} ={\cal D}^+_{ss}\otimes{\cal D}^-_{ss}$
 (see section 2.2.5).
It is straightforward to see that any charge network state $\vert {\bf s}^{\pm}\rangle \in {\cal D}^{\pm}_{ss}$ has charges which satisfy the conditions
\begin{eqnarray}
\pm k^{\pm}_{e_{I^{\pm}}^{\pm}} - \pm  k^{\pm}_{e_{I^{\pm}-1}^{\pm}} &=& \frac{2\pi L}{\hbar A} , I^{\pm}=2,..,N^{\pm}, 
\label{equi}
\\
\pm k^{\pm}_{e_{N^{\pm}}^{\pm}} - \pm  k^{\pm}_{e_{1}^{\pm}}& \leq & \frac{2\pi L}{\hbar } ,
\label{k-kleq2pi}
\\
\pm k^{\pm}_{e_{N^{\pm}}^{\pm}} - \pm  k^{\pm}_{e_{1}^{\pm}} =\frac{2\pi L}{\hbar } &{\rm iff}&
N^{\pm}= A+1 \;\;{\rm and}\;\; l^{\pm}_{e^{\pm}_{N^{\pm}}} = l^{\pm}_{e^{\pm}_{1}}.
\label{k-kequal2pi}
\end{eqnarray}
Here $N^{\pm}$ are the number of edges of  ( the coarsest graphs underlying) ${\bf s}^{\pm}$ and $N^{\pm}=A$ if the 
strict inequality holds in equation (\ref{k-kleq2pi}).

Next, consider any charge network state $\vert{\bf s}^{+},{\bf s}^{-}\rangle \in {\cal D}_{ss}$
and let $\gamma ({\bf s}^{+},{\bf s}^{-})$ be the coarsest graph underlying both ${\bf s}^{+}$ and ${\bf s}^{-}$.
Denote the number of edges of  $\gamma ({\bf s}^{+},{\bf s}^{-})$ by $N$. Each edge $e$ of 
$\gamma ({\bf s}^{+},{\bf s}^{-})$ is labelled by the quadruple $(k^+_e, k^-_e, l^+_e, l^-_e)$ or, alternatively, 
by $(t_e, x_e, l^+_e,l^-_e)$ where $t_e:= \hbar \frac{k^+_e+ k^-_e}{2},  x_e:= \hbar \frac{k^+_e -k^-_e}{2}$.
From equations (\ref{equi})- (\ref{k-kequal2pi}) it follows that  
\begin{eqnarray}
x_{e_I}< x_{e_J} & \rm{iff}& I<J 
\label{xi<xj}
\\
x_{e_N}- x_{e_1} &\leq & 2\pi L 
\label{x-xineq}\\
x_{e_N}- x_{e_1}=2\pi L & \rm{iff}&
k^+_{e_N}- k^+_{e_1}= k^-_{e_1}- k^-_{e_N}= \frac{2\pi L}{\hbar} 
\label{x-xequal}
\\
\Rightarrow t_{e_1} = t_{e_N}, l^{\pm}_{e_1}=l^{\pm}_{e_N} &\rm{when}& x_{e_N}- x_{e_1}=2\pi L .
\label{t-tequal}
\end{eqnarray}
Each pair of embedding charges $x_e,t_e$ defines a spacetime point with inertial coordinates
$(X,T) = (x_e,t_e)$. If equation (\ref{x-xineq}) holds with a strict inequality then
equation (\ref{xi<xj}) implies that the set of pairs $(x_{e_{I}}, t_{e_I}), I=1,..,N$  
define $N$ distinct points in the flat spacetime. If equation (\ref{x-xequal}) holds then 
equations (\ref{xi<xj}), (\ref{x-xequal}) and (\ref{t-tequal}) ensure that this set defines $N-1$ distinct 
spacetime points by virtue of the circular topology of space. 

Denote the set of spacetime points associated to the charge network state 
$\vert{\bf s}^{+},{\bf s}^{-}\rangle\in {\cal D}_{ss}$ 
in this way by ${\cal C}_{{\bf s}^{+},{\bf s}^{-}}$. Next, associate the matter charge labels 
$(l^+_{e_I}, l^-_{e_I})$ to the corresponding spacetime point defined by $(x_{e_{I}}, t_{e_I})$ 
(if equation (\ref{x-xequal}) holds, such an association
is consistent by virtue of equation (\ref{t-tequal})). This association defines the set 
${\cal M}_{{\bf s}^{+},{\bf s}^{-}}$ each element of which is  a spacetime point defined by 
$\vert{\bf s}^{+},{\bf s}^{-}\rangle$ together with the matter charges associated to it.
We shall refer to ${\cal C}_{{\bf s}^{+},{\bf s}^{-}}$ as a discrete Cauchy slice (despite the existence of 
pairs of points which are light like seperated \cite{alokme2}) and 
${\cal M}_{{\bf s}^{+},{\bf s}^{-}}$ as quantum matter on this discrete Cauchy slice

It is then straightforward to see, from the action of ${\hat U}^{\pm}(\phi )$ and the properties of the extensions
of charge network labels to the real line , that \\
\noindent (a) any spatial diffeomorphism preserves both ${\cal C}_{{\bf s}^{+},{\bf s}^{-}}$ and 
${\cal M}_{{\bf s}^{+},{\bf s}^{-}}$,  \\
\noindent (b)if $\vert{\bf s}^{+\prime},{\bf s}^{-\prime}\rangle \in {\cal D}_{ss}$ 
is such that 
${\cal C}_{{\bf s}^{+\prime},{\bf s}^{-\prime}}={\cal C}_{{\bf s}^{+},{\bf s}^{-}}$,   
${\cal M}_{{\bf s}^{+\prime},{\bf s}^{-\prime}}={\cal M}_{{\bf s}^{+},{\bf s}^{-}}$, then 
${\bf s}^{+\prime},{\bf s}^{-\prime}\in [{\bf s}^{+},{\bf s}^{-}]$. \\
It follows that $\eta_{diff} (\vert{\bf s}^{+},{\bf s}^{-}\rangle)$ is in unique correspondence with 
quantum matter data on a discrete Cauchy slice.

Next, consider the group average of $\vert{\bf s}^{+},{\bf s}^{-}\rangle$ with respect to {\em all} finite gauge 
transformations generated by $H_+, H_-$. From section 2.2.5, this yields the distribution 
$\eta^{+} (\vert{\bf s}^{+}\rangle) \otimes\eta^{-} (\vert{\bf s}^{-}\rangle)$:
\begin{equation}
\eta^{+} (\vert{\bf s}^{+}\rangle) \otimes\eta^{-} (\vert{\bf s}^{-}\rangle)
= \sum_{{\bf s}^{+\prime}\in [{\bf s}^{+}]}
\sum_{{\bf s}^{-\prime}\in [{\bf s}^{-}]} \langle {\bf s}^{+\prime}\vert\otimes\langle {\bf s}^{-\prime}\vert
\label{biggrpavg}
\end{equation}
where $[{\bf s}^{\pm}]$ is the orbit of ${\bf s}^{\pm}$ under the action of all finite gauge transformations.
Clearly, the sum (\ref{biggrpavg}) contains the sum (\ref{diffavg}) since every diffeomorphism is a finite 
gauge transformation (see (\ref{udiff})). Moreover since 
$\eta^{+} (\vert{\bf s}^{+\prime}\rangle) \otimes\eta^{-} (\vert{\bf s}^{-\prime}\rangle)
=\eta^{+} (\vert{\bf s}^{+}\rangle) \otimes\eta^{-} (\vert{\bf s}^{-}\rangle)$ if
${\bf s}^{\pm\prime}\in [{\bf s}^{\pm}]$, the sum also contains all the diffeomorphism images of
of any state $\vert{\bf s}^{+\prime}\rangle \otimes \vert{\bf s}^{-\prime}\rangle$ which is gauge related
to $\vert{\bf s}^{+}\rangle \otimes \vert{\bf s}^{-}\rangle$.

It is straightforward to see that generic gauge transformations do not preserve  the discrete Cauchy slice
${\cal C}_{{\bf s}^{+},{\bf s}^{-}}$ and that, in fact, a countable infinity of 
(states corresponding to) 
distinct discrete Cauchy slices
are generated by the action of finite gauge transformations on 
$\vert{\bf s}^{+}\rangle \otimes \vert{\bf s}^{-}\rangle$. In particular it is easy to see that discrete
Cauchy slices which are time translations by $2m \pi L, m\in {\bf Z}$ are in the sum (\ref{biggrpavg}).
\footnote{It is straightforward to check that for $\phi^{\pm}$ chosen to be appropriate rigid translations 
on ${\bf R}$, we have that ${\hat U}(\phi^+){\hat U}(\phi^-) = {\hat T}_n$ with $n=Am$.}
This, together with the orthogonality of diffeomorphism invariant states corresponding to different discrete 
Cauchy slices and condition (\ref{difft}), implies that no solution of \cite{alokme2} is normalizable in
${\cal H}_{diff}$.

\section{The volume and inverse metric operators}

The inverse (square root of the determinant of the) metric operator,
$\widehat{\frac{1}{\sqrt{X^{+ '}X^{- '}}}}$,
 is constructed using a Thiemann- like trick \cite{tthh} to express 
${\frac{1}{\sqrt{X^{+ '}X^{- '}}}}$
in terms of Poisson brackets of the volume with holonomies and then replacing the Poisson brackets with quantum
commutators. We construct the volume operator in section 4.1 and 
 the operator $\widehat{\frac{1}{\sqrt{X^{+ '}X^{- '}}}}$ in section 4.2. 
Calculation details pertaining to section 4.2 are in the Appendix.

\subsection{The Volume operator}
We construct the volume operator corresponding to a small region centered around any point $p_0$ on the circular
Cauchy slice. The volume of any other region can be built out of these. 
As mentioned in section 2, we use angular coordinate systems on $S^1$ whose range is $[0,2\pi]$. 
Let us fix one such coordinate system and denote the shortest
coordinate distance, in the angular coordinates,  between two points  in $S^1$ with coordinates $y_1, y_2 \in [0,2\pi]$
by $d(y_1, y_2)$ i.e. $d(y_1,y_2)$ is the minimum of $(|y_1-y_2|, |y_1-y_2+2\pi |,|y_1-y_2-2\pi|)$.

Let ${\cal U}_{y_{0}}\subset S^1$ be a closed interval of radius ${\epsilon}$ centered around $y_{0}$. Then its 
( one dimensional) volume, $V_{{\cal U}_{y_{0}}}$, as measured by the spatial metric induced on the circular slice from the flat spacetime metric 
is 
given by 
\begin{equation}\label{eq:jun3-1}
V_{{\cal U}_{y_{0}}}=
\int_{{\cal U}_{y_{0}}}dy_{1}\sqrt{|X^{+ '}X^{- '}|}(y_{1})\ =\ \int\kappa_{\epsilon}(y_{1}, y_{0})\sqrt{|X^{+ '}X^{- '}|}(y_{1})
\end{equation}
where $\kappa_{\epsilon}(\cdot,y_{0})$ is the characteristic function on ${S^1}$ defined as,
\begin{equation}
\begin{array}{lll}
\kappa_{\epsilon}(y_{1},y_{0})\ =\ 1\ \textrm{if}\ d(y_1, y_0 )\ <\ \epsilon\\
\vspace*{0.1in}
\kappa_{\epsilon}(y_{1},y_{0})\ =\ \frac{1}{2}\ \textrm{if}\ d(y_1, y_0 )\ =\ \epsilon\\
\vspace*{0.1in}
\hspace*{0.7in}=\ 0\ \textrm{otherwise}
\end{array}
\end{equation}

Let $T$ be a triangulation of $[0,2\pi]$ whose 1-simplices we denote by $\triangle$, each simplex  $\triangle$ being of 
 length $|\triangle |$. 
We orient each simplex in the increasing $x$ direction.
 Now consider the union of 1-simplices $\overline{\triangle}$ which are obtained by joining
the mid-points of $\triangle\ \in T$. We denote the collection of $\overline{\triangle}$ as $T^{*}$ and loosely refer to it as the triangulation dual to $T$. Notice that $T^{*}$ 
does not completely cover $[0,2\pi]$, and that  $|\overline{\triangle}|\ =\ |\triangle|$.
One can now approximate the R.H.S of (\ref{eq:jun3-1}) by a Riemann sum over $T$:
\begin{equation}\label{eq:jun3-2}
\begin{array}{lll}
\int\kappa_{\epsilon}(y_{1}, y_{0})\sqrt{|X^{+ '}X^{- '}|}(y_{1})\\
\vspace*{0.1in}
\hspace*{1.0in}\approx\ \sum_{\triangle\in T} |\triangle|\kappa_{\epsilon}(b(\triangle), y_{0})\sqrt{|X^{+ '}X^{- '}|}(b(\triangle))\\
\end{array}
\end{equation}
where we have implicitly assumed that $|\triangle | <<\epsilon$.\\
The above sum can in turn be approximated by a sum over simplices in $T^{*}$ and the two remaining terms coming from the intervals belonging to $[0,2\pi]-T^{*}$.
\begin{equation}\label{jun3-3a}
\begin{array}{lll}
\int\kappa_{\epsilon}(y_{1}, y_{0})\sqrt{|X^{+ '}X^{- '}|}(y_{1})\\
\vspace*{0.1in}
\hspace*{1.0in}\approx\ \sum_{\overline{\triangle}\in T^{*}} |\overline{\triangle}|\kappa_{\epsilon}(m(\overline{\triangle}), y_{0})\sqrt{|X^{+ '}X^{- '}|}(m(\overline{\triangle})\\
\vspace*{0.1in}
\hspace*{0.6in} + \frac{|\overline{\triangle}|}{2}\kappa_{\epsilon}(0,y_{0})\sqrt{|X^{+ '}X^{- '}|}(0)\ +\ \frac{|\overline{\triangle}|}{2}\kappa_{\epsilon}(0,y_{0})\sqrt{|^{+ '}X^{- '}|}(2\pi)
\end{array}
\end{equation}
The last two terms are infact equal to each other as the characteristic function is periodic and so is $\sqrt{|X^{+ '}X^{- '}|}(x)$.\\
With a further  approximation (which becomes exact in the limit that $|\triangle | \rightarrow 0$), we have that
\begin{equation}\label{jun3-3b}
\begin{array}{lll}
{\int\kappa_{\epsilon}(y_{1}, y_{0})\sqrt{|X^{+ '}X^{- '}|}(y_{1})}\\
\vspace*{0.1in}
\hspace*{1.0in}\approx\ \sum_{\overline{\triangle}\in T^{*}} \kappa_{\epsilon}(m(\overline{\triangle}), y_{0})
\sqrt{|{X}^{+}(f(\overline{\triangle}))-{X}^{+}(b(\overline{\triangle}))|||{X}^{-}(f(\overline{\triangle}))-{X}^{-}(b(\overline{\triangle}))|}\\
\vspace*{0.1in}
\hspace*{0.6in} + \kappa_{\epsilon}(0,y_{0})\sqrt{|{X}^{+}(b(\overline{\triangle}_{1}))-({X}^{+}(f(\overline{\triangle}_{N}))-2\pi)|
|{X}^{-}(b(\overline{\triangle}_{1}))-({X}^{-}(f(\overline{\triangle}_{N}))+2\pi)|}
\end{array}
\end{equation}
where $\overline{\triangle_{1}}$ is the first (left-most) simplex in $T^{*}$ and $\overline{\triangle_{N}}$ the final (right-most) simplex in $T^{*}$.\\

Next, we define the action of the operator corresponding to the right hand side of (\ref{jun3-3b}) (and, eventually, 
its $|\triangle | \rightarrow 0$ limit) on the charge network basis. 
Let $T_{s^+}\otimes T_{s^-}$ be the  embedding charge network state of interest and consider the graphs
$\gamma (s^{\pm}) $ underlying the state.\footnote{We remind the reader  charge networks `live' on $[0,2\pi ]$ and that 
$x=0, x=2\pi$ are always vertices of the graphs; the circular topology is built into the definition of the action of
various operators of interest as detailed in section 2 and in Reference \cite{alokme2}. We also remind the reader that 
the vertex set depends on the specific member of the equivalence class of labelled graphs underlying the charge network 
state.}
Let $v^{\pm}$ be a vertex of $\gamma (s^{\pm}) $. Let
$k^{\pm}_{e_{v^{\pm}}}$ denote the embedding charge in $s^{\pm}$ on an edge terminating at the vertex $v^{\pm}$ and 
$k^{\pm}_{e^{v^{\pm}}}$ denote  the embedding charge in $s^{\pm}$ on an edge originating at vertex $v^{\pm}$. 
If $v^{\pm}= 0$ 
(or $2\pi$) we define $k^{\pm}_{e_{v^{\pm}}}$ (or $k^{\pm}_{e^{v^{\pm}}}$) by the quasiperiodic extension of the 
charge network (see section 2.2.4) i.e. $k^{\pm}_{e_{v^{\pm}=0}}= k^{\pm}_{e_{v^{\pm}=2\pi}} \mp 2\pi$
and $k^{\pm}_{e^{v^{\pm}=2\pi}}=k^{\pm}_{e^{v^{\pm}=0}} \pm 2\pi$.
We shall refer to vertices 
for which $k^{\pm}_{e^{v^{\pm}}}-k^{\pm}_{e_{v^{\pm}}}\neq 0$ as 
{\em nontrivial} embedding vertices of $\gamma (s^{\pm}) $. Note that the set of such 
non-trivial vertices, $V_E(s^{\pm})$ only 
depends on the state
and not on the particular representative of the equivalence class of charge networks which labels the state.

We shall (as in LQG) adapt the triangulation $T$ to the 
graphs underlying the charge network state. Since we defined $T$ in terms of our chosen angular coordinates on the slice,
we shall also adapt our choice of coordinates to the state (see Footnote \ref{footnote1}). 

Note that a one dimensional triangulation naturally defines a graph. We shall
slightly abuse notation and denote the graph defined by $T$, also by $T$. 
We restrict our choice of coordinate system and the resulting choice of $T$ (for small enough $|\triangle |$) to be such that all vertices of the 
graphs  $\gamma (s^{\pm}) $  are vertices of $T$. 
While our final result will be independent of the 
particular choice of  $\gamma (s^{\pm}) $, we find it convenient to choose fine enough graphs so that 
$\gamma (s^+) = \gamma (s^-) =T$. The action of the operator 
corresponding to the right hand side of (\ref{jun3-3b}) is obtained by replacing 
classical embedding variables by the corresponding operators. From equation
(\ref{eq:33b}) it follows that,  
only those $\overline{\triangle} \in T^*$ contribute to the operator action for which 
$m(\overline{\triangle})\in V_E(\gamma^{+})\cap V_E(\gamma^{-})$. Thus the sum over midpoints can be replaced by a sum over 
non-trivial vertices of the graph and we obtain an expression independent of $|\triangle |$ so that the 
$|\triangle | \rightarrow 0$ limit can be taken.
It follows  that 
\begin{equation}\label{eq:jun4-1}
\begin{array}{lll}
{\hat V}_{{\cal U}_{y_{0}}}T_{s^{+}}\otimes T_{s^{-}} =
\widehat{\int_{{\cal U}_{y_{0}}}dy_{1}\sqrt{|X^{+ '}X^{- '}|}}\ T_{s^{+}}\otimes T_{s^{-}} \ =\ \\
\vspace*{0.1in}
\hspace*{0.6in} \hbar
\sum_{v\in V(\gamma^{+})\cap V(\gamma^{-}),v\neq 2\pi}\kappa_{\epsilon}(v,y_{0})\sqrt{|k^{+}_{e^{v}}-k^{+}_{e_{v}}|
|k^{-}_{e^{v}}-k^{-}_{e_{v}}|}  T_{s^{+}}\otimes T_{s^{-}}                       
\end{array}
\end{equation}

For future purposes it is convenient to define the  vertex volume operator, $\hat{V}_{x}\ x\in [0,2\pi]$, as
\begin{equation}
\hat{V}_{x} T_{s^{+}}\otimes T_{s^{-}} \ :=\ \hbar \sqrt{|k^{+}_{e^{x}}-k^{+}_{e_{x}}||k^{-}_{e^{x}}-k^{-}_{e_{x}}|}
T_{s^{+}}\otimes T_{s^{-}},
\end{equation}
where it is understood that we always choose a member of the equivalence class of graphs underlying the state
which has $x$ as a vertex. 
Then the action of the volume operator can also be expressed as:
\begin{equation}
{\hat V}_{{\cal U}_{y_{0}}}\ T_{s^{+}}\otimes T_{s^{-}}                                             
=
\sum_{x\in T,\ x\neq 2\pi}\kappa_{\epsilon}(x,y_{0})\hat{V}_{x}T_{s^{+}}\otimes T_{s^{-}} ,
\label{volume=sumx}
\end{equation}
where we have chosen  the graph naturally defined by $T$ itself as the specific member of the equivalence class of graphs underlying the state and summed over all vertices of $T$ (While the sum is over all vertices of the triangulation (except $x=2\pi$ to avoid double counting the point
on the circle corresponding to $x=0 \equiv x=2\pi$), clearly, only the non-trivial vertices contribute to the sum).

\subsection{The operator $\widehat{\frac{1}{\sqrt{X^{+ '}X^{- '}}}}$}

As above, let ${\cal U}_{y_{0}}$ be a closed interval of radius ${\epsilon}$ centered around $y_{0}$ 
Let $x_{1}\in S^1$ be  such that $x_1\notin {\cal U}_{y_{0}}$. Then one can easily verify the following two identities.\\
\begin{equation}\label{eq:1}
\begin{array}{lll}
-\frac{\textrm{sgn}(X^{+ '}(y_{0}))}{2}\sqrt{|\frac{X^{- '}}{X^{+ '}}}|(y_{0})\ =\ \{\int_{{\cal U}_{y_{0}}}dy_{1}\sqrt{X^{+ '}X^{- '}}(y_{1}),\ \int_{x_{1}}^{y_{0}}\Pi_{+}(y_{2})dy_{2}\}\\
\vspace*{0.1in}
\frac{\textrm{sgn}(X^{+ '}X^{- '})(y_{0})}{4}\frac{1}{\sqrt{|X^{+ '} X^{- '}|}}(y_{0})\ =\nonumber\\
\vspace*{0.1in}
\hspace*{0.4in} \{\{\int_{{\cal U}_{y_{0}}}dy_{1}\sqrt{X^{+ '}X^{- '}}(y_{1}),\ \int_{{\cal U}_{y_{0}}}
dy\int_{x_{1}}^{y}\Pi_{+}(y_{2})dy_{2}\},\ \int_{x_{1}}^{y_{0}}\Pi_{-}(y_{3})dy_{3}\},
\end{array}
\end{equation}
where sgn stand for the signum function. 
The second equation can be written in a manner that treats the (+) and (-) sectors more symmetrically.
\begin{equation}\label{eq:may19-1}
\begin{array}{lll}
\frac{\textrm{sgn}(X^{+ '}X^{- '})(y_{0})}{4}\frac{1}{\sqrt{|X^{+ '} X^{- '}|}}(y_{0})\ =\\
\vspace*{0.1in}
\frac{1}{2}\left(\{\{\int_{{\cal U}_{y_{0}}}dy_{1}\sqrt{X^{+ '}X^{- '}}(y_{1}),\ \int_{{\cal U}_{y_{0}}}
dy\int_{x_{1}}^{y}\Pi_{+}(y_{2})dy_{2}\},\ \int_{x_{1}}^{y_{0}}\Pi_{-}(y_{3})dy_{3}\}\right.\\
\vspace*{0.1in}
\hspace*{0.4in}\left. +\ \{\{\int_{{\cal U}_{y_{0}}}dy_{1}\sqrt{X^{+ '}X^{- '}}(y_{1}),\ \int_{{\cal U}_{y_{0}}}
dy\int_{x_{1}}^{y}\Pi_{-}(y_{2})dy_{2}\},\ \int_{x_{1}}^{y_{0}}\Pi_{+}(y_{3})dy_{3}\}\right)
\end{array}
\end{equation}

We shall turn the classical expression above into a quantum operator. 
Before doing so, we need to define the operator,$\widehat{\textrm{sgn}(X^{+ '}X^{- '})(y_{0})}$, corresponding 
to the signum function 
$sgn(X^{+ '}X^{- '}(y_{0}))$. A natural definition is:
\begin{eqnarray}
\widehat{\textrm{sgn}(X^{+ '}X^{- '})(y_{0})} T_{s^{+}}\otimes T_{s^{-}} &=& 
\textrm{sgn}(k^{+}_{e^{y_{0}}}-k^{+}_{e_{y_{0}}})\textrm{sgn}(k^{-}_{e^{y_{0}}}-k^{-}_{e_{y_{0}}})
 T_{s^{+}}\otimes T_{s^{-}} \nonumber \\
& &\textrm{if}\ y_{0}\in V_E(s^{+})\cap V_E(s^{-})
\end{eqnarray}
We do not specify the action of $\widehat{\textrm{sgn}(X^{+ '}X^{- '})(y_{0})}$ on the state if 
$y_0$ is not a non-trivial vertex of both $\gamma (s^+) $ and $\gamma (s^-)$. As we
shall see, such a specification is not required.

There are a host of quantization ambiguities involved in defining $\frac{1}{4}\frac{1}{\sqrt{|X^{+ '} X^{- '}|}}(y_{0})$, However
in analogy with the inverse volume operator in LQG, we want the operator to be such that\\ 
\noindent {(i)} $\widehat{\frac{1}{\sqrt{X^{+ '}X^{- '}}(y_{0})}}\  T_{s^{+}}\otimes T_{s^{-}}\ =\ 0$, if $y_{0}$ is a vertex of triangulation but does not belong to 
$V_E(\gamma^{+})\cup V_E(\gamma^{-})$.\\
\noindent {(ii)} $\widehat{\frac{1}{\sqrt{X^{+ '}X^{- '}}(y_{0})}}T_{s^{+}}\otimes T_{s^{-}}\ \neq\ 0$ if $y_{0}\in V_E(\gamma^{+})\cup V_E(\gamma^{-})$.\\
\noindent{(iii)} For embedding data $(\vec{k}^{+},\vec{k}^{-})$ which suitably 
approximate  the classical continuum data $(X^{+},X^{-})$, the spectrum of 
$\widehat{\frac{1}{\sqrt{X^{+ '}X^{- '}}(y_{0})}}$ should be well approximated by  the classical expression 
$\frac{1}{\sqrt{|X^{+ '}X^{- '}|}(y_{0})}$.
\footnote{We shall be more precise about the sense in 
which classical data are well approximated at the end of this section}\\
We shall implicitly tune our quantization choices to meet these three requirements.

It is easy to check that 
Equation (\ref{eq:may19-1}) can be rewritten as:
\begin{equation}\label{eq:may19-2}
\begin{array}{lll}
\frac{1}{\sqrt{|X^{+ '} X^{- '}|}}(y_{0})\ = (\frac{\hbar}{a})^2\\
\vspace*{0.1in}
\ \textrm{sgn}(X^{+ '} X^{- '})(y_{0})\left[h_{e_{x_{1},y_{0}}}[\Pi_{-}]^{-1}\{\int_{{\cal U}_{y_{0}}}dy\ h_{e_{x_{1},y}}[\Pi_{+}]
\{\int_{{\cal U}_{y_{0}}}dy_{1}\sqrt{X^{+ '}X^{- '}}(y_{1}), h_{e_{x_{1},y}}[\Pi_{+}]^{-1}\},\right.\\
\hspace*{4.5in} h_{e_{x_{1},y_{0}}}[\Pi_{-}]\}\\
\hspace*{0.4in}\left. - h_{e_{x_{1},y_{0}}}[\Pi_{-}]^{-1}\{\int_{{\cal U}_{y_{0}}}dy\ h_{e_{x_{1},y}}[\Pi_{+}]^{-1}\{\int_{{\cal U}_{y_{0}}}dy_{1}\sqrt{X^{+ '}X^{- '}}(y_{1}), h_{e_{x_{1},y}}[\Pi_{+}]
\}, h_{e_{x_{1},y_{0}}}[\Pi_{-}]\}\right]\\
\vspace*{0.2in}
-\ \textrm{sgn}(X^{+ '} X^{- '})(y_{0})\left[h_{e_{x_{1},y_{0}}}[\Pi_{+}]\{\int_{{\cal U}_{y_{0}}}dy\ h_{e_{x_{1},y}}[\Pi_{-}]\{\int_{{\cal U}_{y_{0}}}dy_{1}\sqrt{X^{+ '}X^{- '}}(y_{1}), h_{e_{x_{1},y}}[\Pi_{-}]^{-1}
\},\right.\\
\hspace*{4.5in} h_{e_{x_{1},y_{0}}}[\Pi_{+}]^{-1}\}\\
\hspace*{0.4in} \left. - h_{e_{x_{1},y_{0}}}[\Pi_{+}]\{\int_{{\cal U}_{y_{0}}}dy\ h_{e_{x_{1},y}}[\Pi_{-}]^{-1}\{\int_{{\cal U}_{y_{0}}}dy_{1}\sqrt{X^{+ '}X^{- '}}(y_{1}), h_{e_{x_{1},y}}[\Pi_{-}]
\}, h_{e_{x_{1},y_{0}}}[\Pi_{+}]^{-1}\}\right].\\
\end{array}
\end{equation}
Here $h_{e_{x,y}}$ refers to the embedding momentum holonomy with unit charge $k_{e_{x,y}}= \frac{a}{\hbar}$ (see section 2.2.1 for the definition of the parameter $a$)
 along the edge
starting at the point $x$ and ending at the point $y$, the edge being oriented
along the orientation of the circle.
%
As we shall see, the  specific choice of the classical expression above  ensures that its quantum correspondent
acts trivially on those vertices of triangulation, which \emph{are not} non- trivial vertices of the acted upon state.

We now  approximate $\int_{{\cal U}_{y_{0}}}dy\ h_{e_{x_{1},y}}[\Pi_{\pm}]$ by a Riemann sum over simplices of the 
 triangulation $T$ (recall that $T$ is a triangulation of the interval $[0,2\pi]$. 
and replace the classical
objects by the corresponding quantum ones.  This yields a finite triangulation approximant to the operator 
$\widehat{\frac{1}{\sqrt{|X^{+ '} X^{- '}|}}}(y_{0})$. We shall refer to this approximant as 
$\widehat{\frac{1}{\sqrt{|X^{+ '} X^{- '}|}}}(y_{0})|_T$. We have that  
\footnote{Due to the negative density weight of the 
inverse metric, the approximant will have an overall factor of $|\triangle |$ but, as in LQG,  this factor
will cancel with other factors which appear in the definition of the Hamiltonian constraint and, as we shall see in the 
next section, yield a well defined $|\triangle | \rightarrow 0$ action of the constraint on diffeomorphism invariant 
states.}
\begin{equation}\label{eq:may21-2a}
\begin{array}{lll}
\widehat{\frac{1}{\sqrt{|X^{+ '} X^{- '}|}}}(y_{0})|_T\ = \\
\vspace*{0.2in}
-a^{-2}\left(\hat{h}_{e_{x_{1},y_{0}}}^{(-) -1}\widehat{\textrm{sgn}(X^{+ '}X^{- '})(y_{0})}\sum_{\triangle}|\triangle|
\kappa_{\epsilon}(b(\triangle),y_{0})\right.\\
\vspace*{0.1in}
\hspace*{2.8in}\left(\left[\hat{h}^{(+)}_{e_{x_{1},b(\triangle)}}\left[\hat{V}_{{\cal U}_{y_{0}}},\hat{h}^{(+) -1}_{e_{x_{1},b(\triangle)}}\right],
\hat{h}^{(-)}_{e_{x_{1},y_{0}}}\right]
\right.\\
\hspace*{2.75in}\left.\left.-\left[\hat{h}^{(+) -1}_{e_{x_{1},b(\triangle)}}\left[\hat{V}_{{\cal U}_{y_{0}}},
\hat{h}^{(+)}_{e_{x_{1},b(\triangle)}}\right],\hat{h}^{(-)}_{e_{x_{1},y_{0}}}\right]
\right)\right)\\
\vspace*{0.1in}
-\left(\hat{h}_{e_{x_{1},y_{0}}}^{(+)}\widehat{\textrm{sgn}(X^{+ '}X^{- '})(y_{0})}\sum_{\triangle}|\triangle|
\kappa_{\epsilon}(b(\triangle),y_{0})\right.\\
\vspace*{0.1in}
\hspace*{2.8in}\left(\left[\hat{h}^{(-)}_{e_{x_{1},b(\triangle)}}\left[\hat{V}_{{\cal U}_{y_{0}}},\hat{h}^{(-) -1}_{e_{x_{1},b(\triangle)}}\right],
\hat{h}^{(+) -1}_{e_{x_{1},y_{0}}}\right]
\right.\\
\hspace*{2.75in}\left.\left.-\left[\hat{h}^{(-) -1}_{e_{x_{1},b(\triangle)}}\left[\hat{V}_{{\cal U}_{y_{0}}},\hat{h}^{(-)}_{e_{x_{1},b(\triangle)}}\right],
\hat{h}^{(+) -1}_{e_{x_{1},y_{0}}}\right]
\right)\right)\\
\end{array}
\end{equation}
As in the previous section we choose the triangulation $T$ to be adapted to the charge network state it acts on 
by requiring that every vertex of the state is a vertex of $T$. 
We further restrict $T$
by requiring that  $x_1, y_0$ be vertices of $T$.
In order to obtain a concise operator action, we
shall also tailor the choice of $\epsilon >> |\triangle |$ to the location of the point $y_0$ as well as the
vertex structure of the graphs underlying the charge network state as follows. 

As before  let the state be $T_{s^+}\otimes T_{s^-}$ and let $V_E(s^+)\cup V_E(s^-)$ be the set of non-trivial vertices.
Given $y_0$
 we choose $\epsilon$ to be small enough that 
${\cal U}_{y_{0}}$ contains no non-trivial vertex of the graph other than one at $y_0$, if  such a vertex exists. 
It is then straightforward to check, using equation (\ref{volume=sumx}) that 
\begin{equation}\label{eq:may21-2}
\begin{array}{lll}
\widehat{\frac{1}{\sqrt{|X^{+ '} X^{- '}|}}}(y_{0})|_T T_{s^+}\otimes T_{s^-}\ =\\
\vspace*{0.2in}
-a^{-2}\left(\hat{h}_{e_{x_{1},y_{0}}}^{(-) -1}\widehat{\textrm{sgn}(X^{+ '}X^{- '})(y_{0})}\sum_{\triangle,\triangle_{1}}|\triangle|
\kappa_{\epsilon}(b(\triangle),y_{0})\kappa_{\epsilon}(b(\triangle_{1}),y_{0})\right.\\
\vspace*{0.1in}
\left(\left[\hat{h}^{(+)}_{e_{x_{1},y_{0}}}\left[\hat{V}_{b(\triangle_{1})},\hat{h}^{(+) -1}_{e_{x_{1},b(\triangle)}}\right],
\hat{h}^{(-)}_{e_{x_{1},y_{0}}}\right]
\right.\\
\left.\left.-\left[\hat{h}^{(+) -1}_{e_{x_{1},y_{0}}}\left[\hat{V}_{b(\triangle_{1})},
\hat{h}^{(+)}_{e_{x_{1},b(\triangle)}}\right],\hat{h}^{(-)}_{e_{x_{1},y_{0}}}\right]
\right)\right)\\
\vspace*{0.1in}
-\left(\hat{h}_{e_{x_{1},y_{0}}}^{(+)}\widehat{\textrm{sgn}(X^{+ '}X^{- '})(y_{0})}\sum_{\triangle,\triangle_{1}}|\triangle|
\kappa_{\epsilon}(b(\triangle),y_{0})\kappa_{\epsilon}(b(\triangle_{1}),y_{0})\right.\\
\vspace*{0.1in}
\left(\left[\hat{h}^{(-)}_{e_{x_{1},y_{0}}}\left[\hat{V}_{b(\triangle_{1})},\hat{h}^{(-) -1}_{e_{x_{1},b(\triangle)}}\right],
\hat{h}^{(+) -1}_{e_{x_{1},y_{0}}}\right]
\right.\\
\left.\left.-\left[\hat{h}^{(-) -1}_{e_{x_{1},y_{0}}}\left[\hat{V}_{b(\triangle_{1})},\hat{h}^{(-)}_{e_{x_{1},b(\triangle)}}\right],
\hat{h}^{(+) -1}_{e_{x_{1},y_{0}}}\right]
\right)\right)T_{s^+}\otimes T_{s^-} \\
\end{array}
\end{equation}

It is easy to show that the double commutators vanish if 
the points $b(\triangle), b(\triangle_{1}), y_{0}$ do not coincide on the circle. This  
simplifies the above expression 
to
\begin{equation}\label{eq:may21-1}
\begin{array}{lll}
\widehat{\frac{1}{\sqrt{|X^{+ '} X^{- '}|}}}(y_{0})|_{T}T_{s^+}\otimes T_{s^-}\ =\\
\vspace*{0.2in}
- a^{-2}\left(\hat{h}_{e_{x_{1},y_{0}}}^{(-) -1}\widehat{\textrm{sgn}(X^{+ '}X^{- '})(y_{0})}|\triangle|\right.\\
\vspace*{0.1in}
\hspace*{1.8in}\left(\left[\hat{h}^{(+)}_{e_{x_{1},y_{0}}}\left[\hat{V}_{y_{0}},\hat{h}^{(+) -1}_{e_{x_{1},y_{0}}}\right],\hat{h}^{(-)}_{e_{x_{1},y_{0}}}\right]
\right.\\
\hspace*{2.75in}\left.\left.-\left[\hat{h}^{(+) -1}_{e_{x_{1},y_{0}}}\left[\hat{V}_{y_{0}},\hat{h}^{(+)}_{e_{x_{1},y_{0}}}\right],\hat{h}^{(-)}_{e_{x_{1},y_{0}}}\right]
\right)\right)\\
\vspace*{0.1in}
-\left(\hat{h}_{e_{x_{1},y_{0}}}^{(+)}\widehat{\textrm{sgn}(X^{+ '}X^{- '})(y_{0})}|\overline{\triangle}|\right.\\
\vspace*{0.1in}
\hspace*{1.8in}\left(\left[\hat{h}^{(-)}_{e_{x_{1},y_{0}}}\left[\hat{V}_{y_{0}},\hat{h}^{(-) -1}_{e_{x_{1},y_{0}}}\right],\hat{h}^{(+) -1}_{e_{x_{1},y_{0}}}\right]
\right.\\
\hspace*{2.75in}\left.\left.-\left[\hat{h}^{(-) -1}_{e_{x_{1},y_{0}}}\left[\hat{V}_{y_{0}},\hat{h}^{(-)}_{e_{x_{1},y_{0}}}\right],\hat{h}^{(+) -1}_{e_{x_{1},y_{0}}}\right]
\right)\right)T_{s^+}\otimes T_{s^-}\\
\end{array}
\end{equation}
Note that the operator
$\hat{h}^{(\pm) -1}_{e_{x_{1},y_{0}}}\left[\hat{V}_{y_{0}},\hat{h}^{(\pm)}_{e_{x_{1},y_{0}}}\right]$ is only sensitive to
that part of $e_{x_{1},y_{0}}$ which overlaps with the 1-simplex which ends at $y_0$ 
(for $y_0= 0\equiv 2\pi$ this would be the `last' simplex ending at $2\pi$). It follows that the operator action 
(\ref{eq:may21-1}) is independent of the choice of $x_1$ (provided, of course, that $x_1\notin {\cal U}_{y_0}$),
just as is the case for the classical expression.

As detailed in the Appendix, a straightforward calculation shows that charge network states are eigen states of the 
inverse metric operator. Specifically, we have that 
\begin{equation}
\widehat{\frac{1}{\sqrt{X^{+ '} X^{- '}}(y_{0})}}|_T T_{s^+}\otimes T_{s^-} 
= |\Delta | \hbar a^{-2}\lambda (s^+, s^-, y_0) T_{s^+}\otimes T_{s^-}
\label{inverseeigen}
\end{equation}
where $\lambda (s^+, s^-, y_0)$ is as follows (see equation... in the Appendix).
$\lambda (s^+, s^-, y_0)$ vanishes 
if $y_0$ is not a non-trivial vertex i.e. if $y_0\notin V_E(s^+)\cup V_E(s^-)$. If $y_0 =v\in V_E(s^+)\cup V_E(s^-)$ then 
we have that 
\begin{equation}
\begin{array}{lll}\label{explicitevalue}
\lambda (s^+, s^-, v)\ =\\
\vspace*{0.1in}
-\textrm{sgn}(k^{+}_{e^{v}}-k^{+}_{e_{v}})\textrm{sgn}(k^{-}_{e^{v}}-(k^{-}_{e_{v}}+ \frac{a}{\hbar}))\\
\Big[\big[\sqrt{|k^{+}_{e^{v}}-(k^{+}_{e_{v}}-\frac{a}{\hbar})||k^{-}_{e^{v}}-(k^{-}_{e_{v}}+\frac{a}{\hbar})|}
-\sqrt{|k^{+}_{e^{v}}-(k^{+}_{e_{v}}-\frac{a}{\hbar})||k^{-}_{e^{v}}-k^{-}_{e_{v}}|}\big]\\
\vspace*{0.1in}
\hspace*{0.2in} -\big[\sqrt{|k^{+}_{e^{v}}-(k^{+}_{e_{v}}+\frac{a}{\hbar})|
|k^{-}_{e^{v}}-(k^{-}_{e_{v}}+\frac{a}{\hbar})|}-\sqrt{|k^{+}_{e^{v}}-(k^{+}_{e_{v}}+\frac{a}{\hbar})|
|k^{-}_{e^{v}}-k^{-}_{e_{v}}|}\big]\Big]\\
\vspace*{0.1in}
+ \textrm{sgn}(k^{+}_{e^{v}}-(k^{+}_{e_{v}}-\frac{a}{\hbar}))\textrm{sgn}(k^{-}_{e^{v}}-k^{-}_{e_{v}})\\
\Big[\big[\sqrt{|k^{+}_{e^{v}}-(k^{+}_{e_{v}}-\frac{a}{\hbar})||k^{-}_{e^{v}}-(k^{-}_{e_{v}}-\frac{a}{\hbar})|}
-\sqrt{|k^{+}_{e^{v}}-(k^{+}_{e_{v}})||k^{-}_{e^{v}}-(k^{-}_{e_{v}}-\frac{a}{\hbar})|}\big]\\
\vspace*{0.1in}
\hspace*{0.2in} -\big[\sqrt{|k^{+}_{e^{v}}-(k^{+}_{e_{v}}-\frac{a}{\hbar})|
|k^{-}_{e^{v}}-(k^{-}_{e_{v}}+\frac{a}{\hbar})|}-\sqrt{|k^{+}_{e^{v}}-k^{+}_{e_{v}}||k^{-}_{e^{v}}-(k^{-}_{e_{v}}+
\frac{a}{\hbar})|}\big]\Big]
\end{array}
\end{equation}

The discrete analog of the classical restrictions $\pm X^{\pm \prime}>0$ are the `positivity' conditions 
$\pm (k^{\pm}_{e^{v}}-k^{\pm}_{e_{v}})\geq 0$. Indeed, the sector of the polymer Hilbert space we shall be interested
in satisfies these conditions and for such states it is straightforward to see that \\
\noindent{(a)} $\lambda (s^+, s^-, v)\ \neq\ 0$ iff $v\in V_E(\gamma^{+}) \cup V_E(\gamma^{-})$. This property is intimately
tied to the operator ordering we have chosen, specifically, the positioning  of the signum operator in the expression.
\footnote{In LQG the inverse triad operator annihilates the
vertices which are annihilated by the volume operator. In contrast we are defining $\widehat{\frac{1}{\sqrt{|X^{+ '} X^{- '}|}}}(y_{0})$ such that it has non-vanishing action on 
\emph{all} non-trivial (embedding)  vertices of the underlying graph. 
Without this property all zero volume states would be in the kernel of the constraint, similar to LQG.
Such a kernel is much larger than the solution space constructed by Group Averaging
in \cite{alokme2}.}\\
\noindent{(b)} $\lambda (s^+, s^-, v=0)\ =\  \lambda (s^+, s^-, v=2\pi)$. \\
\noindent (c)  Whenever the argument of the signum function vanishes so does the factor multiplying it, thus obviating 
the necessity of defining the signum function for vanishing arguments.\\

Finally, it is straightforward to verify that for states  which satisfy the condition 
 $\pm \hbar (k^{\pm}_{e^{v}}-k^{\pm}_{e_{v}}) >> a$ (this condition partially captures the continuum condition
of existence of the second (spatial) derivative of the embedding variables),
 the expression  (\ref{explicitevalue}) has the correct continuum limit. The verification consists in
 showing that it defines a discrete
approximant to the continuum expression  
$-4\frac{d\sqrt{X^{+\prime}}}{dX^{+\prime}}
\frac{d\sqrt{-X^{-\prime}}}{dX^{-\prime}}$ which is just another way to write  $\frac{1}{\sqrt{-X^{+\prime}X^{-\prime}}}$

\section{The Hamiltonian constraint operator}

This section is devoted to the construction of the Hamiltonian constraint as an operator on the space of 
diffeomorphism invariant states. We follow the strategy used in LQG. Our aim is to first define a discrete approximant to
the Hamiltonian constraint on a triangulation of the spatial manifold, promote the expression to an operator on the 
kinematic Hilbert space and then show that its dual action on diffeomorphism invariant distributions, 
admits a well defined continuum limit.

From equation (\ref{hamconstraint}) the smeared Hamiltonian constraint with  lapse $N(x)$ is:
\begin{displaymath}
C_{ham}[N]\ =\ \int N(x)\left[\Pi_{+}(x)X^{+'}(x)\ -\ \Pi_{-}(x)X^{-'}(x)\ +\ \frac{1}{4}(\pi_{f}^{2}+f^{' 2})\right]\frac{1}{\sqrt{X^{+'}(x)X^{-'}(x)}} .
\end{displaymath}
On a triangulation $T$, a discrete approximant to the above expression is given by 
\begin{equation}\label{eq:jun14-1}
\begin{array}{lll}
C_{ham, T}[N]\ =\ \sum_{\triangle\in T}|\triangle|N(b(\triangle))\\
\vspace*{0.1in}
\hspace*{0.6in}\left[\Pi_{+}(b(\triangle))\left(\frac{X^{+}(m(\triangle))-X^{+}(m(\triangle-1) 
+L\delta_{b(\triangle ),0})}{|\triangle|}\right)-\right.\\
 \hspace*{0.9in}\Pi_{-}(b(\triangle))\left(\frac{X^{-}(m(\triangle))-X^{-}(m(\triangle-1)-L\delta_{b(\triangle ),0})}{|\triangle|}\right)\\
\vspace*{0.1in} 
\hspace*{1.2in} \left.+\frac{1}{4}(Y^{+})^{2}(b(\triangle)) + \frac{1}{4}(Y^{-})^{2}(b(\triangle))\right]\frac{1}{\sqrt{X^{+'}X^{-'}}}
(b(\triangle))
\end{array}
\end{equation}
where we have used the notation of section 4 so that $b(\triangle)$ is the beginning vertex of simplex $\triangle$,
$|\triangle|$ is its length and $m (\triangle )$ its midpoint.
The symbol
$\triangle-1$ denotes the simplex to the left of $\triangle$, and 
it is understood that if $\triangle_{1}$ is the left-most simplex with $b(\triangle_{1}) = 0$ then 
$m(\triangle_{1}-1)=m(\triangle_{N})$, with 
$\triangle_{N}$ being the right-most simplex such that $f(\triangle_{N})=2\pi$ where $f(\triangle )$ is the ending
vertex of $\triangle $. The Kronecker delta terms $L\delta_{b(\triangle ),0}$ take into account the quasiperiodic
nature of the embedding variables and, similar to the term  in the second line of equation (\ref{jun3-3b}) come into play
only for the first cell of the triangulation.
Since  only the holonomies of $\Pi_{\pm}, Y^{\pm}$ are well defined operators on ${\cal H}_{kin}$, the local 
fields $\Pi_{\pm}, (Y^{\pm})^2$ need to be approximated on $T$ by appropriate combinations of holonomies.

Recall that the solutions of \cite{alokme2} are obtained via averaging 
over the unitary action of finite gauge transformations.
Since the unitary representation of
gauge transformations is not weakly continuous on the kinematic Hilbert space, we cannot directly define their
putative generators as operators there. Since the classical constraints are the  generators  of such transformations
it seems impossible to define the Hamiltonian constraint in such away that it kills the solutions of \cite{alokme2}.
We get around this potential obstruction by the pursuing the following {\em key} idea.

Note that the solutions of Reference \cite{alokme2}  are invariant 
under the action of any finite gauge transformation and hence would be annihilated by the difference of a {\em finite}
gauge transformation and the identity operator. Note also that  the LQG strategy is to first define an operator on the 
kinematic Hilbert space at {\em finite} triangulation and then take the continuum limit. This suggests that 
we seek finite triangulation holonomy approximants to the various local fields of interest in such a way that 
$C_{ham, T}[N]$ is proportional to a combination of finite gauge transformations minus the identity, with the finite
gauge transformations being parametrized by $|\triangle|$ so that at $|\triangle|=0$ the gauge transformations are
just identity.

We display exactly such  approximants to $\Pi_{\pm}$ in section 5.1. The holonomies turn out to be state
dependent \footnote{As we will see shortly, this dependence involves the eigen-values of the 
embedding operators ${\hat X^{\pm}}$. Since ${\hat X^{\pm}}$ are the analogs of the LQG densitized triad operators, 
this feature is reminiscent of the $\bar{\mu}$ scheme employed in the improved dynamics in LQC\cite{aplimproved}.},
 the approximants yield $\Pi_{\pm} (b( \triangle ))$ to leading order in $|\triangle|$ and the resulting 
`$\Pi_{\pm} X^{\pm \prime }$' terms are proportional to a the difference of a finite gauge transformation and identity.
It is then easy to guess the form of the matter terms which contributes the correct finite gauge transformation
on the matter sector. 
However the tricky part is to realise these terms
as  approximants to the $(Y^{\pm})^2$  terms.
As far as we can see, the terms we need do not correspond to (the replacements by quantum operators of) classical 
functions which yield $(Y^{\pm})^2 (b(\triangle))$ upto higher order terms in $|\triangle |$.
Instead they may be derived as operators whose action  approximates that
 of the Hamiltonian vector fields of $(Y^{\pm})^2$.
We present these derivations in section 5.2. In our opinion this  provides a useful lesson for LQG: We may have to be
similarly open minded there in our search for finite triangulation approximants to local fields such as the 
curvature of the Ashtekar- Barbero connection if we want a satisfactory definition of the quantum dynamics of LQG.

We use the approximants defined in sections 5.1 and 5.2 to define  finite triangulation 
approximants to the Hamiltonian constraint in 
section 5.3  and show the resulting operator action has a continuum limit on diffeomorphism invariant states.

\subsection{Embedding Momenta Approximants}
We shall, as usual, focus on the left moving variables.
Let $v$ be a vertex of $T$ such that $v=b (\triangle )$. 
The traditional choice in LQG corresponds to the embedding momentum approximant $\Pi^{\triangle , k }_+ (v)$ defined
through the holonomy, $h^{(+),k}_{\triangle}$ with charge $k$ over the edge $\triangle$ as
\begin{equation}
\hat{\Pi}_{+}^{\triangle, k}(v)\ =\ \frac{i}{|\triangle|k}(\hat{h}_{\triangle}^{+,k}-1).
\label{tradpi}
\end{equation}
where $\hat{h}_{\triangle}^{+ k}\ =\ \widehat{e^{-ik\int_{\triangle}\Pi_{+}}}$. Clearly equation (\ref{tradpi}) can be
promoted to an operator on ${\cal H}_{kin}$.

Another possible choice is to allow the charge label of the holonomy to depend on that of the embeding charge
network state it acts on.
Accordingly we  fix the state dependent triangulation $T$ of section 4 (recall that every non-trivial embedding 
vertex of the state is a vertex of $T$). We further restrict $T$ to be fine enough that for any pair of 
successive vertices of $T$ only one, at most, is non- trivial and define ${\hat\Pi}^{\triangle}_+ (v)$ as:
\begin{equation}\label{newpi}
\begin{array}{lll}
\hat{\Pi}_{+}^{\triangle}(v)T_{s^{+}} =\ \frac{i}{|\triangle|(k^{+}_{e_{v}}-k^{+}_{e^{v}})}(\hat{h}_{\triangle}^{(k^{+}_{e_{v}}-k^{+}_{e^{v}})}-1)T_{s^{+}}
\textrm{if}\ v \in V_E(\gamma^{+})\\
\vspace*{0.1in}
\hat{\Pi}_{+}^{\triangle}(v)T_{s^{+}}\ =\ \frac{i}{|\triangle|}(\hat{h}^{(+)}_{\triangle}-1)T_{s^{+}}\ 
\textrm{if}\ v\notin V_E(\gamma^{+})\\.
\end{array}
\end{equation}
Here, as in section 4.1, ${e_v}$ $({e^v})$ refer to the edges which terminate (originate) at $v$,
 $V(\gamma^{+})$ refers to the set of non-trivial vertices where $k^+_{e_v}- k^+_{e^v}\neq 0$
where $k^+_{e_v}$ $(k^+_{e^v})$ for the first (last) edge are defined through the quasi- periodic extension
of the state (see section 2.2.4) and we have set
$\hat{h}^{k^{+}_{e_{v}}-k^{+}_{e^{v}}}_{\triangle}\ :=\ \widehat{e^{-i(k^{+}_{e_{v}}-k^{+}_{e^{v}})
\int_{\triangle}\Pi_{+}}}$, 
$\hat{h}_{\triangle}^{+}\ :=\ \widehat{e^{-i\int_{\triangle}\Pi_{+}}}$.

Next, we show that the above choice (\ref{newpi}) directly leads to an operator action of the `$+$' embedding part of the 
$C_{ham, T}$ which is a finite diffeomorphism on the $+$ part of the embedding state. As we shall see,
this will finally lead to a satisfactory definition of the Hamiltonian constraint in section 5.3.
With the above definition, the approximant to the $\Pi_+X^{+\prime}$ term is,
\begin{equation}
\begin{array}{lll}
\hat{\Pi}_{+}\hat{X}^{+ '}(b (\triangle ))|_TT_{s^{+}} :=\ \frac{1}{|\triangle|}\hat{\Pi}_{+}^{\triangle}(b(\triangle))
\big(\hat{X}^{+}(m(\triangle))-\hat{X}^{+}(m(\triangle -1 ))\big)T_{s^{+}}\\
\vspace*{0.1in}
=\ \frac{-i\hbar}{|\triangle|^{2}}(\hat{h}_{\triangle}^{k^{+}_{e_{v}}-k^{+}_{e^{v}}}-1)T_{s^{+}}\ \textrm{if}\ 
b (\triangle )\in V_E(\gamma^{+})\\
\vspace*{0.1in}
=\ 0\ \textrm{if}\ b (\triangle )\notin\ V_E(\gamma^{+})  
\end{array}
\end{equation}

Whence for $x\in V_E(\gamma^{+})$,
\begin{equation}
\begin{array}{lll}\label{phideltaemb+}
\hat{\Pi}_{+}\hat{X}^{+ '}( b (\triangle ) )|_TT_{s^{+}}\ =\\
\vspace*{0.1in} 
\hspace*{0.7in}\frac{-i\hbar}{|\triangle|^{2}}(\hat{h}_{\triangle}^{k^+_{e_{v}}-k^{+}_{e^{v}}}-1)T_{s^{+}}=\\
\vspace*{0.1in}
\hspace*{0.7in} \frac{-i\hbar}{|\triangle|^{2}}(T_{s^{+}_{\phi_{\triangle}}} - T_{s^{+}})
= \frac{-i\hbar}{|\triangle|^{2}}({\hat U}^{+,E}( \phi_{\triangle}) -1) T_{s^{+}}
\end{array}
\end{equation}

Here,$\phi_{\triangle}$ is a diffeomorphism of the circle 
(more precisely,  $\phi_{\triangle}$ is a periodic diffeomorphism of the real line) which is identity 
in the neighbourhood of all 
the vertices of $T$ except $b(\triangle ), f(\triangle )$.
\footnote{In addition, $\phi_{\triangle}$ 
also differs from identity at $x=2\pi$ (or $x=0$) if $b(\triangle )=0$ (or $f(\triangle )= 2\pi$); this is just a 
consequence of the circular topology of space.\label{fnphidelta}}
 Further,$\phi_{\triangle}$  maps $b (\triangle )$ to $f(\triangle )$  and its action on the charge network label $s^+$
is denoted by 
$s^{+}_{\phi_{\triangle}}$ as in  section 2.2.4. 
${\hat U}^{+,E}(\phi_{\triangle})$ is the restriction of the unitary action of the finite gauge transformation 
${\hat U}^{+}(\phi^+=\phi_{\triangle} )$ (see section 2.2.4) to the left- moving embedding Hilbert space. 
Finally note that we could as well have chosen
$\hat{\Pi}_{+}^{\triangle}(v^{+})T_{s^{+}}\ =\ \frac{-i}{|\triangle|(k^{+}_{e_{v}}-k^{+}_{e^{v}})}
(({\hat{h}_{\triangle}^{(k^{+}_{e_{v}}-k^{+}_{e^{v}})}})^{\dagger}-1)T_{s^{+}} $ and we would have obtained
the inverse diffeomorphism with a negative sign. We shall use this flexibility according to our convenience.

The analysis of the right  moving mode proceeds in a similar way.

\subsection{Matter field approximants}

Let $T$ be the triangulation of section 4.1 with the restriction that every nontrivial  matter vertex of the charge 
network state on which ${\hat C}_{ham, T}$ acts is a vertex of $T$ and no two succesive vertices of $T$ are nontrivial.
A nontrivial left moving (or right moving) matter vertex $v$ is one for which 
$l^+_{e_v}- l^+_{e^v}\neq 0$ (or $l^-_{e_v}- l^-_{e^v}\neq 0$).
Here, similar to section 4.1, ${e_v}$ $({e^v})$ refer to the edges which terminate (originate) at $v$ 
and $l^{\pm}_{e_v}$ $(l^{\pm}_{e^v})$ for the first (last) edge are defined through the periodic extension
of the matter charge network (see section 2.2.4).
We shall refer to the set of non-trivial matter vertices as $V_M(\gamma^{\pm})$.

The traditional LQG type approximant, similar to how curvature terms are quantized in lattice gauge theories, 
is of the form:
\begin{equation}
(Y^{\pm})^{2}(x=b(\triangle))|_T\:
 =\ \frac{e^{im\int_{\triangle}Y^{+}}+e^{-im\int_{\triangle}Y^{+}}-2}{m^{2}|\triangle|^{2}} .
\end{equation}
We are unable to see how such a choice could lead to an operator action which is  that of 
a finite diffeomorphism. Hence, it seems
unlikely that with this choice,  the physical states of \cite{alokme2} are in the kernel of the 
Hamiltonian constraint. 

Instead, as we shall see, the following choice yields a satisfactory definition of the Hamiltonian constraint:
\begin{equation}
\widehat{(Y^{+})^{2}}(b (\triangle ))_TW_{s^{+}} =\ \frac{-4i\hbar}{|\triangle |^{2}}[\ e^{-i\frac{\hbar}{2}(l_{e_{I}^{+}}^{+}-l_{e_{I+1}^{+}}^{+})^{2}}\hat{h}_{\triangle}^{l_{e_{I}^{+}}^{+}-l_{e_{I+1}^{+}}^{+}}\ - 1\ ]W_{s^+}
\label{y2=h}
\end{equation}
which can also be rewritten as  
\begin{equation}
\widehat{(Y^{+})^{2}}(b(\triangle ))_T = \ \frac{-4i\hbar}{|\triangle |^{2}} ({\hat U}^{+,M}({\phi_{\triangle}}) -1),
\label{y2=u}
\end{equation}
where $\phi_{\triangle}$ has been defined in the previous section and ${\hat U}^{+,M}({\phi_{\triangle}})$ is the 
restriction of the finite gauge transformation operator  ${\hat U}^{+}(\phi^+={\phi_{\triangle}})$ 
to the matter Hilbert space.
This choice does not arise straightforwardly as the evaluation of some finite $\triangle$ approximant
to $(Y^+(x))^2$ as in the case of the embedding momentum. Rather, as we now argue,
its justification lies in an analysis of the Hamiltonian vector field generated by 
the corresponding classical quantity. In what follows we use the notation  $A \approx B$ to indicate that $A$ and
$B$ agree to leading order in $|\triangle |$.

Let ${\bar{\triangle}}$ be the interval obtained by joing the points $m(\triangle -1 )$ and $m (\triangle )$ on the 
circle. Let $x$ be the coordinate system which defines $T$ so that the 1-simplices of $T$ are of equal length 
$|\triangle |$. Let $s^+$ be a matter charge network label such that every vertex of $s^+$ is a vertex of $T$ and
no pair of successive vertices of $T$ are nontrivial vertices of $s^+$ (Since we are ultimately interested 
in the $|\triangle |\rightarrow 0$ limit, this is not an unreasonable restriction).

Then it is straightforward to check the following:
\begin{eqnarray}
Y^{+2}(b (\triangle )) & {\approx} &  \frac{1}{|\triangle |}\int_{\bar{\triangle}}(Y^+)^2dx , \\
\{\int_{\bar{\triangle}}(Y^+)^2dx , W({s^+})(Y^+)\}& {\approx} &  \frac{-4}{|\triangle |}
                                             (W(s^+_{\phi_{\triangle}})(Y^+)- W(s^+)(Y^+))\nonumber \\ 
 {\approx}  & \frac{-4}{|\triangle |}
                                             (W(s^+_{\phi_{\triangle}})- W(s^+)) &
                                                  ( 1- \frac{|\triangle |^2}{4i\hbar}       
                                                 Y^{+2}(b (\triangle ))) ,
\label{approxpb}
\end{eqnarray}
where the validity of the last line lies in the fact that $\hbar$ remains non- zero in the continuum limit defined
by $|\triangle |\rightarrow 0$. 

If we define $\widehat{(Y^{+}(b (\triangle)))^2_T}$ by 
\begin{equation}
\widehat{(Y^{+}(b (\triangle)))^2_T}:= 
\frac{-4i\hbar}{|\triangle |^2} ({\hat U}_{\phi_{\triangle}} -1),
\end{equation}
it is easy to check that 
\begin{equation}
[\widehat{(Y^{+}(b (\triangle)))^2_T}, {\hat W}(s^+) ]= \frac{-4i\hbar}{|\triangle |}
({\hat W}(s^+_{\phi_{\triangle}})- {\hat W}(s^+)) ( 1- \frac{|\triangle |^2}{4i\hbar}       
                                                \widehat{(Y^{+}(b (\triangle)))^2_T}.
\label{y255}
\end{equation}
Thus, the definition (\ref{y2=h}) yields a representation of the approximant to the Poisson bracket
$\{(Y^{+}(b (\triangle))^2_T, W (s^+) \}$ and this is the justification for the choice (\ref{y2=h}).

The following observations offer further evidence that as far as the 
continuum limit of the ensuing Hamiltonian constraint operator is concerned,
 our choice (\ref{y2=h}) is a reasonable one. First, 
consider the one parameter family of gauge transformations
labelled by the one parameter family of diffeomorphims 
$\phi^+ (\lambda )$ and generated by the function $\int_{S^1} N_+(x) (Y^+(x))^2$ for some smooth smearing function $N_+$.
Then we have that $\lim_{\lambda \rightarrow 0}\frac{U_{\phi^+ (\lambda )}-1}{\lambda} |\circ \rangle =0$, 
where $|\circ \rangle$ is the state with  vanishing  matter charges.
This is supportive of the putative operator identity $({\hat Y}^+)^2 (x) |\circ \rangle =0$ which implies that it 
is reasonable to impose the condition
\begin{equation}
\widehat{(Y^{+}(b (\triangle))^2_T}|\circ \rangle =0 \;\;\forall \; \triangle \in T .
\label{y255.1}
\end{equation}
Next, restrict attention to matter charge nets whose underlying graph can be chosen to be 
(coarser than or equal to)
$T$. Let the Weyl algebra of matter holonomies labelled by such charge nets be ${\cal W}^M_T$. 
Let the Hilbert space of states labelled by such charge nets be ${\cal H}^M_T$.
Clearly, ${\cal H}^M_T$ supports a cyclic representation of 
${\cal W}^M_T$ with the cylic state $|\circ\rangle$.
By virtue of the cyclicity of $|\circ\rangle$, it is easy to see that 
$\widehat{(Y^{+}(b (\triangle))^2_T}$ is uniquely specified as an operator on ${\cal H}^M_T$ by 
equations (\ref{y255}) and (\ref{y255.1}).

\subsection{The Hamiltonian constraint operator at finite triangulation}

Recall that our aim is to write the Hamiltonian constraint operator in terms of the 
difference of a finite gauge transformation and identity.  From equation (\ref{hamconstraint}), the Hamiltonian
constraint is proportional to the difference, $H_+- H_-$, of the generators of the `+' and`-' gauge transformations.
From considerations similar to those of section 3.1, it follows that the transformations generated by the combination
$H_+- H_-$ correspond to gauge transformations for which $\phi^+= (\phi^{-})^{-1}$. Given that the left moving operators
of equations (\ref{phideltaemb+}), (\ref{y2=u}) are associated with diffeomorphisms which displace  vertices to their
right (i.e. anticlockwise on the circle), 
this suggests that we construct the right moving operators in terms of diffeomorphisms which displace vertices to their
left (i.e. clockwise on the circle). 

Accordingly, we define:
\begin{equation}\label{newpi-}
\begin{array}{lll}
\hat{\Pi}_{-}^{\triangle}(v)T_{s^{-}} =\ \frac{-i}{|\triangle|(k^{-}_{e_{v}}-k^{-}_{e^{v}})}
((\hat{h}_{\triangle}^{(k^{-}_{e_{v}}-k^{-}_{e^{v}})})^{\dagger}-1)T_{s^{-}}
\textrm{if}\ v \in V_E(\gamma^{-})\\
\vspace*{0.1in}
\hat{\Pi}_{-}^{\triangle}(v)T_{s^{-}}\ =\ \frac{i}{|\triangle|}(\hat{h}^{(-)}_{\triangle}-1)T_{s^{-}}\ 
\textrm{if}\ v\notin V_E(\gamma^{-}),\\
\end{array}
\end{equation}
and
\begin{equation}
\widehat{(Y^{-})^{2}}(b(\triangle ))_T = \ \frac{-4i\hbar}{|\triangle |^{2}} 
\ ({\hat U}^{-,M}({\phi^{-1}_{\triangle -1}}) -1).
\label{y2=u-}
\end{equation}
It is straightforward to derive equations (\ref{newpi-}), (\ref{y2=u-}) along the lines of section 5.1 and 5.2 and to see 
that for $x\in V_E(\gamma^{-})$,
\begin{equation}
\hat{\Pi}_{-}\hat{X}^{- '}( b (\triangle ) )|_TT_{s^{-}} =
\frac{i\hbar}{|\triangle|^{2}}(T_{s^-_{ \phi^{-1}_{\triangle -1}}  } - T_{s^{-}})
=\frac{i\hbar}{|\triangle|^{2}} ({\hat U}^{-,E}({\phi^{-1}_{\triangle -1}}) -1)  T_{s^{-}}.
\label{phideltaemb-}
\end{equation}

Here, $(\triangle -1) \in T $ refers to the edge immediately preceding $\triangle$ with $\triangle -1 = \triangle_N$ if
$b (\triangle )= 0$. The diffeomorphism $\phi_{\triangle}$ for any ${\triangle} \in T$ has already been defined
in section 5.1 and $\phi^{-1}_{\triangle }$ denotes the inverse of $\phi_{\triangle}$. Thus,
$\phi^{-1}_{\triangle -1}$ maps $b(\triangle )$ to $b(\triangle -1 )$ and is identity on all vertices other 
$b(\triangle ), b(\triangle -1 )$ (modulo the identifications $x=0 \sim x=2\pi$, see Footnote \ref{fnphidelta}).
Finally, ${\hat U}^{-,M}({\phi^{-1}_{\triangle -1}}), {\hat U}^{-,E}({\phi^{-1}_{\triangle -1}})$ refer to the 
restriction of the unitary action of the finite gauge transformation
${\hat U}^{-}(\phi^-= {\phi^{-1}_{\triangle -1}})$ to the matter and embedding sectors.

Next, recall that the finite gauge transformation labelled by $\phi^+$ moves the right moving embedding fields
and the right moving matter fields {\em together}. The same is true for the corresponding objects in the `$-$' sector.
If the classical quantities in (\ref{eq:jun14-1}) are replaced by their corresponding operators through
equations (\ref{newpi}),(\ref{y2=u}),(\ref{newpi-}),(\ref{y2=u-}), it is immediate to see that the action of the 
resulting 
constraint operator on a charge network state 
is the {\em sum} of gauge transformations each acting only on the matter part of the state or only on the embedding
part of the state. This is not what we desire and is remedied as follows.

Equation (\ref{eq:jun14-1}) is a discrete approximant to the continuum expression and we may modify it by terms which 
vanish in the continuum limit, $|\triangle |\rightarrow 0$. It is straightforward to see that the following expression
is one such modification:
\begin{eqnarray}
C_{ham, T}[N] &=& \sum_{\triangle\in T}\frac{-i\hbar N(b(\triangle))}{|\triangle |\sqrt{X^{+'}X^{-'}}} \nonumber\\
&\; &
\big(
[1+\frac{|\triangle |^2}{-i\hbar}\Pi_{+}(b(\triangle))X^{+\prime}(b (\triangle )) ]
[1+ \frac{|\triangle |^2}{i\hbar}\Pi_{-}(b(\triangle))(b (\triangle )) ] \nonumber \\
&\; & 
[1 + \frac{|\triangle |^2}{-4i\hbar}(Y^{+})^{2}(b(\triangle))]
[1 + \frac{|\triangle |^2}{-4i\hbar}(Y^{-})^{2}(b(\triangle)) ]\big)
\label{hamconstrproduct}
\end{eqnarray}
Replacing the classical quantities in the above equation by their quantum operators through equations
(\ref{phideltaemb+}),(\ref{y2=u}),(\ref{y2=u-}) and (\ref{phideltaemb-}) and  ordering the constraint operator so that
the inverse metric is rightmost, the action of the quantum Hamiltonian constraint at finite triangulation
on the state $\vert \bf{s^+},\bf{s^-}\rangle$ is:
\begin{eqnarray}
{\hat C}_{ham, T}[N]\vert {\bf s^+},{\bf s^-}\rangle&=& 
\sum_{\triangle\in T, b(\triangle )\in V_E(s^+)\cup V_E(s^-)}
 N(b(\triangle))\nonumber \\
& &
[{\hat U}^{+,E}(\phi_{\triangle})\otimes{\hat U}^{-,E}(\phi^{-1}_{\triangle -1 })\otimes{\hat U}^{+,M}(\phi_{\triangle})\otimes{\hat U}^{-,M}(\phi^{-1}_{\triangle -1 }) -1] \nonumber \\
& & \frac{-i\hbar}{|\triangle |\hat{\sqrt{X^{+'}X^{-'}}}}\vert {\bf s^+},{\bf s^-}\rangle
\end{eqnarray}
Using equation (42) and the fact that the unitary operators in the above equation are just restricted actions of 
unitary operators associated with finite gauge transformations, we obtain:
\begin{eqnarray}
{\hat C}_{ham, T}[N]\vert {\bf s^+},{\bf s^-}\rangle&=& 
\sum_{\triangle\in T, b(\triangle )\in V_E(s^+)\cup V_E(s^-)}
N(b(\triangle)) \frac{-i\hbar}{a^2}\lambda (s^+, s^-, b(\triangle ))\nonumber \\
& &
[{\hat U}^{+}(\phi_{\triangle})\otimes{\hat U}^{-}(\phi^{-1}_{\triangle -1 })-1]
\vert {\bf s^+},{\bf s^-}\rangle .
\label{qhamt}
\end{eqnarray}
Clearly, the above action kills any state invariant under all finite gauge transformations generated by $H_+, H_-$
and thus provides a satisfactory definition of the Hamiltonian constraint at finite triangulation.
We now show that the above action admits a continuum limit on the space of diffeomorphism invariant distributions.

\subsection{The continuum limit of the action of ${\hat C}_{ham, T}$ on ${\cal H}_{kin }$}
Let us summarise the properties of the triangulation $T$:\\
\noindent (i) $T$ depends on the (coarsest) graph $\gamma :=\gamma ({\bf s^+}, {\bf s^-})$ underlying the state
$\vert {\bf s^+},{\bf s^+}\rangle$ on which  ${\hat C}_{ham, T}$ acts.\\
\noindent (ii) Every vertex of the graph is a vertex of $T$.\\
\noindent (iii) No two successive vertices of $T$ are non- trivial (matter or embedding) vertices of 
$\gamma ({\bf s^+}, {\bf s^-})$.\\
\noindent (iv) There is a coordinate system in which every edge $\triangle $ of $T$ has the same length
$|\triangle |$.

We shall often emphasise (i) and (iv) above by setting  $\delta :=|\triangle |$ and denoting $T$ by 
$T(\gamma, \delta )$. Consider a 1 parameter family of triangulations $T(\gamma, \delta )$, parameterised by 
$\delta >0$ for fixed $\gamma $. The continuum limit of any quantity defined on $T(\gamma, \delta )$ is its
limiting beaviour as $|\triangle |\rightarrow 0$.
In what follows it is convenient to change our notation for $\phi_{\triangle}, \phi^{-1}_{\triangle -1}$.
Accordingly, for $v:=b(\triangle )$
we set 
$\phi_{\triangle}=: \phi_{v,\delta}$ and $\phi^{-1}_{\triangle -1}=: \phi_{v,-\delta}$. The notation signifies
that $\phi_{v,\delta}$ moves the point $v$ to the point $v+\delta $ on the circle and 
$\phi_{v,-\delta}$ moves the point  $v$ to the point $v-\delta$ on the circle.

Let $\Psi \in {\cal H}_{diff}$ be a diffeomorphism invariant distribution.
From equation (\ref{qhamt}), in our new notation, we have that 
\begin{eqnarray}
&\Psi({\hat C}_{ham, T(\gamma, \delta) }[N]\vert {\bf s^+},{\bf s^-}\rangle ) \nonumber \\
&=\sum_{v\in V_E(s^+)\cup V_E(s^-)}
N(v) \frac{-i\hbar}{a^2}\lambda (s^+, s^-, v)
\Psi([{\hat U}^{+}(\phi_{v,\delta})\otimes{\hat U}^{-}(\phi_{v,-\delta })-1]
\vert {\bf s^+},{\bf s^-}\rangle) \nonumber \\
&=\sum_{v\in V_E(s^+)\cup V_E(s^-)}
N(v) \frac{-i\hbar}{a^2}\lambda (s^+, s^-, v)
[\Psi (\vert {\bf s^+}_{\phi_{v,\delta}},{\bf s^-}_{\phi_{v,-\delta}}\rangle)
- \Psi (\vert {\bf s^+},{\bf s^-}\rangle)].\nonumber\\
\label{qhamtpsi}
\end{eqnarray}
It is easy to see, for any two triangulations $T(\gamma, \delta_1 ), T(\gamma, \delta_2 )$ and 
any vertex $v \in  V_E(s^+)\cup V_E(s^-)$, that there exists a diffeomorphism 
$\phi(v, \delta_1, \delta_2)$ such that
\begin{equation}
\vert {\bf s^+}_{\phi_{v,\delta_1}},{\bf s^-}_{\phi_{v,-\delta_1}}\rangle
= {\hat U}(\phi(v, \delta_1, \delta_2))
\vert {\bf s^+}_{\phi_{v,\delta_2}},{\bf s^-}_{\phi_{v,-\delta_2}}\rangle
\end{equation}
where ${\hat U}(\phi(v, \delta_1, \delta_2))$ is the unitary operator corresponding to the spatial diffeomorphism
$\phi(v, \delta_1, \delta_2)$ so that, from section 3.1, 
${\hat U}(\phi(v, \delta_1, \delta_2)):= {\hat U}^+(\phi(v, \delta_1, \delta_2))\otimes
{\hat U}^-(\phi(v, \delta_1, \delta_2))$.
It is then immediate from equation (\ref{qhamtpsi}) that 
\begin{equation}
\Psi({\hat C}_{ham, T(\gamma, \delta_1) }[N]\vert {\bf s^+},{\bf s^-}\rangle ) 
=\Psi({\hat C}_{ham, T(\gamma, \delta_2) }[N]\vert {\bf s^+},{\bf s^-}\rangle ) 
\end{equation}
so that 
\begin{equation}
\lim_{\delta\rightarrow 0}
\Psi({\hat C}_{ham, T(\gamma, \delta ) }[N]\vert {\bf s^+},{\bf s^-}\rangle )
=\Psi({\hat C}_{ham, T(\gamma, \delta_0) }[N]\vert {\bf s^+},{\bf s^-}\rangle )
\label{contlimhamdiff}
\end{equation}
Here,  the choice of $\delta_0 >0 $ is arbitrary (subject, of course, to the restriction that $T(\gamma, \delta_0)$
satsifies properties (i)- (iv) above).
Equation (\ref{contlimhamdiff}) shows that the dual action of the 
Hamiltonian constraint operator (\ref{qhamt}) possesses a well defined
continuum limit.
 Indeed, this conclusion is unchanged if $\Psi$ is any diffeomorphism invariant 
distribution  i.e. $\Psi$ has a finite action on the (dense) space of finite linear combinations of charge network states
and $\Psi$ is invariant under the (dual) action of the unitary operators of section  3.1
 which implement spatial diffeomorphisms on 
${\cal H}_{kin}$. In particular it follows that the continuum limit of the Hamiltonian constraint operator 
annihilates the physical states of \cite{alokme2} (see section 2.2.5)
 which are obtained by group averaging over the action of $H_{\pm}$.

Finally, we note that as explained beautifully  by Thiemann in \cite{bigbook}, equation (\ref{contlimhamdiff}) 
shows that the one parameter family of triangulated operators ${\hat C}_{ham T(\gamma, \delta)}[N]$ 
converges to a (non- unique) densely defined operator ${\hat C}_{ham}[N]$ on the {\em kinematic} Hilbert space
in the so called
Uniform Rovelli- Smolin (URS) topology. Specifically, in the notation used above, we may choose the limit of the 
one parameter family ${\hat C}_{ham T(\gamma, \delta)}[N]$ to be the operator ${\hat C}_{ham}[N]$ where
\begin{equation}\label{urst}
{\hat C}_{ham}[N]|{\bf s}^{+},{\bf s}^{-}\rangle\ :=\ 
{\hat C}_{ham T(\gamma, \delta_{0})}[N]|{\bf s}^{+},{\bf s}^{-}\rangle
\end{equation}

\section{The constraint algebra and the arena of  diffeomorphism invariant distributions}
Given a distribution $\Psi$ (more precisely,  an element of the algebraic dual to the superselected sector 
${\cal D}_{ss}$ of section 2.2.5) and a charge network $\vert {\bf s^+},{\bf s^-}\rangle$, we would like to check
if 
\begin{equation}
\Psi ([\hat{C}_{ham}[N_2],\hat{C}_{ham}[N_1]]\vert {\bf s^+},{\bf s^-}\rangle) = \Psi( \widehat{C_{diff}[\vec{\beta}(N,M)]}\vert {\bf s^+},{\bf s^-}\rangle .
\label{hathhdiff}
\end{equation}
where the embedding dependent structure function $\vec{\beta}(N,M)$ has been defined in (\ref{eq:structure}). Section 6.1 
is devoted  to the definition and evaluation of the left hand side of the above equation and  
section 6.2 to the right hand side when $\Psi$ is diffeomorphism invariant.
We find that both sides of the equation vanish as is the case in LQG \cite{tthh}.
We structure our computations so that they are of use for evaluations in which $\Psi$ is not diffeomorphism invariant but 
lies in  a suitable `habitat'. We shall explore the constraint algebra on such habitats in sections 7 and 8.
\subsection{The commutator of 2 Hamiltonian constraints}
In this section we compute the continuum limit of the commutator between 2 Hamiltonian constraints on the space of
diffeomorphism invariant distributions. Since the Hamiltonian constraint does not map the space of such distributions
to itself, we proceed along the lines of Thiemann's seminal work \cite{tthh}. Specifically, we define the left hand side
of equation (\ref{hathhdiff}) through:
\begin{eqnarray}
\Psi ([\hat{C}_{ham}[N_2],\hat{C}_{ham}[N_1]]\vert {\bf s^+},{\bf s^-}\rangle):=
\;\;\;\;\;\;\;\;\;\;\;\;\;\;\;\;\;\;\;\;\;\;\;\;\;\;\;\;\;\;\;\;\;\;\;\;\;\;&  \nonumber \\
\lim_{\delta^{\prime}\rightarrow 0}
\lim_{\delta\rightarrow 0}
\Psi({\hat C}_{ham, T^{\prime}(\delta^{\prime}) }[N_2]{\hat C}_{ham, T(\delta) }[N_1]
-{\hat C}_{ham, T^{\prime}(\delta^{\prime}) }[N_1]{\hat C}_{ham, T(\delta) }[N_2]
\vert {\bf s^+},{\bf s^-}\rangle ). &\nonumber\\
\label{h,hcont}
\end{eqnarray}
Here $T(\delta ):= T(\gamma, \delta )$ is a triangulation adapted to $\vert {\bf s^+},{\bf s^-}\rangle$
$T^{\prime}(\delta^{\prime} )$ is a refinement of $T(\delta )$ and has 1- cells of size $\delta^{\prime}<<\delta$ 
in the same 
coordinate system in which $T(\delta )$ has 1- cells of size $\delta$. Further  $T^{\prime}(\delta^{\prime} )$
is adapted to the charge networks which appear on the right hand side of equation (\ref{qhamt}) 
(with the appropriate replacement of $N$ by
$N_1$ or $N_2$). It is easy to see that for small enough $\delta^{\prime}<<\delta$, such triangulations always exist.

Recall that $T, T^{\prime}$ are subject to the conditions (i)- (iv) of section 5.4. In addition we shall, for simplicity,
require that $\delta$ be small enough that (iii) is strengthened to the condition that 
non- trivial vertices of $\vert {\bf s^+},{\bf s^-}\rangle$ are seperated by a large number of 1- cells of $T$.

Using the notation of section 5.4 in conjunction with equation (\ref{qhamt}), a straightforward computation yields:
\begin{eqnarray}
{\hat C}_{ham, T^{\prime}(\delta^{\prime}) }[N_2]{\hat C}_{ham, T(\delta) }[N_1] \vert {\bf s^+},{\bf s^-}\rangle
=
(\frac{-i\hbar^2}{a^2})^2\sum_{v\in V_E(s^+)\cup V_E(s^-)}N_1(v) 
\lambda (s^+, s^-, v) &\nonumber \\
{[} \sum_{v^{\prime}\in V_E(s^+_{\phi_{v,\delta}})\cup V_E(s^-_{\phi_{v,-\delta}})}N_2(v^{\prime}) 
\lambda (s^+_{\phi_{v,\delta}}, s^-_{\phi_{v,-\delta}}, v^{\prime})
(\vert 
({\bf s^+}_{\phi_{v,\delta}})_{\phi_{v^{\prime},\delta^{\prime}}}, 
({\bf s^-}_{\phi_{v,-\delta}})_{\phi_{v^{\prime},\delta^{\prime}}}
\rangle
- \vert {\bf s^+}_{\phi_{v,\delta}},{\bf s^+}_{\phi_{v,-\delta}}\rangle)&
\nonumber \\
\sum_{v^{\prime}\in V_E(s^+)\cup V_E(s^-)}N_2(v^{\prime}) 
\lambda (s^+, s^-, v^{\prime})
(\vert 
{\bf s^+}_{\phi_{v^{\prime},\delta^{\prime}}}, 
{\bf s^-}_{\phi_{v^{\prime},-\delta^{\prime}}}
\rangle
- \vert {\bf s^+},{\bf s^-}\rangle)\;\;\;\; {]}  &\nonumber\\
\label{masterhh}
\end{eqnarray}

Next, we restrict attention to $N_i, i=1,2$ of compact support. Specifically, let $N_i$ be supported in a neighbourhood
$U_i(v_i)$ of the vertex $v_i \in  V_E(s^+)\cup V_E(s^-)$ such that 
$U_i(v_i)\cap (V_E(s^+)\cup V_E(s^-))= v_i, i=1,2$. The linear dependence of $C_{ham}(N_i)$ on the lapse $N_i$ 
together with the fact 
that an arbitrary lapse function can be obtained by linear combinations of ones which have the above compact
support property imply that the restriction to lapses of compact support entail no loss of generality.
In section 6.1.1 we consider the case $v_1\neq v_2$ and in section 6.1.2, the case $v_1=v_2$.

\subsubsection{The case $v_1\neq v_2$.}
Note that :
\begin{eqnarray}
{[}\phi_{v_i,\pm \delta}, \phi_{v_j, \pm \delta^{\prime}}{]}&=&0 , \;\;i\neq j\label{v1neqv21} \\
U_i(v_i)\cap (V(s^+_{\phi_{v_j,\delta}})\cup V(s^-_{\phi_{v_j,-\delta}})) &=& v_i, \;\;\;i\neq j \label{v1neqv22}\\
\lambda (s^+_{\phi_{v_i,\delta}}, s^-_{\phi_{v_i,-\delta}}, v_j) &=&
\lambda (s^+, s^-, v_j) \;\;\;i\neq j
\label{v1neqv23}
\end{eqnarray}

Using this in conjunction with equation (\ref{masterhh}), it is straightforward to see that 
\begin{eqnarray}
{\hat C}_{ham, T^{\prime}(\delta^{\prime}) }[N_2]{\hat C}_{ham, T(\delta) }[N_1] \vert {\bf s^+},{\bf s^-}\rangle
=
(\frac{-i\hbar^2}{a^2})^2N_1(v_1) 
\lambda (s^+, s^-, v_1) 
N_2(v_2) \lambda (s^+, s^-, v_2 )&\nonumber \\
{[}( \vert 
({\bf s^+}_{\phi_{v_1,\delta}})_{\phi_{v_2^{\prime},\delta^{\prime}}}, 
({\bf s^-}_{\phi_{v_1,-\delta}})_{\phi_{v_2^{\prime},-\delta^{\prime}}}
\rangle
- \vert {\bf s^+}_{\phi_{v_1,\delta}},{\bf s^-}_{\phi_{v_1,-\delta}}\rangle)
-(\vert 
{\bf s^+}_{\phi_{v_2,\delta^{\prime}}}, 
{\bf s^-}_{\phi_{v_2,-\delta^{\prime}}}
\rangle
- \vert {\bf s^+},{\bf s^-}\rangle)\;\;\;\; {]}.  &\nonumber\\
\label{masterhhv1neqv2}
\end{eqnarray}
The second term in the commutator is obtained by interchanging $N_1(v_1), v_1$ with $N_2(v_2), v_2$ in the above equation
so that the commutator evaluates to
\begin{eqnarray}
{\hat C}_{ham, T^{\prime}(\delta^{\prime}) }[N_2]{\hat C}_{ham, T(\delta) }[N_1] 
-{\hat C}_{ham, T^{\prime}(\delta^{\prime}) }[N_1]{\hat C}_{ham, T(\delta) }[N_2] 
\vert {\bf s^+},{\bf s^-}\rangle
= & \nonumber \\
(\frac{-i\hbar^2}{a^2})^2N_1(v_1) 
\lambda (s^+, s^-, v_1) 
N_2(v_2) \lambda (s^+, s^-, v_2 )&\nonumber \\
{[}    ( \vert 
({\bf s^+}_{\phi_{v_1,\delta}})_{\phi_{v_2^{\prime},\delta^{\prime}}}, 
({\bf s^-}_{\phi_{v_1,-\delta}})_{\phi_{v_2^{\prime},-\delta^{\prime}}}
\rangle
-\vert 
({\bf s^+}_{\phi_{v_2,\delta}})_{\phi_{v_1^{\prime},\delta^{\prime}}}, 
({\bf s^-}_{\phi_{v_2,-\delta}})_{\phi_{v_1^{\prime},-\delta^{\prime}}}
\rangle )&
\nonumber \\
-( \vert {\bf s^+}_{\phi_{v_1,\delta}},{\bf s^-}_{\phi_{v_1,-\delta}}\rangle
- \vert {\bf s^+}_{\phi_{v_2,\delta}},{\bf s^-}_{\phi_{v_2,-\delta}}\rangle )&
\nonumber\\
-(\vert 
{\bf s^+}_{\phi_{v_2,\delta^{\prime}}}, 
{\bf s^-}_{\phi_{v_2,-\delta^{\prime}}} \rangle
-\vert 
{\bf s^+}_{\phi_{v_1,\delta^{\prime}}}, 
{\bf s^-}_{\phi_{v_1,-\delta^{\prime}}}\rangle ) 
{]}    &
\label{masterh,hv1neqv2}
\end{eqnarray}
From equation (\ref{h,hcont}), the continuum limit of the commutator on the distribution $\Psi$ is:
\begin{eqnarray}
\Psi ([\hat{C}_{ham}[N_2],\hat{C}_{ham}[N_1]]\vert {\bf s^+},{\bf s^-}\rangle)&:=& 
(\frac{-i\hbar^2}{a^2})^2N_1(v_1) 
\lambda (s^+, s^-, v_1) 
N_2(v_2) \lambda (s^+, s^-, v_2 )) \nonumber \\
& & \lim_{\delta^{\prime}\rightarrow 0}\lim_{\delta\rightarrow 0}
(\Psi_1(\delta, \delta^{\prime}) +\Psi_2(\delta, \delta^{\prime})+\Psi_3(\delta, \delta^{\prime})),
\label{psi123v1v2}
\end{eqnarray}
where, using (\ref{v1neqv21}), 
\begin{equation}
\Psi_1(v_1,v_2,\delta, \delta^{\prime}):=
\Psi ( \vert{\bf s^+}_{\phi_{v_1,\delta}})_{\phi_{v_2^{\prime},\delta^{\prime}}}, 
({\bf s^-}_{\phi_{v_1,-\delta}})_{\phi_{v_2^{\prime},-\delta^{\prime}}}
\rangle
-\vert 
({\bf s^+}_{\phi_{v_1,\delta^{\prime}}})_{\phi_{v_2,\delta}}, 
({\bf s^-}_{\phi_{v_1,-\delta^{\prime}}})_{\phi_{v_2,-\delta}}
\rangle     ) 
\label{psi1v1v2}
\end{equation}
\begin{eqnarray}
\Psi_2(v_1, v_2, \delta, \delta^{\prime})&:=&
-\Psi ( \vert {\bf s^+}_{\phi_{v_1,\delta}},{\bf s^-}_{\phi_{v_1,-\delta}}\rangle
- \vert {\bf s^+}_{\phi_{v_1,\delta^{\prime}}},{\bf s^-}_{\phi_{v_1,-\delta^{\prime}}}\rangle )
\label{psi2v1v2}
\\
\Psi_3(v_1, v_2, \delta, \delta^{\prime}) &:=&
-\Psi(\vert 
{\bf s^+}_{\phi_{v_2,\delta^{\prime}}}, 
{\bf s^-}_{\phi_{v_2,-\delta^{\prime}}} \rangle
-\vert 
{\bf s^+}_{\phi_{v_2,\delta}}, 
{\bf s^-}_{\phi_{v_2,-\delta}}\rangle ) .
\label{psi3v1v2}
\end{eqnarray}
It is easy to see that irrespective of the nature of the non- trivial vertices $v_1, v_2$
(i.e. whether $v_i \in V(s^+) \cap V(s^-)$ or not)
each of the two charge network states in  equations (\ref{psi1v1v2})- (\ref{psi3v1v2})
are diffeomorphic. Thus if $\Psi$ is a diffeomorphism invariant distribution, 
we have that 
$\Psi_I (v_1, v_2, \delta, \delta^{\prime}) =0, I=1,2,3$ for all $\delta, \delta^{\prime}$ under consideration
which, in turn, implies that the commutator (\ref{psi123v1v2}) vanishes.

\subsubsection{The case $v_1=v_2=v$.}
We note that the set $U_i(v)\cap V({\bf s}^+_{\phi_{v,\delta}})$ is either empty or 
consists of the single point
$v+\delta$ (recall that $U_i(v)$ is the support of the lapse function $N_i$). 
Similarly, the set 
$U_i(v)\cap V({\bf s}^-_{\phi_{v,-\delta}})$ is either empty or 
consists of the single point
$v-\delta$. This implies that 
\begin{eqnarray}
\lambda (s^+_{\phi_{v,\delta}}, s^-_{\phi_{v,-\delta}}, v+\delta )
&=& \lambda (s^+_{\phi_{v,\delta}}, s^-, v+\delta )\\
\lambda (s^+_{\phi_{v,\delta}}, s^-_{\phi_{v,-\delta}}, v-\delta )
&=& 
\lambda (s^+, s^-_{\phi_{v,-\delta}}, v-\delta ) .
\end{eqnarray}
Using the  remarks above, together with equation (\ref{masterhh}), a straightforward computation leads to the result:
\begin{eqnarray}
\Psi ([\hat{C}_{ham}[N_2],\hat{C}_{ham}[N_1]]\vert {\bf s^+},{\bf s^-}\rangle)
=(\frac{-i\hbar^2}{a^2})^2 & \nonumber \\
\lim_{\delta^{\prime}\rightarrow 0}\lim_{\delta \rightarrow 0} (\Psi_1 (N_1, N_2, v, \delta, \delta^{\prime})
+\Psi_2 (N_1, N_2, v, \delta, \delta^{\prime}) ) ,&
\label{psi12v}
\end{eqnarray}
where 
\begin{eqnarray}
\Psi_1 (N_1, N_2, v, \delta, \delta^{\prime})
&:=&\lambda (s^+_{\phi_{v,\delta}}, s^-, v+\delta )
\lambda (s^+, s^-, v ) (N_1(v)N_2(v+\delta )- N_1(v+\delta ) N_2(v)) \nonumber \\
& &\Psi (\vert ({\bf s}^+_{\phi_{v,\delta}})_{\phi_{v+\delta, \delta^{\prime}}},
{\bf s}^-_{\phi_{v,-\delta}}\rangle - 
\vert {\bf s}^+_{\phi_{v,\delta}},
{\bf s}^-_{\phi_{v,-\delta}}\rangle ) , 
\label{psi1v}
\end{eqnarray}
\begin{eqnarray}
\Psi_2 (N_1, N_2, v, \delta, \delta^{\prime})
&:=&\lambda (s^+, s^-_{\phi_{v,-\delta}}, v-\delta )
\lambda (s^+, s^-, v ) (N_1(v)N_2(v-\delta )- N_1(v-\delta ) N_2(v)) \nonumber \\
& &\Psi (\vert ({\bf s}^+_{\phi_{v,\delta}}),
({\bf s}^-_{\phi_{v,-\delta}})_{\phi_{v-\delta,- \delta^{\prime}}}\rangle - 
\vert {\bf s}^+_{\phi_{v,\delta}},
{\bf s}^-_{\phi_{v,-\delta}}\rangle ).
\label{psi2v}
\end{eqnarray}
It is easy to see that the 2 charge networks in each of the above equations are diffeomorphic.
Hence, if $\Psi$ is a diffeomorphism invariant distribution, $\Psi_I (N_1, N_2, v, \delta, \delta^{\prime})=0, I=1,2$
for all $\delta, \delta^{\prime}$ under consideration. 
This , in turn, implies that the commutator (\ref{psi123v1v2}) vanishes.

\subsection{The diffeomorphism constraint $\widehat{C_{diff}[\beta(N,M)]}$}
In this section we analyse the right hand side of
(\ref{hathhdiff}). Recall that 
\begin{equation}
\begin{array}{lll}
C_{diff}[\beta(N_1,N_2)]\ =\\
\vspace*{0.1in}
\hspace*{0.6in}\int_{\Sigma} (\Pi_{+}X^{+ '} + \Pi_{-}X^{- '} + \frac{1}{4}((Y^{+})^{2} - (Y^{-})^{2}))q^{xx}(N_1\partial_{x} N_2 - N_2\partial_{x} N_1) .
\label{cbetaeqn}
\end{array}
\end{equation}
From equation (\ref{qab}) it follows that 
$q^{xx}(x)= -(X^{+\prime}(x) X^{-\prime}(x))^{-1}$.
The operator corresponding to $C_{diff}[\beta(N_1,N_2)]$ 
can be obtained  by using the same ideas that we employed for the 
Hamiltonian constraint. Thus, we first define action of the operator at finite triangulation $T$ on 
$\vert {\bf s^+},{\bf s^-}\rangle$ where, as before $T= T(\gamma, \delta )$ satisfies conditions (i)- (iv) of section 5.4.
The operator correspondent of $q^{xx}$ is obtained by squaring equation (\ref{inverseeigen}).  The operators for 
the other fields in equation (\ref{cbetaeqn}) can be constructed along the lines of section 5.1 and 5.2  and chosen 
in such a way that both the `+' and the `-' parts of the constraint  are replaced by unitary operators labelled
by the  {\em same} gauge transformation so that the constraint operator at finite triangulation kills diffeomorphism 
invariant states. Specifically, 
it is easy to show that:
\begin{eqnarray}
\widehat{C_{diff}[\vec{\beta}(N_1,N_2)]}|_{T}(|{\bf s}^{+}\rangle\otimes|{\bf s}^{-}\rangle =
(-i\hbar) \sum_{v\in V_E(s^+)\cup V_E(s^-)} (\frac{\hbar}{a^2})^2 (\lambda (s^+, s^-, v))^2  & \nonumber \\
\delta (N_1(v)N_2^{\prime} (v) - N_2(v)N_1^{\prime}(v)) 
[{\hat U}^{+}(\phi_{v,\delta})\otimes{\hat U}^{-}(\phi_{v,\delta })-1]
\vert {\bf s^+},{\bf s^+}\rangle &\nonumber\\
\label{hatbetadiff1}
\end{eqnarray}
The action of this operator on the distribution $\Psi$ is then:
\begin{eqnarray}
\lim_{\delta \rightarrow 0}
\Psi (\widehat{C_{diff}[\vec{\beta}(N_1,N_2)]}|_{T}(|{\bf s}^{+}\rangle\otimes|{\bf s}^{-}\rangle )=
(-i\hbar)\sum_{v\in V_E(s^+)\cup V_E(s^-)} (\frac{\hbar}{a^2})^2 (\lambda (s^+, s^-, v))^2  & \nonumber \\
\lim_{\delta \rightarrow 0}\delta (N_1(v)N_2^{\prime} (v) - N_2(v)N_1^{\prime}(v)) (\Psi  (N_1, N_2, v, \delta ) &\nonumber\\
\label{masterdiff1}
\end{eqnarray}
where
\begin{equation}
\Psi (N_1, N_2, v, \delta )=\Psi (\vert {\bf s}^+_{\phi_{v,\delta}} ,{\bf s}^-_{\phi_{v,\delta}}\rangle)
                                             -\Psi (\vert {\bf s}^+,{\bf s}^-\rangle ).
\label{defpsidiff}
\end{equation}
Clearly, $\Psi (N_1, N_2, v, \delta )$ vanishes if $\Psi$ is a diffeomorphism invariant distribution so that action
of operator $\widehat{C_{diff}[\vec{\beta}(N_1,N_2)]}|_{T}$ at any {\em finite} triangulation of the type under consideration
vanishes. Thus for diffeomorphism invariant distributions the continuum limit of this operator, although trivial, 
 exists in the same sense as for the Hamiltonian constraint (see equation (\ref{contlimhamdiff})).
Note also that, by virtue of the factor of $\delta$, 
the right hand side of (\ref{masterdiff1}) vanishes, 
for a large class of non- diffeomorphism invariant distributions $\Psi$. 
This is in exact analogy to what happens in LQG \cite{lmhabitat2}.

\section{The algebra of quantum constraints on the LM habitat}

In section 7.1 we define Lewandowski- Marolf habitat \cite{donjurek} for PFT.
In the LQG context, 
the LM habitat is a specific  enlargement of the space of spatial diffeomorphism group averages of charge networks 
(see section 3.2) such that the (continuum limit of the triangulated) Hamiltonian constraint operator maps
the habitat into itself. In section 7.2 we show that the same is true here. We also show that the commutator
of a pair of smeared Hamiltonian constraints, $[{\hat C}_{ham}(N_1), {\hat C}_{ham}(N_2)]$, 
as well as the operator corresponding to their classical
Poisson bracket,$C_{diff}[\vec{\beta}(N_1,N_2)]$, annihilate all states in the habitat. This is the exact 
analog of the result \cite{lmhabitat2} for LQG. As we shall see, these operators kill states in the habitat for 
a very trivial reason stemming from the density weight 1 character of the Hamiltonian constraint: at finite 
triangulation, these operators do not have enough factors of $\delta$ in the denominator
to obtain non- trivial action on the habitat (Note that 
this is already apparent for ${\hat C}_{diff}[\beta(N_1,N_2)]$ from the 
discussion at the end of section 6.2.). This motivates the exploration, in section 7.3, of slightly more singular 
constraint operators, namely those correponding to the smeared density weight 2 Hamiltonian constraint, $H_+- H_-$,
their commutator and the operator corresponding to their Poisson bracket. We show that while the last
is a well defined operator on the habitat, neither the smeared density 2 constraint operators, nor their
commutator is well defined on the habitat. Our calculations indicate that a key role is played by states of 
non- zero volume in this discrepancy. In section 7.4 we shrink both the habitat as well as the space of charge networks
by removing such states from their construction and show that the constraint algebra is represented in anomaly free 
manner on this smaller set of states.

\subsection{The LM habitat}
Let $V_E ({\bf s}^+,{\bf s}^-)$ be the set of non-trivial embedding vertices of the state 
$\vert {\bf s}^+,{\bf s}^-\rangle$ so that 
\begin{equation}
V_E ({\bf s}^+,{\bf s}^-) = V_E (s^+) \cup V_E(s^-) , 
\end{equation}
where $s^+, s^-$ are the embedding charge network labels of
the state (see section 4.1 for a definition of $V_E (s^{\pm})$ ).
Note that $v=0$ is a non-trivial vertex iff $v=2\pi$ is a non-trivial vertex. This is simply a consequence
of the circular topology of space. Note also that the elements of $V_E (s^{\pm})$ are points in the 
interval $[0,2\pi ]$ and  hence can be mapped to points on the circle via the identification $x=0\sim x=2\pi$.
Let the set of images in $S^1$ of the elements of $V_E (s^{\pm}),V_E ({\bf s}^+,{\bf s}^-) $ be denoted by 
$V^{S^1}_E (s^{\pm}),V^{S^1}_E ({\bf s}^+,{\bf s}^-)$ so that 
\begin{equation}
V^{S^1}_E ({\bf s}^+,{\bf s}^-)= V_E^{S^1} (s^+) \cup V_E^{S^1}(s^-).
\end{equation}

It is easy to check that if $V_E (s^{\pm})$ define $m^{\pm}$ points on the circle so that 
$V^{S^1}_E (s^{\pm}) = \{p^{\pm}_i, p^{\pm}_2,..., p^{\pm}_{m^{\pm}} \in S^1\}$ then 
$V^{S^1}_E (s_{\phi^{\pm}}^{\pm})$ also defines $m^{\pm}$ points on the circle and is given by 
\begin{equation}
V^{S^1}_E (s_{\phi^{\pm}}^{\pm}) = \{\phi^{\pm}p_i, \phi^{\pm}p_2,..., \phi^{\pm}p_{m^{\pm}} \in S^1\}.
\label{ggeinvvertices}
\end{equation}
Here $s_{\phi^{\pm}}^{\pm}$ denotes the embedding charge network label of the gauge related 
state $\vert s_{\phi^{\pm}}^{\pm}\rangle$ for a gauge transformation labelled by $\phi^{\pm}$ and
$\phi^{\pm}(p)$ denotes the image of $p\in S^1$ under $\phi^{\pm}$.\footnote{Recall that $\phi^{\pm}$ is a periodic  diffeomorphism of the real line and hence can be naturally 
identified with a diffeomorphism of the circle.}
It is also easy to see that, since the charge nets are in the superselected sector ${\cal D}_{ss}$, 
the $\pm$ embedding charges can be arranged in increasing/decreasing order, thus inducing an ordering of 
vertices. This implies a unique identification of vertices in  
$V^{S^1}_E (s^{\pm})$ with those in $V^{S^1}_E (s_{\phi^{\pm}}^{\pm})$.

Let $V^{S^1}_E ({\bf s}^+,{\bf s}^-)$  consist of the points $q_i, i=1,.., n$ i.e.
\begin{equation}
V^{S^1}_E ({\bf s}^+,{\bf s}^-) = \{q_1, q_2,..,q_{n} \in S^1\}
\end{equation}
and let $f$ be a smooth (real valued) function of $n$ points on the circle. 
Then the LM habitat, ${\cal V}_{LM}$, is defined
as the linear span of the distributions $\Psi_{ f,[{\bf s}^+,{\bf s}^-]}$, where
\begin{equation}
\Psi_{ f,[{\bf s}^+,{\bf s}^-]}:= \sum_{{\bf s}^{\prime +},{\bf s}^{\prime -} \in [{\bf s}^+,{\bf s}^-]}
                                  f (V^{S^1}_E ({\bf s}^{\prime +},{\bf s}^{\prime -}))
\langle {\bf s}^+,{\bf s}^- \vert
.
\label{lmhabbasis}
\end{equation}
Note that by virtue of the discussion centering on  equation (\ref{ggeinvvertices}), 
the cardinality of $V^{S^1}_E ({\bf s}^{\prime +},{\bf s}^{\prime -})$
is independent of ${\bf s}^{\prime +},{\bf s}^{\prime -}$ if 
${{\bf s}^{\prime +},{\bf s}^{\prime +} \in [{\bf s}^+,{\bf s}^-]}$ (we remind the reader that 
$[{\bf s}^+,{\bf s}^-]$ is the orbit of ${\bf s}^+,{\bf s}^-$ under diffeomorphisms).
Further, that discussion also indicates that we can uniquely define the orbit of points 
in $V^{S^1}_E ({\bf s}^{\prime +},{\bf s}^{\prime -})$ under the action of some 1 paramter set of
diffeomorphisms. This fact will be implicitly used in our considerations below.

\subsection{Density one constraints}

\subsubsection{Continuum limit of the Hamiltonian constraint on ${\cal V}_{LM}$}
We show that equation (\ref{qhamtpsi}) has a well defined continuum limit if  $\Psi \in {\cal V}_{LM}$.
As described in section 6.1, without loss of generality, 
we restrict attention to lapses $N$ of compact support around the non- trivial  
vertex $v$ of the state  $\vert {\bf s^+},{\bf s^-}\rangle$. From equation (\ref{qhamtpsi})
we have that 
\begin{eqnarray}
\Psi_{f, [{\bf s^+}^{\prime},{\bf s^-}^{\prime}]}({\hat C}_{ham, T(\gamma, \delta) }[N]\vert {\bf s^+},{\bf s^-}\rangle  ) =
\sum_{v\in V_E(s^+)\cup V_E(s^-)}
N(v) \frac{-i\hbar}{a^2}\lambda (s^+, s^-, v) & \nonumber \\
\sum_{{\bf s}^{\prime\prime +},{\bf s}^{\prime\prime -} \in [{\bf s}^{+\prime},{\bf s}^{-\prime}]}
f (V^{S^1}_E ({\bf s}^{\prime \prime +},{\bf s}^{\prime \prime -}))
{[}\delta_{{\bf s^+}^{\prime\prime}, {\bf s^+}_{\phi_{v,\delta}}}
\delta_{{\bf s^-}^{\prime\prime}, {\bf s^-}_{\phi_{v,-\delta}}}
-
\delta_{{\bf s^+}^{\prime\prime}, {\bf s^+}}
\delta_{{\bf s^-}^{\prime\prime}, {\bf s^-}}
{]}.  &\nonumber\\
\label{qhamtlm}
\end{eqnarray}
If $v\in V(s^+) \cap V(s^-)$ and $[{\bf s}^{+\prime},{\bf s}^{-\prime}]=[{\bf s}^{+},{\bf s}^{-}]$, we have that 
\begin{eqnarray}
\Psi_{f, [{\bf s^+}^{\prime},{\bf s^-}^{\prime}]}({\hat C}_{ham, T(\gamma, \delta) }[N]\vert {\bf s^+},{\bf s^-}\rangle  )
= -N(v) \frac{-i\hbar}{a^2}\lambda (s^+, s^-, v) f (V^{S^1}_E ({\bf s}^{+},{\bf s}^{-})) &
\nonumber\\
=\lim_{\delta \rightarrow 0}
\Psi_{f, [{\bf s^+}^{\prime},{\bf s^-}^{\prime}]}({\hat C}_{ham, T(\gamma, \delta) }[N]\vert {\bf s^+},{\bf s^-}\rangle ). &\nonumber\\
\end{eqnarray}
If $v\in V(s^+) \cap V(s^-)$ and 
$[{\bf s}^{+\prime},{\bf s}^{-\prime}]=[{\bf s^+}_{\phi_{v,\delta}},{\bf s^-}_{\phi_{v,-\delta}}]$, we have that 
\begin{equation}
\lim_{\delta \rightarrow 0}
\Psi_{f, [{\bf s^+}^{\prime},{\bf s^-}^{\prime}]}({\hat C}_{ham, T(\gamma, \delta) }[N]\vert {\bf s^+},{\bf s^-}\rangle ).
=  \lim_{\delta \rightarrow 0}f ({\vec v}^{\prime},  v+\delta, v-\delta )= f({\vec v}^{\prime}, v, v)
\end{equation}
where ${\vec v}^{\prime}$ denotes all the non-trivial  vertices of $\vert {\bf s^+},{\bf s^-}\rangle$
outside the support of $N$.

If $v\notin V(s^+) \cap V(s^-)$ then 
$[{\bf s^+}_{\phi_{v,\delta}},{\bf s^-}_{\phi_{v,-\delta}}]=[{\bf s}^{+},{\bf s}^{-}]$. It follows that
\begin{eqnarray}
\lim_{\delta \rightarrow 0}
\Psi_{f, [{\bf s^+}^{\prime},{\bf s^-}^{\prime}]}({\hat C}_{ham, T(\gamma, \delta) }[N]\vert {\bf s^+},{\bf s^-}\rangle )&
\nonumber \\
=  -N(v) \frac{-i\hbar}{a^2}\lambda (s^+, s^-, v) \lim_{\delta \rightarrow 0}(f ({\vec v}^{\prime},  v\pm\delta)
- f ({\vec v}^{\prime},  v))=0, &
\end{eqnarray}
where the $\pm$ signs refer to the cases $v\in V(s^{\pm})$.

In all the above the continuum limit of the action of Hamiltonian constraint
 is well defined. It is also straightforward to see that, due to the diffeomorphism covariance of the 
operator ${\hat C}_{ham, T }[N]$, the continuum limit of the  Hamiltonian constraint operator maps 
${\cal V}_{LM}$ into itself.\\

Note that, as emphasized by Thiemann \cite{bigbook}, 
the continuum limit of ${\hat C}_{ham, T(\gamma, \delta) }[N]$ on the LM habitat as defined above, 
and the continuum limit of ${\hat C}_{ham, T(\gamma, \delta) }[N]$ in the URS topology as defined in section 6 are
distinct from each other in that the latter  is implemented via uniform convergence in ${\cal H}_{kin}$ whereas the 
latter is implemented via pointwise convergence in ${\cal V}_{LM}$.

That 
the convergence of the one parameter family of operators ${\hat C}_{ham, T(\gamma, \delta) }[N]$ on ${\cal H}_{kin}$ 
defined in the URS topology is uniform, follows directly from equation (\ref{contlimhamdiff}) by virtue of the fact that,
with repect to the URS topology, the sequence is a constant one. We now show through an example that 
the the convergence of the one parameter family of operators on the LM habitat is pointwise i.e. that
given
$\mu\ >\ 0$, $\Psi\ \in\ {\cal V}_{LM}$, $|{\bf s}^{+},{\bf s}^{-}\rangle\ \in {\cal D}_{ss}$ $\exists\ \delta(\mu,\Psi,({\bf s}^{+},{\bf s}^{-}))$ such that 
\begin{equation}\label{pointwise}
|(\hat{C}_{ham}[N]\Psi)|{\bf s}^{+},{\bf s}^{-}\rangle\ -\ \Psi(\hat{C}_{ham, T(\gamma, \delta)}[N]|{\bf s}^{+},{\bf s}^{-}\rangle)|\ <\ \mu
\end{equation}
$\forall\ \delta\ <\ \delta(\mu, \Psi, ({\bf s}^{+},{\bf s}^{-}))$.

Consider the charge network state $|{\bf s}^{+},{\bf s}^{-}\rangle$ with a vertex $v$ such that 
$v\ \in\ V_E({s}^{+})$ but
 $v\notin V_E(s^-)$ so that  
$[{\bf s^+}_{\phi_{v,\delta}},{\bf s^-}_{\phi_{v,-\delta}}]=[{\bf s}^{+}_{\phi_{v,\delta}},{\bf s}^{-}]=[{\bf s}^{+},{\bf s}^{-}]$. It follows that:
\begin{eqnarray}
\Psi_{f, [{\bf s^+},{\bf s^-}]}({\hat C}_{ham, T(\gamma, \delta) }[N]\vert {\bf s^+},{\bf s^-}\rangle )&
\nonumber \\
=  -N(v) \frac{-i\hbar}{a^2}\lambda (s^+, s^-, v)(f ({\vec v}^{\prime},  v + \delta)
- f ({\vec v}^{\prime},  v)).
\end{eqnarray}
\begin{equation}
\Rightarrow \Psi_{f,[{\bf s}^{+},{\bf s}^{-}]}(\hat{C}_{ham}[N]|{\bf s}^{+},{\bf s}^{-}\rangle)\ =\ 0.
\end{equation}
It follows that 
\begin{equation}\label{pointwise1}
\begin{array}{lll}
|\Psi_{f,[{\bf s}^{+},{\bf s}^{-}]}(\hat{C}_{ham}[N]|{\bf s}^{+},{\bf s}^{-}\rangle)\ -\ 
\Psi_{f,[{\bf s}^{+},{\bf s}^{-}]}(\hat{C}_{ham, T(\gamma, \delta)}[N]|{\bf s}^{+},{\bf s}^{-}\rangle)|=\\
\vspace*{0.1in}
\hspace*{1.5in} N(v) \frac{\hbar}{a^2}\lambda (s^+, s^-, v)\ |f ({\vec v}^{\prime},  v + \delta)
- f ({\vec v}^{\prime},  v))|.
\end{array}
\end{equation}
Equation (\ref{pointwise1}) implies that equation (\ref{pointwise}) (with $\Psi:=\Psi_{f,[{\bf s}^{+},{\bf s}^{-}]}$) is
satisfied 
for $\delta$ which depends on $\Psi$ (through the function $f$) and on $({\bf s}^{+},{\bf s}^{-})$
(through 
the position of arguement $v$), thus indicating pointwise convergence. 

Despite the notion of 
convergence on ${\cal V}_{LM}$ being (seemingly) much weaker than that  with respect to the URS topology, 
our considerations below illustrate the usefulness of habitats such as ${\cal V}_{LM}$ in exploring the off shell
closure of the quantum constraint algebra.

\subsubsection{The constraint algebra}
Consider, first, the commutator of two Hamiltonian constraints (\ref{h,hcont}) with 
$\Psi= \Psi_{f, [{\bf s^+}^{\prime},{\bf s^-}^{\prime}]}\in{\cal V}_{LM}$. 
Let the cardinality of $V(s^+) \cup V(s^-)$ be $n^{\prime}$.
Recall that the function $f (V^{S^1}_E ({\bf s}^{\prime +},{\bf s}^{\prime -}))$ is a smooth function from 
$(S^1)^{n^{\prime}}$ (i.e. $n^{\prime}$ copies of the circle) to the complex numbers.
Next, note that, whenever non- trivial, 
the terms in equations (\ref{psi1v1v2})- (\ref{psi3v1v2}), and 
(\ref{psi1v})-(\ref{psi2v}) consist of the difference of the evaluation of the function
$f (V^{S^1}_E ({\bf s}^{\prime +},{\bf s}^{\prime -}))$ at nearby points in $(S^1)^{n^{\prime}}$ which coincide
in the continuum limit. Hence by virtue of the smoothness of $f (V^{S^1}_E ({\bf s}^{\prime +},{\bf s}^{\prime -}))$
all these terms vanish in the continuum limit. Clearly, the commutator trivialises due to the absence of 
factors of $\delta, \delta^{\prime}$ in the denominator. Had such factors been present there could be the possibility
that the terms which vanished now yield derivatives of $f (V^{S^1}_E ({\bf s}^{\prime +},{\bf s}^{\prime -}))$.
Such factors could arise if we considered higher density constraints. This motivates the analysis of the density two
constraints in section...

What about the left hand side of equation (\ref{hathhdiff}) with $\Psi= \Psi_{f, [{\bf s^+}^{\prime},{\bf s^-}^{\prime}]}\in{\cal V}_{LM}$?  It is easy to see, from section 6.2 and from arguments identical to those above 
 that the continuum limit of the diffeomorphism constraint
$\widehat{C_{diff}[\beta(N_1,N_2)]}|_{T}$  vanishes on  ${\cal V}_{LM}$. 
Indeed it vanishes ``doubly''': first,  due to the extra factor of $\delta$ in equation 
(\ref{masterdiff1}) and second, by virtue of the fact that, similar to the case of the Hamiltonian constraint commutator
discussed above, equation (\ref{defpsidiff}) consists of the evaluation of the habitat state on the difference of 
a pair of charge networks related by a small diffeomorphism which approaches the identity in the continuum limit.

Thus both sides of equation (\ref{hathhdiff}) vanish on the LM habitat in exact analogy, and, in fact, for exactly the
same reasons as in LQG: namely the absence of suitable factors of $\delta$ in the denominator. The considerations of 
sections 8 and .. will make this remark precise.

For later use,
we conclude this section with an explicit evaluation of the commutator on the LM habitat for 2 specific cases
outlined below. \\

\noindent Case 1: See section 6.1.1. Let $v_1\neq v_2$. Set 
$\Psi= \Psi_{f, [{\bf s^+}^{\prime},{\bf s^-}^{\prime}]}\in{\cal V}_{LM}$. Let $v_1,v_2 \in V(s^+) \cap V(s^-)$  and
let $[{\bf s^+}^{\prime},{\bf s^-}^{\prime}]= [
({\bf s}^+_{\phi_{v_1,\delta}})_{\phi_{v_2,\delta^{\prime}}},
({\bf s}^-_{\phi_{v_1,-\delta}})_{\phi_{v_2,-\delta^{\prime}}}]$
for sufficiently small $\delta, \delta^{\prime}$. Note that for 
sufficiently small $\delta, \delta^{\prime}$,
$[({\bf s}^+_{\phi_{v_1,\delta}})_{\phi_{v_2,\delta^{\prime}}},
({\bf s}^-_{\phi_{v_1,-\delta}})_{\phi_{v_2,-\delta^{\prime}}}]$
is independent of $\delta, \delta^{\prime}$. Also note that the vertices $v_i,i=1,2$ of 
$\vert {\bf s}^+,{\bf s}^-\rangle$ each split into 2 vertices around $v_i$, a `+' vertex and a `-' vertex,
to yield 
$\vert({\bf s}^+_{\phi_{v_1,\delta}})_{\phi_{v_2,\delta^{\prime}}},
({\bf s}^-_{\phi_{v_1,-\delta}})_{\phi_{v_2,-\delta^{\prime}}}\rangle$. This immediately implies that 
$\Psi_2 (v_1, v_2, \delta , \delta^{\prime}) = \Psi_3 (v_1, v_2, \delta , \delta^{\prime}) = 0$.
Further, we have that 
\begin{equation}
\Psi_3 (v_1, v_2, \delta , \delta^{\prime}) =
f({\vec v}^{\prime}, v_1+\delta, v_1-\delta, v_2+\delta^{\prime}, v_2- \delta^{\prime})
-
f({\vec v}^{\prime}, v_1+\delta^{\prime}, v_1-\delta^{\prime}, v_2+\delta, v_2- \delta )
\end{equation}
which vanishes in the continuum limit.\\

\noindent Case 2: See section 6.1.2. Let $v_1=v_2=v \in  V(s^+) \cap V(s^-)$ and set 
$\Psi= \Psi_{f, [{\bf s^+},{\bf s^-}]}\in{\cal V}_{LM}$ so that we are interested in the case where the 
diffeomorphism class which labels the habitat state is the same as that of the charge network state 
on which the commutator acts. Since the terms in 
equations (\ref{psi1v})- (\ref{psi2v}) involve the charge nets in which the joint $+,-$ vertex at $v$ splits
into `+' one and a `-' one we have that $\Psi_1 (N_1, N_2, v, \delta, \delta^{\prime})
=\Psi_2 (N_1, N_2, v, \delta, \delta^{\prime})= 0$.

\subsection{Density two constraints}

Rescaling the density weight one Hamiltonian constraint $C_{ham}$ (given in equation (\ref{hamconstraint})) by 
the square root of the determinant of the spatial metric yields the density weight 2 Hamiltonian constraint
$H:=H_+- H_-$ which on smearing with the density weight -1 lapse, $N$ yields
\begin{equation}
H (N):=\int dx
\left[\Pi_{+}(x)X^{+'}(x)\ -\ \Pi_{-}(x)X^{-'}(x)\ +\ \frac{1}{4}(\pi_{f}^{2}+f^{' 2})\right].
\label{defh}
\end{equation}
In 1 spatial dimension, a scalar of density weight -1 transforms in the same way as vector field. Thus, $N$
in the above equation can equally well be thought of as a vector field. We shall use this equivalence
to denote $N$ by ${\vec N}$ whenever it is convenient.
Replacing $C_{ham}$ by $H$ in the Dirac algebra (\ref{eq:structure}) yields the Lie algebra:
\begin{equation}
\begin{array}{lll}
\{C_{diff}(\vec{N}_1),\ C_{diff}(\vec{N}_2)\}\ =\ C_{diff}([\vec{N}_1,\vec{N}_2])
\label{diffdiff}
\\
\vspace*{0.1in}
\{C_{diff}(\vec{N_1}),\ H(N_2)\}\ =\ H[{\it L}_{\vec{N}_1}N_2] 
\label{diffham}\\
\vspace*{0.1in}
\{H (N_1),\ H (N_2)\}\ =\ C_{diff}([\vec{N_1}, \vec{N_2}]),
\label{hamham}
\end{array}
\end{equation}
where in the last equation we have used the equivalence $N_i\equiv {\vec N}_i$ between density weight -1 scalars and
vectors.

In section 7.3.1 we show that ${\hat C}_{diff}(\vec{N})$ is a well defined operator on ${\cal V}_{LM}$ and that the 
Poisson bracket (\ref{diffdiff}) is represented in an anomaly free manner on ${\cal V}_{LM}$. In section 7.3.2
we construct the semared density two hamiltonian constraint operator at finite triangulation, ${\hat H}_T(N)$. 
and show that neither ${\hat H}_T (N)$ nor the commutator between a pair such operators admits a continuum limit on 
all of ${\cal V}_{LM}$.  We also show, through an example that there exist states in ${\cal V}_{LM}$ on which the 
action of the commutator admits a continuum limit but is anomalous. The example shows that the anomaly can be 
traced to the existence of charge network states with non- vanishing volume and motivates the considerations of section
7.4.

\subsubsection{The diffeomorphism constraint and its commutator}

The analysis of ${\hat C}_{diff}(\vec{N})$ parallels that of section 6.2. It is straightforward to see that, 
due to the absence  of the metric dependent factor, there are now no factors of $\lambda$ and an overall factor of
$\delta^{-1}$ instead of $\delta$ (see equation (\ref{hatbetadiff1}) ). In detail, we have that
\begin{equation}
{\hat C}_{diff, T(\delta)}(\vec{N})|{\bf s}^{+},{\bf s}^{-}\rangle=
(-i\hbar) \sum_{v\in V_E(s^+)\cup V_E(s^-)}N^x (v)
\frac{\vert ({\bf s}^+_{\phi_{v,\delta}},{\bf s}^-_{\phi_{v,\delta}}\rangle
-|{\bf s}^{+},{\bf s}^{-}\rangle}{\delta},
\label{cdiff0}
\end{equation}
where $N^x$ is the component of the shift vector in the coordinate system $\{x\}$ for which the length of each 
edge of $T$ is $\delta$.\footnote{In the interest of clarity, we denote $T(\gamma, \delta)$ by $T(\delta)$ from now on. We hope to have conveyed to the reader by now that the triangulation is graph dependent and hence hope that ommiting the label $\gamma$ will not create any confusion.}
\begin{equation}
\Rightarrow 
\Psi_{f, [{\bf s^+}^{\prime},{\bf s^-}^{\prime}]}({\hat C}_{diff, T(\delta)}(\vec{N})|{\bf s}^{+},{\bf s}^{-}\rangle)
=0 \;\;{\rm  if \;} [{\bf s^+}^{\prime},{\bf s^-}^{\prime}]\neq [{\bf s^+},{\bf s^-}],
\label{cdiff1}
\end{equation}
and
\begin{equation}
\Psi_{f, [{\bf s^+},{\bf s^-}]}({\hat C}_{diff, T(\delta)}(\vec{N})|{\bf s}^{+},{\bf s}^{-}\rangle)
= -i\hbar \sum_{v\in V_E(s^+)\cup V_E(s^-)} N^x(v) \frac{f({\vec v}^{\prime}, v+\delta)- f({\vec v}^{\prime}, v)}{\delta}
\end{equation}
\begin{equation}
\lim_{\delta\rightarrow 0}\Psi_{f, [{\bf s^+},{\bf s^-}]}({\hat C}_{diff, T(\delta)}(\vec{N})|{\bf s}^{+},{\bf s}^{-}\rangle)
= -i\hbar \sum_{v\in V_E(s^+)\cup V_E(s^-)} N^x(v) \partial_xf({\vec v}^{\prime}, v),
\label{cdiff2}
\end{equation}
so that $\Psi_{f, [{\bf s^+},{\bf s^-}]} \in {\cal V}_{LM}$ is mapped to 
$\Psi_{g_{\vec N}, [{\bf s^+},{\bf s^-}]} \in {\cal V}_{LM}$ with 
\begin{equation}
g_{\vec N}(V^{S^1}_E ({\bf s}^+,{\bf s}^{-}))
:= -i\hbar \sum_{v\in V_E(s^+)\cup V_E(s^-)} N^x(v) (\partial_xf({\vec v}^{\prime}, x))|_{x=v}.
\end{equation}
Here (and in an obvious fashion, below) the argument ${\vec v}^{\prime}$ of 
$f({\vec v}^{\prime}, x)$ indicates the set of non-trivial vertices other than the vertex $x$ under consideration.
This immediately implies that 
\begin{equation}
\Psi_{f, [{\bf s^+}^{\prime},{\bf s^-}^{\prime}]}({\hat C}_{diff}(\vec{N_1}){\hat C}_{diff}(\vec{N_2})|{\bf s}^{+},{\bf s}^{-}\rangle)
=0 \;\;{\rm  if \;} [{\bf s^+}^{\prime},{\bf s^-}^{\prime}]\neq [{\bf s^+},{\bf s^-}],
\label{cdiffdiff1}
\end{equation}
\begin{eqnarray}
& \Psi_{f, [{\bf s^+},{\bf s^-}]}({\hat C}_{diff}(\vec{N_1}){\hat C}_{diff}(\vec{N_2})|{\bf s}^{+},{\bf s}^{-}\rangle)
=\Psi_{g_{\vec N_1}, [{\bf s^+},{\bf s^-}]}({\hat C}_{diff}(\vec{N_2})|{\bf s}^{+},{\bf s}^{-}\rangle)\nonumber \\
&= (-i\hbar)
\sum_{v\in V_E(s^+)\cup V_E(s^-)}N_2^x\partial_x g_{\vec N_1}({\vec v}^{\prime}, x)|_{x=v}\nonumber \\
&=(-i\hbar)^2\sum_{v,{\bar v}\in V_E(s^+)\cup V_E(s^-), {\bar v}\neq v}
N_2^x(v)N_1^{\bar x}({\bar v})\partial_x\partial{{\bar x}}f ({\vec v}^{\prime}, x, {\bar x})|_{x=v,{\bar x}={\bar v}}
\nonumber\\
&+(-i\hbar )^2
\sum_{v\in V_E(s^+)\cup V_E(s^-)}(N_2^x(v)\partial_x (N^x_1(x) \partial_x f ({\vec v}^{\prime}, x)))|_{x=v}.
\label{cdiffdiff2}
\end{eqnarray}
From equations (\ref{cdiffdiff1}), (\ref{cdiffdiff2}) it is easy to see that 
\begin{equation}
\Psi_{f, [{\bf s^+}^{\prime},{\bf s^-}^{\prime}]}([{\hat C}_{diff}(\vec{N_1}),{\hat C}_{diff}(\vec{N_2}])|{\bf s}^{+},{\bf s}^{-}\rangle)
=(-i\hbar )
\Psi_{f, [{\bf s^+}^{\prime},{\bf s^-}^{\prime}]}({\hat C}_{diff}([\vec{N_2},\vec{N_1}] )
|{\bf s}^{+},{\bf s}^{-}\rangle),
\label{diff,difflm}
\end{equation}
which is  an antirepresentation of the Poisson bracket algebra (\ref{diffdiff}).

\subsubsection{The Hamiltonian constraint and its commutator}
The smeared density weight 2 Hamiltonian constraint ${\hat H}_{T}$ at finite triangulation is  constructed 
along the lines of section 6. As for the diffeomorphism constraint, the rescaling of the density 1 constraint and the 
consequent absence of factors of the square root of determinant of the spatial metric imply that there are no longer any
factors of $\lambda$ (see equation (\ref{qhamt}) and that there is now an overall factor of $\delta^{-1}$.
More in detail, it is straightforward to see that for $\Psi\in {\cal V}_{LM}$, we have that
\begin{equation}
\Psi({\hat H}_{T(\delta) }[N]\vert {\bf s^+},{\bf s^-}\rangle )
= (-i\hbar)\sum_{v\in V_E(s^+)\cup V_E(s^-)}N(v)
\Psi (\frac{\vert {\bf s}^+_{\phi_{v,\delta}},{\bf s}^-_{\phi_{v,-\delta}}\rangle
-|{\bf s}^{+},{\bf s}^{-}\rangle}{\delta}) .
\label{hlm}
\end{equation}
Let $v,{\bf s}^{+},{\bf s}^{-}$ be such that $v\in V(s^+)\cap V(s^-)$.
Let $N$ be compactly supported around $v$ with support of the type discussed in section 6.1 and
let $\Psi= \Psi_{f, [{\bf s^+},{\bf s^-}]}$. Then we equation (\ref{hlm}) implies that
\begin{equation}
\Psi_{f, [{\bf s^+},{\bf s^-}]}({\hat H}_{T(\delta) }[N]\vert {\bf s^+},{\bf s^-}\rangle )
=i\hbar N(v)  \frac{f (V^{S^1}_E ({\bf s}^{\prime +},{\bf s}^{\prime -}))}{\delta},
\end{equation}
which does not admit a $\delta\rightarrow 0$ continuum limit. Thus ${\hat H}[N]$ is not well defined on
(all of) ${\cal V}_{LM}$. 

Can we make sense of the commutator of a pair of density 2 Hamiltonian constraints on (all of) ${\cal V}_{LM}$?
The example below shows that the answer is in the negative.
Let $N_1, N_2, v_1, v_2$ be as in section 6.1.1 and let $\Psi\in {\cal V}_{LM}$. 
It is straightforward to see, from equations (\ref{hlm}) and
(\ref{masterh,hv1neqv2}), that
\begin{eqnarray}
\Psi({\hat H}_{T^{\prime}(\delta^{\prime}) }[N_2]{\hat H}_{ T(\delta) }[N_1] 
-{\hat H}_{ T^{\prime}(\delta^{\prime}) }[N_1]{\hat H}_{T(\delta) }[N_2] 
\vert {\bf s^+},{\bf s^-}\rangle)
=& \nonumber \\
(-i\hbar)^2 N_1(v_1)N_2(v_2) \frac{\sum_{i=1}^3 \Psi_i(v_1, v_2,\delta, \delta^{\prime} )}{\delta\delta^{\prime}}&
\end{eqnarray}
where $\Psi_i(v_1, v_2,\delta, \delta^{\prime} ), i=1,2,3$ are given by equations (\ref{psi1v1v2}),
(\ref{psi2v1v2}) and (\ref{psi3v1v2}). Clearly, the existence of the continuum limit is tied to that of the 
limit 
$\lim_{\delta\rightarrow 0}\lim_{\delta^{\prime}\rightarrow 0}
\frac{\sum_{i=1}^3 \Psi_i(v_1, v_2,\delta, \delta^{\prime} )}{\delta\delta^{\prime}}$.
Now, consider Case 1 of section 7.2.2. Clearly,
\begin{equation}
\lim_{\delta^{\prime}\rightarrow 0}\lim_{\delta\rightarrow 0}
\frac{\sum_{i=1}^3 \Psi_i(v_1, v_2,\delta, \delta^{\prime} )}{\delta\delta^{\prime}}
=\lim_{\delta^{\prime}\rightarrow 0}\lim_{\delta\rightarrow 0}
\frac{\Psi_1(v_1, v_2,\delta, \delta^{\prime} )}{\delta\delta^{\prime}}.
\end{equation}
Since the limit\\ 
$\lim_{\delta^{\prime}\rightarrow 0}\frac
{f({\vec v}^{\prime}, v_1+\delta, v_1-\delta, v_2+\delta^{\prime}, v_2- \delta^{\prime})
-
f({\vec v}^{\prime}, v_1+\delta^{\prime}, v_1-\delta^{\prime}, v_2+\delta, v_2- \delta )}{\delta^{\prime}}$\\
does not exist for generic $f$, the commutator does not admit a continuum limit on (all of)
${\cal V}_{LM}$.

Nevertheless, as the following calculation suggests, such a limit may exist for a subset of 
states in ${\cal V}_{LM}$. Consider the setting of section 6.1.2 where $v_1=v_2=v$. 
It is straightforward to see that
\begin{eqnarray}
\Psi({\hat H}_{T^{\prime}(\delta^{\prime}) }[N_2]{\hat H}_{ T(\delta) }[N_1] 
-{\hat H}_{ T^{\prime}(\delta^{\prime}) }[N_1]{\hat H}_{T(\delta) }[N_2] 
\vert {\bf s^+},{\bf s^-}\rangle)
=& \nonumber \\
_(-i\hbar )^2 \frac{\Psi_1 (N_1, N_2, v, \delta, \delta^{\prime})
+\Psi_2 (N_1, N_2, v, \delta, \delta^{\prime})}{\delta\delta^{\prime}} ,&
\label{anomaly}
\end{eqnarray}
where $\Psi_i (N_1, N_2, v, \delta, \delta^{\prime}), i=1,2$ are defined in equations (\ref{psi1v}),(\ref{psi2v}).
Now let us consider Case 2 of section 7.2.2. From the discussion there we have that 
the right hand side of the above equation vansihes. However, from equation (\ref{cdiff2}) we see that for generic
$f$ that the particular evaluation of the commutator (\ref{anomaly}), while possessing a continuum limit, is 
anomalous. More generally, if we restrict attention to habitat states 
$\Psi_{f, [{\bf s^+}^{\prime},{\bf s^-}^{\prime}]}$ for which $\vert{\bf s^+}^{\prime},{\bf s^-}^{\prime}\rangle$
is such that 
$V(s^{+\prime})= V(s^{-\prime})$, the commutator always vanishes, and, for generic $f$ is anomalous.

The three sets of calculations above all involve states for which $V(s^+)\cap V(s^-)$ is non- empty.
This suggests that perhaps the problems with ill definednes and the presence of anomalies could disappear by 
removing such states from our considerations. This is the subject of the next section.

\subsection{The zero volume sector}

From section 4.1, it follows that 
given a charge network state $\vert {\bf s}^+,{\bf s}^-\rangle$, 
the operator corresponding to the volume of some spatial region ${\cal R}\subset S^1$ acts non- trivially only on 
those vertices which are in the set $V_E(s^+)\cap V_E(s^-) \cap {\cal R}$. 
Hence we shall refer to a charge network $\vert {\bf s}^+,{\bf s}^-\rangle$ as a zero volume charge network
iff $V_E(s^+)\cap V_E(s^-)$ is empty.
We define the zero volume sector, ${\cal D}^0_{ss}$, of
${\cal D}_{ss}$ to be  the finite span of all `zero volume' charge networks in ${\cal D}_{ss}$.

It is easy to see that if $V_E(s^+)\cap V_E(s^-)$ is empty then
$V_E(s^{+\prime})\cap V_E(s^{-\prime})$ is also empty for any ${\bf s^+}^{\prime},{\bf s^-}^{\prime}\in 
[{\bf s}^{+},{\bf s}^{-}]$ so that the zero volume property extends to spatial diffeomorphism classes of 
charge networks.
We define the ``zero volume'' sector ${\cal V}^0_{LM} \subset {\cal V}_{LM}$ as the finite span of those
states $\Psi_{ f,[{\bf s}^+,{\bf s}^-]}\in {\cal V}_{LM}$ for which $V_E(s^+)\cap V_E(s^-)$ is empty.

In the rest of this section, we shall restrict attention to charge nets in ${\cal D}^0_{ss}$ and distributions in 
${\cal V}^0_{LM}$. Thus we shall think of ${\cal V}^0_{LM}$ as a subset of $({\cal D}^{0}_{ss})^*$, 
where $({\cal D}^{0}_{ss})^*$ is the algebraic dual to 
${\cal D}^0_{ss}$.

Next, note that, for $\vert {\bf s^+},{\bf s^-}\rangle \in {\cal D}^0_{ss}$, $v\in V_E(s^{+})\cap V_E(s^{-})$
and sufficiently small $\delta$, the charge network states
$\vert {\bf s}^+_{\phi_{v,\delta}},{\bf s}^-_{\phi_{v,-\delta}}\rangle$ ,
$\vert {\bf s}^+_{\phi_{v,\delta}},{\bf s}^-_{\phi_{v,\delta}}\rangle$, and
$\vert {\bf s^+},{\bf s^-}\rangle$, are related to each other by the action of spatial diffeomorphisms so that 
\begin{equation}
[{\bf s}^+_{\phi_{v,\delta}},{\bf s}^-_{\phi_{v,-\delta}}]=
[{\bf s}^+_{\phi_{v,\delta}},{\bf s}^-_{\phi_{v,\delta}}]=
[{\bf s^+},{\bf s^-}] .
\label{diffclass0vol}
\end{equation}
Since $\vert ({\bf s}^+_{\phi_{v,\delta}},{\bf s}^-_{\phi_{v,-\delta}}\rangle$ and
$\vert ({\bf s}^+_{\phi_{v,\delta}},{\bf s}^-_{\phi_{v,\delta}}\rangle$ are generated from 
$\vert {\bf s^+},{\bf s^-}\rangle$ by the action of the Hamiltonian and diffeomrphism constraints,
and since the zero volume property holds for diffeomorphism classes of charge networks, it follows that 
the constraints at finite triangulation map ${\cal D}^0_{ss}$ to itself so that it is consistent to restrict attention
to ${\cal D}^0_{ss}$. 

We have already shown, in section 7.3.1, that the diffeomorphism constraint has a well defined continuum limit 
on ${\cal V}_{LM}$ and hence also on ${\cal V}^0_{LM}$. We now show that 
${\hat H}_T[N]$ also has a well defined continuum limit on ${\cal V}^0_{LM}$.
We shall denote a vertex of $\vert {\bf s}^+,{\bf s}^-\rangle$ which is in $V_E(s^+)$ by $v^+$
and one which is in $V_E(s^+)$ by $v^-$.
Then from equations (\ref{hlm}) and (\ref{diffclass0vol}), it follows that , for 
$\Psi_{f, [{\bf s^+}^{\prime},{\bf s^-}^{\prime}]}\in {\cal V}^0_{LM}$,
\begin{equation}
\Psi_{f, [{\bf s^+}^{\prime},{\bf s^-}^{\prime}]}({\hat H}_{ T(\delta)}(N)|{\bf s}^{+},{\bf s}^{-}\rangle)
=0 \;\;{\rm  if \;} [{\bf s^+}^{\prime},{\bf s^-}^{\prime}]\neq [{\bf s^+},{\bf s^-}],
\label{h1}
\end{equation}
and that, in obvious notation, 
\begin{eqnarray}
\lim{\delta\rightarrow 0}\Psi_{f, [{\bf s^+},{\bf s^-}]}({\hat H}_{ T(\delta)}(N)|{\bf s}^{+},{\bf s}^{-}\rangle)
&=& (-i\hbar)
(\sum_{v^+\in V_E(s^+)}N (v^+)\frac{\partial f}{\partial v^+}
-\sum_{v^-\in V_E(s^-)}N (v^-)\frac{\partial f}{\partial v^-})\nonumber\\
&=:& \Psi_{f, [{\bf s^+},{\bf s^-}]}({\hat H}(N)|{\bf s}^{+},{\bf s}^{-}\rangle)
\label{h2}
\end{eqnarray}
It is straightforward to see, similar to the case of the diffeomorphism constraint (\ref{cdiff2}),
that the state $\Psi_{f, [{\bf s^+},{\bf s^-}]}$ is mapped into 
$\Psi_{g_N, [{\bf s^+},{\bf s^-}]}\in {\cal V}^0_{LM}$ where
\begin{equation}\label{zvol1}
g_{N}(V^{S^1}_E ({\bf s}^+,{\bf s}^{-}))
:= -i\hbar (\sum_{v^+\in V_E(s^+)}N (v^+)\frac{\partial f}{\partial v^+}
-\sum_{v^-\in V_E(s^-)}N (v^-)\frac{\partial f}{\partial v^-})
\end{equation}

It is then also straightforward
to see that 
a calculation, almost identical to that for the commutator of the diffeomorphism constraint (\ref{diff,difflm}) 
then yields the following
result:
\begin{equation}
\Psi_{f, [{\bf s^+},{\bf s^-}]}([{\hat H}(N_2),{\hat H}(N_1)] |{\bf s}^{+},{\bf s}^{-}\rangle)
= (-i\hbar )\Psi_{f, [{\bf s^+},{\bf s^-}]}({\hat C}_{diff}[{\vec N}_2, {\vec N}_1]|{\bf s}^{+},{\bf s}^{-}\rangle),
\label{h,hlm0}
\end{equation}
where as in equation (\ref{hamham}), we have used the equivalence between density weight -1 scalars and vectors to denote
$N_1, N_2$ by ${\vec N}_1,{\vec N}_2$.
It is easy to see that equations (\ref{h1}) and (\ref{h,hlm0}) imply that the Poisson- Lie algebra (\ref{hamham}) 
of density 2
constraints is represented in an anomaly free manner on 
${\cal V}^0_{LM}\subset ({\cal D}^{0}_{ss})^*$.\\
We now determine the kernel of ${\hat H}[N]$ inside ${\cal V}^{0}_{LM}$. From (\ref{h2}) and (\ref{zvol1}) it follows rather straightforwardly that diffeomorphism invariant distribution which lie inside ${\cal V}^{0}_{LM}$are certainly in the kernel of ${\hat H}[N]$. We now show that these are the only states in the kernel.\\
{\bf Lemma}\\
$\Psi_{f, [{\bf s^+},{\bf s^-}]}$ is in the kernel of ${\hat H}[N]$ iff $f$ is a constant function.\\

{\bf Proof} : \\

In light of (\ref{h1}) we want to show that 

\begin{equation}\label{zvol2}
\Psi_{f, [{\bf s}^{+},{\bf s}^{-}]}({\hat H}[N]|{\bf s}^{+ '},{\bf s}^{- '}\rangle)\ =\ 0
\end{equation}
$\forall\ N$ and $\forall\ |{\bf s}^{+ '},{\bf s}^{- '}\rangle$ for which $[{\bf s}^{+ '},{\bf s}^{- '}]\ =\ [{\bf s}^{+},{\bf s}^{-}]$.\\
(\ref{zvol1}) essentially implies that this will be true iff
\begin{equation}
g_{N}(V^{S^1}_E ({\bf s}^+_{\phi},{\bf s}^{-}_{\phi}))
:= -i\hbar (\sum_{v^+\in V_E(s^+)}N (\phi(v^+))\frac{\partial f}{\partial \phi(v^+)}
-\sum_{v^-\in V_E(s^-)}N (\phi(v^-))\frac{\partial f}{\partial \phi(v^-)}))\ =\ 0
\end{equation}
$\forall\ N$ and $\forall\ \phi$.\\
Clearly this will be true iff $f$ are constant functions.

Whence the kernel of density two Hamiltonian constraint inside ${\cal V}^{0}_{LM}$ is analogous to the kernel of Hamiltonian constraint in LQG when the domain of the operator is restricted to planar spin-networks.

\section{The algebra of quantum constraints on the new habitat}

The computations of section 6 and 7.2 indicate that 
due to their density one character, the quantum constraints $\hat{C}_{ham}[N]$ are 
``too non-singular"  to give rise to a non- trivial commutator algebra 
either on ${\cal H}_{kin}$ (when working in the URS) , or on the LM Habitat 
(where one looks at a net of regulated dual operators).

This motivates the consideration of the density 2 constraints in  sections 7.3 and 7.4. Section 7.3 
throws up an apparent paradox. On the one hand, it is easy to see that 
the ``correct'' physical states (see section 2.2.5)  lie in the 
kernel of the constraints at any finite triangulation, thus indicating that the constraints have been correctly
constructed. On the other,
the continuum limit of the the smeared density 2 constraint is  ill-defined on states in the  LM habitat 
\footnote{This naturally implies that in the URS, 
the continuum limit will certainly not be well defined on ${\cal H}_{kin}$ either).} and that of its commutator,
anomalous. Section 7.4 shows that, 
if one throws states of non- zero volume
out of the description, there does exist a smaller habitat on which the density 2 constraints are well defined
and their commutator anomaly free. However, even this is not completely satisfactory for two reasons:
(i) our aim
is to preserve contact with the physical states constructed in \cite{alokme2} (and reviewed in section 2.2.5), and 
the  states of non- zero volume are retained in their construction, 
(ii) our aim is to uncover lessons for LQG and in LQG,
a key role is played by states of non- vanishing volume in semiclassical considerations at the kinematic level
\cite{weavelit}. 

For these reasons, in this section we construct a new habitat where
the continuum limit of density two (Hamiltonian and diffeomorphism) constraint operators is well defined,
their Poisson- Lie algebra is faithfully represented and 
their kernel in 
this new habitat is precisely   the set of physical states of section 2.2.5.

In section 8.1 we define the new habitat. In section 8.2 we show that the diffeomorphism constraint at 
finite triangulation has a well defined continuum limit on the habitat and that its  commutator is anomaly free.
In section 8.3 we prove identical results for the density 2 Hamiltonian constraint.
In section 8.4 we show that the (joint) kernel of the density 2 Hamiltonian constraint (and the diffeomorphism constraint)
is precisely the set of physical states of section 2.2.5.

\subsection{The new habitat}

Given a pair of charge-networks $({\bf s}^{+}, {\bf s}^{-})$ let 
\begin{equation}
[{\bf s}^{+}, {\bf s}^{-}]_{+-}\ =\ \{ ({\bf s}^{+ '},{\bf s}^{- '})|\  ({\bf s}^{+ '},{\bf s}^{- '})\ =\ ({\bf s}^{+}_{\phi^{+}}, {\bf s}^{-}_{\phi^{-}})\ \textrm{for some}\ \phi^{\pm}\},
\end{equation}
so that $[{\bf s}^{+}, {\bf s}^{-}]_{+-}$  is the set of all charge networks related by the finite gauge transformations
generated by $H_+$ and $H_-$.
\footnote{
The astute reader will recognise that a marginally 
simpler treatement of the material in this section would ensue if we worked with 
$H_{\pm}$ and  the sets $[{\bf s}^{\pm}] $ of section 2.2.5,  and derived
the results for the density two constraints $H= H_+-H_-, C_{diff}= H_++ H_-$ as immediate consequences. 
The reason for our presentation of $H, C_{diff}$ (and hence, $[{\bf s}^{+}, {\bf s}^{-}]_{+-}$) as primary
structures is to preserve, as far as possible, structural similarity with LQG.}

We define the new habitat, ${\cal V}_{+-}$, as 
 the finite linear span of distributions (over ${\cal D}_{ss}$) of the type, 
\begin{equation}
\Psi_{f^{+},f^{-},[{\bf s}^{+},{\bf s}^{-}]_{+-}}\ =\  \sum_{({\bf s}^{+ '},{\bf s}^{- '})\in\ [{\bf s}^{+},{\bf s}^{-}]_{+-}}f^{+}(V_{E}^{S^{1}}({\bf s}^{+ '}))f^{-}(V_{E}^{S^{1}}({\bf s}^{- '}))
\langle {\bf s}^{+ '}, {\bf s}^{- '}|.
\end{equation}
Note that as claimed in section 7.1 the cardinality of the  sets 
$V_{E}^{S^{1}}({ s}^{+}),  V_{E}^{S^{1}}({ s}^{+}_{\phi^{+}})$ is identical for any $\phi^+$, a similar result being
true for the `$-$' sector and that each $v^{\pm}\in V_{E}^{S^{1}}({ s}^{\pm})$
has the unique image $(\phi^{\pm}( v^{\pm}))\in V_{E}^{S^{1}}({ s}^{\pm}_{\phi^{\pm}})$. 
This implies that we can uniquely define  the orbit of points in $V_{E}^{S^{1}}({ s}^{\pm})$
under the action of some 1 parameter family of gauge transformations $\phi^{\pm}$.
We shall implicitly use this fact in our considerations below.

\subsection{The diffeomorphism constraint and its commutator}

Since the computations here are very similar to those encountered in section 7.3.1, we shall be brief in our presentation.
Using the fact that finite diffeomorphisms are gauge tranformations (see equation (\ref{udiff})) in conjunction with 
equation (\ref{cdiff0}), it follows that 
\begin{equation}
\Psi_{f^{+},f^{-},[{\bf s}^{+\prime},{\bf s}^{-\prime}]_{+-}}
(\hat{C}_{diff}[\vec{N}]|{\bf s}^{+},{\bf s}^{-}\rangle)\ =\ 0
\;\;{\rm if} \;\;({\bf s}^{+},{\bf s}^{-})\notin [{\bf s}^{+\prime},{\bf s}^{-\prime}]_{+-}. 
\end{equation}
Next, it is straightforwad to see that from equation (\ref{cdiff0}) we have, in obvious notation,  
\begin{equation}
\lim_{\delta\rightarrow 0}\Psi_{f^+,f^-, [{\bf s^+},{\bf s^-}]_{+-}}({\hat C}_{diff, T(\delta)}(\vec{N})|{\bf s}^{+},{\bf s}^{-}\rangle)
= -i\hbar \sum_{v\in V_E(s^+)\cup V_E(s^-)} N^x(v) \partial_x(f^+f^-)|_{x=v}.
\label{cdiffnew1}
\end{equation}
It is easy to see that the above equation implies that 
$\Psi_{f^+,f^-, [{\bf s^+},{\bf s^-}]_{\pm}}\in {\cal V}_{+-}$ is mapped to the linear combination 
$(\Psi_{f^+,g^-_{\vec N}, [{\bf s^+},{\bf s^-}]_{+-}} +
\Psi_{g^+_{\vec N},f^-, [{\bf s^+},{\bf s^-}]_{+-}}) \in {\cal V}_{+-}$ where
\begin{equation}
g^{\pm}_{\vec N}= (-i\hbar) \sum_{v\in V_E(s^{\pm})} N^x(v) \partial_x(f^{\pm})|_{x=v}
\label{defgnew}
\end{equation}
It is then straightforward to compute  action of the commutator on the habitat state 
$\Psi_{f^+,f^-, [{\bf s^+},{\bf s^-}]_{+-}}$ along the lines of section 7.3.1 and verify that 
\begin{eqnarray}
&\Psi_{f^+, f^-, [{\bf s^+}^{\prime},{\bf s^-}^{\prime}]_{+-}}([{\hat C}_{diff}(\vec{N_1}),{\hat C}_{diff}(\vec{N_2}])|{\bf s}^{+},{\bf s}^{-}\rangle) &\nonumber \\
&=(-i\hbar )
\Psi_{f^+,f^-, [{\bf s^+}^{\prime},{\bf s^-}^{\prime}]_{+-}}({\hat C}_{diff}([\vec{N_2},\vec{N_1}] )
|{\bf s}^{+},{\bf s}^{-}\rangle),&
\label{diff,diffnew}
\end{eqnarray}
which is an antirepresentation of the corresponding Poisson brackets.

\subsection{The density 2 Hamiltonian constraint and its commutator}
The computations parallel that of the previous section.

It is easy to see that equation (\ref{hlm}) holds for any distribution $\Psi$, and in particular, for
$\Psi \in {\cal V}_{+-}$. Using the fact that the charge nets on the right hand side of 
equation (\ref{hlm}) are related by the action of finite gauge transformations, it follows that
\begin{equation}
\Psi_{f^{+},f^{-},[{\bf s}^{+\prime},{\bf s}^{-\prime}]_{+-}}
(\hat{H}[N]|{\bf s}^{+},{\bf s}^{-}\rangle)\ =\ 0
\;\;{\rm if} \;\;({\bf s}^{+},{\bf s}^{-})\notin [{\bf s}^{+\prime},{\bf s}^{-\prime}]_{+-}. 
\end{equation}
Next, it is straightforwad to see that from equation (\ref{hlm}) that we have, in obvious notation,  
\begin{eqnarray}
&\lim_{\delta\rightarrow 0}
\Psi_{f^+,f^-, [{\bf s^+},{\bf s^-}]_{+-}}({\hat H}_{ T(\delta)}({N})|{\bf s}^{+},{\bf s}^{-}\rangle)& \nonumber \\
&= -i\hbar \sum_{v\in V_E(s^+)\cup V_E(s^-)} 
N(v)(f^- \frac{\partial f^+}{\partial v}-f^+ \frac{\partial f^-}{\partial v})&
\label{hnew1}
\end{eqnarray}

It is easy to see that the above equation implies that 
$\Psi_{f^+,f^-, [{\bf s^+},{\bf s^-}]_{\pm}}\in {\cal V}_{+-}$ is mapped to the linear combination 
$(\Psi_{f^+,h^-_{ N}, [{\bf s^+},{\bf s^-}]_{+-}} +
\Psi_{h^+_{N},f^-, [{\bf s^+},{\bf s^-}]_{+-}}) \in {\cal V}_{+-}$ where
\begin{equation}
h^{\pm}_{N}= (-i\hbar) \sum_{v\in V_E(s^{\pm})} \pm N(v) \frac{\partial f^{\pm}}{\partial v}
\label{defhnew}
\end{equation}
It is then straightforward to compute  action of the commutator on the habitat state 
$\Psi_{f^+,f^-, [{\bf s^+},{\bf s^-}]_{+-}}$ along the lines of section 7.3.1 (or 7.4) and verify that 
\begin{eqnarray}
&\Psi_{f^+, f^-, [{\bf s^+}^{\prime},{\bf s^-}^{\prime}]_{+-}}([{\hat H}(\vec{N_1}),{\hat H}(\vec{N_2}])|{\bf s}^{+},{\bf s}^{-}\rangle) &\nonumber \\
&=(-i\hbar )
\Psi_{f^+,f^-, [{\bf s^+}^{\prime},{\bf s^-}^{\prime}]_{+-}}({\hat C}_{diff}([\vec{N_2},\vec{N_1}] )
|{\bf s}^{+},{\bf s}^{-}\rangle),&
\label{h,hnew}
\end{eqnarray}
which is an antirepresentation of the corresponding Poisson brackets (Recall that the density weight -1 lapses can
equally well be thought of as vector fields). 

It is also easy, using equations 
(\ref{cdiffnew1}),  (\ref{hnew1}), (\ref{defgnew}), (\ref{defhnew}), to see that the Poisson bracket between the
diffeomorphism constraint and the density weight 2 Hamiltonian constraint is represented faithfully on the new
habitat.

\subsection{The kernel of the density 2 constraints on the new habitat.}

{\bf Lemma} :  Given any state $\Psi_{f^{+},f^{-},[{\bf }s^{+},{\bf s}^{-}]_{+-}}$ it will be in the kernel of $\hat{H}[N]\ \forall\ N$ iff $f^{+},\ f^{-}$ are constant functions.\\

{\bf Proof} : ``If side'' is trivial in light of (\ref{hnew1}), we now prove the ``only-if'' side.\\

Let $\Psi_{f^{+}, f^{-}, [{\bf s}^{+ '},{\bf s}^{- '}]_{+-}}$ 
be in the kernel of $\hat{H}[N]$ $\forall\ N$. 
Notice that given any $[{\bf s}^{+ '},{\bf s}^{- '}]_{+-}$, there exists infinitely many $({\bf s}^{+},{\bf s}^{-})$ such that\\
\noindent {(i)} $V_{E}({\bf s}^{+})\cap V_{E}({\bf s}^{-})\ =\ \Phi$ , where $\Phi$ denotes an empty set and\\
\noindent {(ii)} $[{\bf s}^{+ '},{\bf s}^{- '}]_{+ -}\ =\ [{\bf s}^{+},{\bf s}^{-}]_{+ -}$.\\
We choose one such $({\bf s}^{+},{\bf s}^{-})$. As 
\begin{equation}
\begin{array}{lll}
 \Psi_{f^{+}, f^{-}, [{\bf s}^{+},{\bf
 s}^{-}]_{+-}}\big(\hat{H}[N]|{\bf s}^{+},{\bf s}^{-}\rangle\big)\ =\
 (-i\hbar)\sum_{v\in V_{E}({\bf s}^{+})\cup V_{E}({\bf
 s}^{-})}N(v)(f^{-}\frac{\partial f^{+}}{\partial v}\ -\
 f^{+}\frac{\partial f^{-}}{\partial v}\ \big)\\
\vspace*{0.1in}
=(-i\hbar)\big(\ \sum_{v\in V_{E}({\bf s}^{+})}N(v)\ f^{-}(V_{E}({\bf
   s}^{-}))\frac{\partial f^{+}}{\partial v}\ -\ \sum_{v^{'}\in
   V_{E}({\bf s}^{-})}N(v^{'})\ f^{+}(V_{E}({\bf
  s}^{+}))\frac{\partial f^{-}}{\partial v^{'}}\ \big)\\
\vspace*{0.1in}
\hspace*{3.7in}=\ 0
\end{array}
\end{equation}

As above equation is true for all $N$, it implies that $f^{\pm}$ are
constant in the neighbourhood of each vertex $v\in V_{E}({\bf
  s}^{+})\cup V_{E}({\bf s}^{-})$.\\

Now consider the set $\{({\bf s}^{+}_{\phi},{\bf s}^{-}_{\phi})\}$ for all periodic diffeomorphisms $\phi$ which do not keep $({\bf s}^{+},{\bf s}^{-})$ invariant.\\

As, $V_{E}({\bf s}^{+}_{\phi})\cap V_{E}({\bf s}^{-}_{\phi})\ =\ \{\Phi\}$ and as $({\bf s}^{+}_{\phi},{\bf s}^{-}_{\phi})\ \in\ [{\bf s}^{+},{\bf s}^{-}]_{+-}$ for all $\phi$, 

\begin{equation}
 \Psi_{f^{+}, f^{-}, [{\bf s}^{+},{\bf s}^{-}]_{+-}}\big(\hat{H}[N]|{\bf s}^{+}_{\phi},{\bf s}^{-}_{\phi}\rangle\big)\ =\ 0
\end{equation}
for all $N$ implies that,\\
$\frac{\partial f^{\pm}(V_{E}^{S^{1}}({\bf s}^{\pm}_{\phi}))}{\partial \phi\cdot v^{\pm}}\ =\ 0\ \forall\ (\phi\cdot v^{\pm})\in\ V_{E}({\bf s}^{\pm}_{\phi})$. These conditions for all diffeomorphisms $\phi$ show that
$f^{\pm}$ are constant everywhere.\\
This completes the proof.

\section{Discussion}

A key open issue in canonical LQG relates to the definition of the 
Hamiltonian constraint operator. This operator is constructed as the continuum
limit of its finite triangulation approximant. The latter is the quantum 
correspondent of a classical approximant which is uniquely defined only upto
terms which vanish in the classical continuum limit of infinitely fine triangulation.
In contrast to the classical continuum limit, the continuum limit of the quantum operator
is not independent of the choice of finite triangulation approximant thus implying an 
unacceptable (infinitely manifold) choice in the definition of the quantum dynamics of LQG.
On the other hand, a necessary condition for the very consistency of the quantum theory is an anomaly free
representation of the constraint algebra. Therefore, one possible way to restrict the choice of quantum dynamics
is to demand that the ensuing algebra of quantum constraints is free from anomalies.
Unfortunately, 
irrespective of the specific choice of quantum dynamics made in the current state of art in LQG, 
the quantum constraint algebra 
trivialises i.e. the commutator of a pair of Hamiltonian constraints as well as the operator corresponding to
their classical Poisson bracket vanish.
The question then is: Is there any way out i.e. is it still possible to use the anomaly free requirement on the 
constraint algebra to single out a (hopefully almost) unique choice  for the quantum dynamics?
Below we argue that the PFT results derived in this work suggest a strategy to answer this question.

Let us first summarise the state of art in LQG in more technical terms. In what follows we shall refer to the 
commutator between a pair of Hamiltonian constraints as the ``Left Hand Side'' (LHS) and the operator correspondent 
of their Poisson bracket (which is proportional to the diffeomorphism constraint) as the ``Right Hand Side'' (RHS).

The quantum dynamics of LQG was first defined rigorously in the seminal work of Thiemann \cite{tthh} in terms of
the smeared density weight one Hamiltonian constraint. The continuum limit of its action is defined either with respect
to the URS topology \cite{tthh} or directly on an appropriate habitat of distributions \cite{donjurek}. In both cases the
continuum limit of the Hamiltonian constraint operator is well defined and its commutator (i.e. the LHS)
trivialises as does the RHS \cite{tthh}. In the URS topology the RHS vanishes ``doubly'', first due to a  factor of
`$\delta$' and second by virtue of the fact that the RHS is proportional to the diffeomorphism constraint.
The RHS vanishes ``doubly'' also on the LM habitat, first due to an overall factor of $\delta$ and second due to the 
absence of an additional factor of $\delta^{-1}$ which could have converted 
the difference in the evaluation of vertex smooth functions at points seperated by $\delta$ into a 
a derivative in the continuum limit (here $\delta$ is a parameter which measures the fineness of the triangulation,
$\delta \rightarrow 0$ being the continuum limit). As we have seen, {\em exactly} the same situation prevails in PFT.

In PFT additional factors of $\delta^{-1}$ can be introduced by replacing the density 1 constraints by density 2 
constraints. As we have seen in sections 7.4 and section 8, this leads to a non- trivial representation of the 
ensuing constraint algebra on appropriate spaces of distributions. The lesson we draw from this is that the choice of 
density 1 constraints in PFT and LQG hides the underlying non- triviality of the constraint algebra. The key issue 
is then: Can we handle the algebra of appropriately chosen higher density weight Hamiltonian constraints in LQG?
We first discuss the RHS and then the LHS.\\

\noindent {\bf The RHS}: In PFT the RHS corresponding to the Poisson bracket between a pair of density 2 Hamiltonian 
constraints is just the diffeomorphism constraint smeared with a c- number, metric independent shift
constructed out of the lapses. In the terminology used in LQG, the RHS can no longer be defined as a finite operator
on the kinematic Hilbert space. Nevertheless, as we have seen, the operator is perfectly well defined on the LM habitat
in terms of (Lie) derivatives of vertex smooth functions. Moreover the commutator between a pair of diffeomorphism 
constraints is anomaly free on this habitat. We take this is as indicative of being on the right track with reference to 
the 
definition of the various finite triangulation approximants to the local fields which comprise the constraint.
In LQG most of the ambiguities in the Hamiltonian constraint arise from those involved in the choice of finite 
triangulation approximant to the curvature, $F_{ab}^i$, of the Ashtekar- Barbero connection.
The question then arises as to whether we can define the curvature operator at finite triangulation in such a way
that the diffeomorphism constraint (smeared with a c- number shift) has a well definded continuum limit on the 
LM habitat in such a way that 
the Poisson- Lie algebra of diffeomorphism constraints is represented in an anomaly free manner
on the LM habitat. This is the subject of work now in progress which suggests that the answer may indeed be in the 
affirmative. That we have even contemplated such a possibility is already evidence of the usefulness of PFT.\\

\noindent {\bf The LHS}:
In PFT the density 2 Hamiltonian constraint does not admit a continuum limit (both with respect to the URS topology as
well as on the LM habitat). Nevertheless, a sign that the strategy of using density 2 constraints may be a profitable 
one is provided by the existence of the zero volume habitat of section 7.4 which is closely related to the 
LM habitat. The final solution requires the new habitat of section 8 which is geared to the physical state space
of section 2.2.5.  In LQG one can check that, modulo some subtelities, if the Hamiltonian constraint is rescaled by 
a factor of the determinant of the metric to the power $\frac{1}{6}$, the RHS obtains 2 factors of $\delta^{-1}$
which then could perhaps yield a nontrivial action of the RHS on the LM habitat. However these higher density
(smeared) Hamiltonian constraints are not themselves well defined on the LM habitat by virtue of the extra factor
of $\delta^{-1}$ at finite triangulation. Further, as indicated in the beautiful analysis of Reference \cite{lmhabitat2},
wherein the authors simply rescale the action of the density weight 1 Hamiltonian constraint and attempt to evaluate the 
action of rescaled commutator on the habitat, there is no way to obtain the diffeomorphism generated by the RHS
unless the Hamiltonian constraint moves the vertices of the state on which it acts. The current proposals for the 
Hamiltonian constraint do not involve movement of vertices and this can again be traced to the inadequacy of the choice
of approximant to $F_{ab}^i$ at finite triangulation. 

In PFT despite the fact that the density 2 Hamiltonian constraint does move vertices, the LHS is not well defined
on ${\cal V}_{LM}$. However it is well defined on ${\cal V}^0_{LM}$ or ${\cal V}_{+-}$. The lesson we draw from this 
is to \\
\noindent (i) search for a better finite triangulation approximant to $F_{ab}^i$ which can move vertices around; our work
in progress \cite{alokmelqg} should feed into this.

\noindent (ii) search for some analog of ${\cal V}^0_{LM}$ or ${\cal V}_{+-}$. Of course 
${\cal V}_{+-}$ could be constructed precisely because we already know the correct physical state space through
Reference \cite{alokme2}. In LQG the physical state space is presumably only known once the Hamiltonian constraint 
is defined so the situation is far more involved. However, the aim, at least is clear, namely that we need a satisfactory
habitat and a definition of the hamiltonian constraint such that the higher density constraints are well defined on this
habitat and the constraint algebra, anomaly free. A more modest question, relevant to see if a candidate definition
of the constraint operator may be viable, is to look for the analog of ${\cal V}^0_{LM}$. The direct analog 
does not help at all because, in contrast to our PFT construction (see footnote 9),
 the present construction of the ``inverse volume'' operator in LQG is such that all the 
zero volume states are also annihilated by the inverse volume operator. 
A natural question is: Is it possible to find
an alternate construction of the inverse in LQG so that it does  not annihilate zero volume states?
This also brings us to a natural question in PFT: Can the restriction to zero volume 
states still yield a physically sensible theory? This actually may be the case because 
(a) zero volume states are gauge related to non- zero volume ones and (b) as can easily be checked, 
the Dirac observables of \cite{alokme2} preserve the space of zero volume charge nets.

In conclusion, while the situation in LQG is far more complicated, we are convinced that the 
structures which 
permit the construction of a non- trivial representation of the constraint algebra of PFT (higher density constraints,
alternate habitats, operator definitions through an analysis of Hamiltonian vector fields, holonomies in representations
attuned to the edge labels of the state on which they act, 
the coexistence of discontinuous unitary operators on the kinematic Hilbert space with the well definedness of the action
of their generators on a suitable space of distributions) will 
open up new directions with regard to the problem of a 
consistent definition of the quantum dynamics of LQG.\\

\noindent{\bf Acknowledgements}: We are indebted to Abhay Ashtekar for his constant encouragement. One of us (AL) thanks Prof. Seshadri and Prof. Sreedhar at Chennai Mathematical Institute for the kind hospitality during summer of 2010 where part of this work was done. Work of AL was supported in part by NSF grant, Phy-08-54743 and by the 
Eberly Endowment fund.

\section*{Appendix}
\appendix
\section{Spectrum of the inverse metric operator}
Our starting point is equation (\ref{eq:may21-1}). We adopt the following notation in this section.
Let $y_0$ be located at the vertex $v$ of the graph (which is naturally associated with the triangulation) $T$. 
We remind the reader that we have set 
$\gamma (s^{\pm}) = T$. 
The charges $k^{\pm}_{e_v}, k^{\pm}_{e^v}$ are defined
as in section 4.1  (see the discussion after equation (\ref{jun3-3b})).
Let the edge (i.e. the 1 simplex of the triangulation) which ends at $v$ be $\triangle$. 
If $v=0$, $\triangle$ denotes the edge which ends at $2\pi$. 
We shall denote signum operator at the vertex $v$ by  $\widehat{\rm{sgn}}(v)$ i.e.
$\widehat{\rm{sgn}}(v):=\widehat{{\rm{sgn}}X^{+\prime}X^{-\prime}}(v)$.
It is also convenient to change the notation of embedding charge network states. We shall denote
the  charge network state 
$T_{s^{\pm}}$ by $\vert s^{\pm}\rangle$ and, whenever required, expand out the charge network label
$s^{\pm}$ in terms of its defining data i.e. its underlying graph and charge labels (see section 2.2.1 and 
\cite{alokme2}). Finally, in this section we choose units in which $\hbar = a =1$.

As noted in section 4.2, 
the operator
$\hat{h}^{(\pm) -1}_{e_{x_{1},y_{0}}}\left[\hat{V}_{y_{0}},\hat{h}^{(\pm)}_{e_{x_{1},y_{0}}}\right]$ is only sensitive to
that part of $e_{x_{1},y_{0}}$ which overlaps with $\triangle$. Setting $y_0= v$ and 
expanding the double commutators in the right hand side of equation (\ref{eq:may21-1}), it is straightforward to see that,
\begin{equation}\label{eq:may24-2}
\begin{array}{lll}
\widehat{\frac{1}{\sqrt{|X^{+ '} X^{- '}|}}}(v)|_T(|s^{+}\rangle\otimes|s^{-}\rangle)\ =\\
\vspace*{0.1in}
-|\triangle|\hat{h}^{(-) -1}_{\triangle}\widehat{\rm{sgn}}(v)\\
\vspace*{0.1in}
\hspace*{0.2in}\big[\hat{h}_{\triangle}^{(+)}\hat{V}_{v}\hat{h}_{\triangle}^{(+) -1}\hat{h}_{\triangle}^{(-)}-
\hat{h}_{\triangle}^{(-)}\hat{h}_{\triangle}^{(+)}\hat{V}_{v}\hat{h}_{\triangle}^{(+) -1} -
\hat{h}_{\triangle}^{(+) -1}\hat{V}_{v}\hat{h}_{\triangle}^{(+)}\hat{h}_{\triangle}^{(-)}+ 
\hat{h}_{\triangle}^{(-)}\hat{h}_{\triangle}^{(+) -1}\hat{V}_{v}\hat{h}_{\triangle}^{(+)}\big]\\
\hspace*{4.5in}|s^{+}\rangle\otimes|s^{-}\rangle\\
\vspace*{0.1in}
+|\overline{\triangle}|\hat{h}^{(+)}_{\overline{\triangle}}\widehat{\rm{sgn}}(v)\\
\vspace*{0.1in}
\hspace*{0.2in}\big[\hat{h}_{\triangle}^{(-)}\hat{V}_{v}\hat{h}_{\triangle}^{(-) -1}\hat{h}_{\triangle}^{(+) -1}-
\hat{h}_{\triangle}^{(+) -1}\hat{h}_{\triangle}^{(-)}\hat{V}_{v}\hat{h}_{\triangle}^{(-) -1}-
\hat{h}_{\triangle}^{(-) -1}\hat{V}_{v}\hat{h}_{\triangle}^{(-)}\hat{h}_{\triangle}^{(+) -1}+ 
\hat{h}_{\triangle}^{(+) -1}\hat{h}_{\triangle}^{(-) -1}\hat{V}_{v}\hat{h}_{\triangle}^{(-)}\big]\\
\hspace*{4.5in}|s^{+}\rangle\otimes|s^{-}\rangle
\end{array}
\end{equation}
Since the edge $\triangle$ ends at $v$, the action of  the embedding holonomies in the expression above is to change 
$k^{\pm}_{e_{v}}$ by unity. It is then
straightforward to obtain:
\begin{equation}
\begin{array}{lll}
\widehat{\frac{1}{\sqrt{|X^{+ '} X^{- '}|}}}(v)|_T(|s^{+}\rangle\otimes|s^{-}\rangle)\ =\\
\vspace*{0.1in}
-|\triangle|\hat{h}^{(-) -1}_{\triangle}\widehat{\rm{sgn}}(v)\\
\vspace*{0.1in}
\Big[\big[\ \sqrt{|k^{+}_{e^{v}}-(k^{+}_{e_{v}}-1)||k^{-}_{e^{v}}-(k^{-}_{e_{v}}+1)|}-\sqrt{|k^{+}_{e^{v}}-(k^{+}_{e_{v}}-1)||k^{-}_{e^{v}}-k^{-}_{e_{v}}|}\ \big]\\
\vspace*{0.1in}
\hspace*{0.2in} -\big[\ \sqrt{|k^{+}_{e^{v}}-(k^{+}_{e_{v}}+1)||k^{-}_{e^{v}}-(k^{-}_{e_{v}}+1)|}-\sqrt{|k^{+}_{e^{v}}-(k^{+}_{e_{v}}+1)||k^{-}_{e^{v}}-k^{-}_{e_{v}}|}\ \big]
\big]\\
\vspace*{0.1in}
\hspace*{0.5in}|s^+ \rangle \otimes
\vert \gamma (s^-)=T , (...k_{e_{v}}^{-},k_{e_{v}}^{-}+1,k^{-}_{e^{v}},...)\rangle\\
\vspace*{0.1in}
+|\triangle|\hat{h}^{(+)}_{\triangle}\widehat{\rm{sgn}}(v)\\
\vspace*{0.1in}
\Big[\big[\ \sqrt{|k^{+}_{e^{v}}-(k^{+}_{e_{v}}-1)||k^{-}_{e^{v}}-(k^{-}_{e_{v}}-1)|}-\sqrt{|k^{+}_{e^{v}}-(k^{+}_{e_{v}})||k^{-}_{e^{v}}-(k^{-}_{e_{v}}-1)|}\ \big]\\
\vspace*{0.1in}
\hspace*{0.2in} -\big[\ \sqrt{|k^{+}_{e^{v}}-(k^{+}_{e_{v}}-1)||k^{-}_{e^{v}}-(k^{-}_{e_{v}}+1)|}-\sqrt{|k^{+}_{e^{v}}-k^{+}_{e_{v}}||k^{-}_{e^{v}}-(k^{-}_{e_{v}}+1)|}\ \big]\Big]
\\
\vspace*{0.1in}
\hspace*{0.5in} |\gamma (s^+)=T,(...k_{e_{v}}^{+},k_{e_{v}}^{+}-1,k^{+}_{e^{v}},...)\rangle
\otimes \vert s^- \rangle
\end{array}
\end{equation}
Using
\begin{equation}
\begin{array}{lll}
\widehat{\rm{sgn}(X^{+ '}X^{- '})}(v)|s^{+}\rangle\otimes|s^{-}\rangle = \rm{sgn}(k^{+}_{e^{v}}-k^{+}_{e_{v}})\rm{sgn}(k^{-}_{e^{v}}-k^{-}_{e_{v}})|s^{+}\rangle\otimes|s^{-}\rangle
\end{array}
\end{equation}
we get,
\begin{equation}
\begin{array}{lll}
\widehat{\frac{1}{\sqrt{|X^{+ '} X^{- '}|}}}(v)|_T(|s^{+}\rangle\otimes|s^{-}\rangle)\ =\\
\vspace*{0.1in}
-\textrm{sgn}(k^{+}_{e^{v}}-k^{+}_{e_{v}})\textrm{sgn}(k^{-}_{e^{v}}-(k^{-}_{e_{v}}+1))\\
\vspace*{0.1in}
\Big[\big[\ \sqrt{|k^{+}_{e^{v}}-(k^{+}_{e_{v}}-1)||k^{-}_{e^{v}}-(k^{-}_{e_{v}}+1)|}-\sqrt{|k^{+}_{e^{v}}-(k^{+}_{e_{v}}-1)||k^{-}_{e^{v}}-k^{-}_{e_{v}}|}\ \big]\\
\vspace*{0.1in}
\hspace*{0.2in} -\big[\ \sqrt{|k^{+}_{e^{v}}-(k^{+}_{e_{v}}+1)||k^{-}_{e^{v}}-(k^{-}_{e_{v}}+1)|}-\sqrt{|k^{+}_{e^{v}}-(k^{+}_{e_{v}}+1)||k^{-}_{e^{v}}-k^{-}_{e_{v}}|}\ 
\big]\Big]
|s^{+}\rangle\otimes|s^{-}\rangle\\
\vspace*{0.1in}
+\textrm{sgn}(k^{+}_{e^{v}}-(k^{+}_{e_{v}}-1))sgn(k^{-}_{e^{v}}-k^{-}_{e_{v}})\\
\vspace*{0.1in}
\Big[\big[\ \sqrt{|k^{+}_{e^{v}}-(k^{+}_{e_{v}}-1)||k^{-}_{e^{v}}-(k^{-}_{e_{v}}-1)|}-\sqrt{|k^{+}_{e^{v}}-(k^{+}_{e_{v}})||k^{-}_{e^{v}}-(k^{-}_{e_{v}}-1)|}\ \big]\\
\vspace*{0.1in}
\hspace*{0.2in} -\big[\ \sqrt{|k^{+}_{e^{v}}-(k^{+}_{e_{v}}-1)||k^{-}_{e^{v}}-(k^{-}_{e_{v}}+1)|}-\sqrt{|k^{+}_{e^{v}}-k^{+}_{e_{v}}||k^{-}_{e^{v}}-(k^{-}_{e_{v}}+1)|}\ \big]\Big]
|s^{+}\rangle\otimes|s^{-}\rangle
\end{array}
\end{equation}
Whence,
\begin{equation}\label{evalueinverse}
\begin{array}{lll}
\lambda (s^+, s^-, v)\ =\\
\vspace*{0.1in}
-\textrm{sgn}(k^{+}_{e^{v}}-k^{+}_{e_{v}})sgn(k^{-}_{e^{v}}-(k^{-}_{e_{v}}+1))\\
\Big[\big[\sqrt{|k^{+}_{e^{v}}-(k^{+}_{e_{v}}-1)||k^{-}_{e^{v}}-(k^{-}_{e_{v}}+1)|}-\sqrt{|k^{+}_{e^{v}}-(k^{+}_{e_{v}}-1)||k^{-}_{e^{v}}-k^{-}_{e_{v}}|}\big]\\
\vspace*{0.1in}
\hspace*{0.2in} -\big[\sqrt{|k^{+}_{e^{v}}-(k^{+}_{e_{v}}+1)||k^{-}_{e^{v}}-(k^{-}_{e_{v}}+1)|}-\sqrt{|k^{+}_{e^{v}}-(k^{+}_{e_{v}}+1)||k^{-}_{e^{v}}-k^{-}_{e_{v}}|}\big]\Big]\\
\vspace*{0.1in}
+\textrm{sgn}(k^{+}_{e^{v}}-(k^{+}_{e_{v}}-1))\textrm{sgn}(k^{-}_{e^{v}}-k^{-}_{e_{v}})\\
\Big[\big[\sqrt{|k^{+}_{e^{v}}-(k^{+}_{e_{v}}-1)||k^{-}_{e^{v}}-(k^{-}_{e_{v}}-1)|}-\sqrt{|k^{+}_{e^{v}}-(k^{+}_{e_{v}})||k^{-}_{e^{v}}-(k^{-}_{e_{v}}-1)|}\big]\\
\vspace*{0.1in}
\hspace*{0.2in} -\big[\sqrt{|k^{+}_{e^{v}}-(k^{+}_{e_{v}}-1)||k^{-}_{e^{v}}-(k^{-}_{e_{v}}+1)|}-\sqrt{|k^{+}_{e^{v}}-k^{+}_{e_{v}}||k^{-}_{e^{v}}-(k^{-}_{e_{v}}+1)|}\big]\Big]
\end{array}
\end{equation}

\end{document}